\documentclass[a4paper,12pt]{article}

\pdfoutput=1 

\usepackage[hmargin=.7in,vmargin=1.1in]{geometry}
\usepackage{indentfirst}
\usepackage{amsfonts}
\usepackage{mathrsfs}
\usepackage{amsmath}
\usepackage{amssymb}
\usepackage{authblk}
\usepackage{cite}
\usepackage{xcolor}
\usepackage{mathtools}
\usepackage{tensor}
\usepackage{graphicx}
\usepackage{bm}
\usepackage{upgreek}
\usepackage{braket}
\usepackage{color,soul}
\usepackage{csquotes}
\usepackage{caption}
\usepackage{subcaption}
\usepackage[section]{placeins}

\newcommand\ii{\mathcal{I}}
\newcommand\ca{\mathcal{A}}
\newcommand\mi{\mathrm{i}}
\newcommand\me{\mathrm{e}}
\newcommand\const{\text{const}}

\newcommand\pp{\uppi}
\newcommand{\dif}{\mathrm{d}}

\DeclareMathOperator{\diag}{diag}

\def\boxit#1{%
  \smash{\color{red}\fboxrule=1pt\relax\fboxsep=2pt\relax%
  \llap{\rlap{\fbox{\vphantom{0}\makebox[#1]{}}}~}}\ignorespaces
}

\usepackage[bookmarksnumbered=true,bookmarksopen=true]{hyperref}
 \hypersetup{colorlinks,
             linkcolor=[rgb]{0,0.3,0.6}, 
             citecolor=[rgb]{0,0.3,0.6}, 
             urlcolor=[rgb]{0,0.3,0.6}}

\title{\Large\textbf{A regular black hole as the final state of evolution of a singular black hole}}

\author[1]{Han-Wen Hu\thanks{huhanwen@mail.nankai.edu.cn}}
\author[2]{Chen Lan\thanks{stlanchen@126.com}}
\author[1]{Yan-Gang Miao\thanks{Corresponding author, miaoyg@nankai.edu.cn.}}

\affil[1]{\normalsize{\em School of Physics, Nankai University, 94 Weijin Road, Tianjin 300071, China.}}		
\affil[2]{\normalsize{\em Department of Physics, Yantai University, 30 Qingquan Road, Yantai 264005, China.}}

\date{}

\begin{document}

\maketitle

\begin{abstract}
We propose a novel black hole model in which singular and regular black holes are combined as a whole and more precisely singular and regular black holes are regarded as different states of parameter evolution. 
We refer to them as {\em singular} and {\em regular states}, respectively. 
Furthermore, the regular state is depicted by the final state of parameter evolution in the model. 
We also present the sources that can generate such a black hole spacetime in the framework of $F(R)$ gravity. 
This theory of modified gravity is adopted because it offers a possible resolution to a tough issue in the thermodynamics of regular black holes,
namely the discrepancy between the thermal entropy and Wald entropy.
The dynamics and thermodynamics of the novel black hole model are also discussed when a singular state evolves into a regular state during the change of charge or horizon radius from its initial value to its extreme value.
\end{abstract}

\tableofcontents

\section{Introduction}
\label{sec:intr}

Regular black holes (RBHs) are a kind of spacetime that includes \cite{Dymnikova:1992ux,bogojevic:1998ma,Hayward:2005gi,Bronnikov:2006fu,Ansoldi:2008jw,Nicolini:2008aj,Frolov:2016pav} only coordinate singularities as black hole (BH) horizons. 
They can be regarded as either classical objects or quantum corrections to classical singular black holes (SBHs).
As classical objects, RBHs are solutions of Einstein's equations coupled with matters or gravitational equations of a modified gravity \cite{Berej:2006cc}, where the matter source could be phantom fields \cite{Bronnikov:2005gm}, magnetic monopoles in nonlinear electrodynamics \cite{Ayon-Beato:1998hmi,Bronnikov:2000vy,Ayon-Beato:2004ywd,Balart:2014cga,Fan:2016hvf,Toshmatov:2018cks,Vrba:2019vqh,Vrba:2020ijv}, or a combination of more complex substances \cite{Bronnikov:2021uta}. 
There also exists a conception that RBHs are the consequence of certain quantum effects \cite{Greenwood:2008ht,Wang:2009ay,Modesto:2011kw,Saini:2014qpa}, such as the loop quantum gravity \cite{Ashtekar:2023cod}, the gravitational theory with asymptotic safety \cite{Bonanno:2000ep}, etc., in which RBHs are not necessary to satisfy gravitational equations.

The definition of an RBH is twofold. One is based on the coordinate-independent approach, 
known as {\em finiteness of curvature invariants} \cite{Dymnikova:1992ux,Ayon-Beato:1998hmi,Bronnikov:2000vy}, i.e., the RBH is defined as a BH spacetime with finite curvature invariants. This definition is related to Markov's limited curvature conjecture \cite{Markov:1982ld}. 
The other is based on the coordinate-dependent approach, 
called {\em geodesic completeness} \cite{Carballo-Rubio:2019fnb,Carballo-Rubio:2021wjq}, 
i.e., the RBH is defined as a BH spacetime with complete null and timelike geodesics.
These two definitions are not equivalent generally, which means that some models satisfy one of the definitions but violate the other~\cite{Geroch:1968ut,Olmo:2015bya}.

In addition, we have to clarify two concepts of geodesic completeness so that the reader can understand our subsequent discussions unambiguously.
At first, the ``completeness'' implies that a test (null or timelike) particle  arriving at any point of spacetime requires an infinite affine parameter.
Secondly, the ``coordinate dependence'' means that it is possible for the affine parameter to be finite for a certain choice of coordinates even if this spacetime is regular everywhere.

In order to clarify specifically the ``coordinate dependence'' mentioned above, we take two examples. One is the Rindler spacetime which can be described \cite{Wald:1984rg,Carroll:2004st} by the line element, $\dif s^2 = -z^2\dif t^2 +\dif x^2+\dif y^2 +\dif z^2$.
The affine parameter is finite as a test particle approaches to $z=0$, i.e., it seems that the Rindler spacetime is ``incomplete'' at $z=0$ in this choice of coordinates. 
Nevertheless, it is known that the Rindler spacetime is a part of the Minkowski spacetime,
thus it cannot include any incomplete points. 
In fact, the Rindler spacetime will convert to the Minkowski spacetime via an appropriate choice of coordinates.
The other example is the Hayward BH spacetime  \cite{Zhou:2022yio}, where
the affine parameter is finite at $r=0$ in a certain choice of coordinates, 
seeming that the Hayward BH is ``geodesic incomplete". 
However, the Hayward BH is regular  if the Painlev\'e–Gullstrand coordinate is chosen.  In Sec.\ \ref{sec:geodesic}, we will demonstrate this point and note that such a coordinate should be adopted when we analyze the geodesic completeness at the center of RBHs with the spherical symmetry and one shape function.

The differences in dynamics and/or thermodynamics between RBHs and SBHs are a topic of great interest in the field of RBHs \cite{Bronnikov:2012ch,Fernando:2012yw,Flachi:2012nv,Toshmatov:2015wga,Abdujabbarov:2016hnw,Kumar:2019pjp,Stuchlik:2019uvf,Liu:2020ola,Myung:2007av,Lan:2021ngq,Konoplya:2022hll,Vagnozzi:2022moj,Konoplya:2023aph}.
By investigating these differences, one expects to study the effects of singularities on the dynamics and/or thermodynamics of BHs. 
In addition, RBHs are generally considered to be the product of quantum gravity. From this point of view, if a BH with a singularity at the center is regarded as a classical state, the BH evolves to its final state through self-heating radiation, 
where the final state is a quantum state without a singularity at the center. 
That is to say, RBHs can be regarded as the final state of parameter evolution of SBHs.
To this end, we construct a novel model which combines RBHs and SBHs into one whole. This proposal can highlight the differences in dynamics and/or thermodynamics between RBHs and SBHs, where
an RBH appears as the final state of parameter evolution of an SBH.
It provides a basis for the study of singularities' effects when an SBH evolves into an RBH.
The idea to combine RBHs and SBHs into one whole is essentially similar to the situation in  Refs.~\cite{Fan:2016hvf,Toshmatov:2018cks,Vrba:2019vqh,Vrba:2020ijv}. In these references, the BHs are in singular states when $M\neq 0$ and in regular states when $M=0$, i.e., the entire contribution to the ADM mass of BHs comes from electromagnetic interactions.

The paper is organized as follows.
At first, we prove in Sec.\ \ref{sec:criterion} that it is equivalent to determine RBHs by using complete geodesics or finite curvature invariants for the BHs that are spherically symmetric and have only one shape function.
Then, in 
Sec.\ \ref{sec:construction} we propose a spherically symmetric BH model with one shape function, which combines singular and regular BHs as a whole. 
In particular, an RBH can be regarded as the final state of an SBH when the charge or horizon radius goes to its extreme value.

The matter sources that generate such a BH are established in terms of modified gravity  $F(R)$ rather than Einstein's gravity in Sec.\ \ref{sec:sources}.
The motivation comes from the need for the thermodynamics of RBHs, i.e., we want to make the Wald entropy in $F(R)$ gravity consistent with the entropy calculated by the normal thermodynamic formula, $\displaystyle\int \dif M/T$, where $M$ and $T$ are BH mass and temperature, respectively.

As physical applications, we investigate the dynamics and thermodynamics of our BH model in 
Sec.\ \ref{sec:qnf}
and Sec.\ \ref{sec:thermody}, respectively. 
The purpose of these two sections is to analyze the essential changes in physical properties that take place during the evolution of an SBH into an RBH.
In addition, as a by-product, we also discuss the differences in the interpretation of RBHs under $F(R)$ gravity and Einstein's gravity.

The conclusions with future outlooks are summarized in Sec.\ \ref{sec:conclusion}, 
whereas App.\ \ref{app:monodromy} includes the calculations of monodromy that are applied in Sec.\ \ref{sec:aqnf} and App.\ \ref{sec:Analytical-matter} gives the actions of matter sources Eq.~(\ref{eq:action-matter}) explicitly.

\section{Criterion of regular black holes}
\label{sec:criterion}

The condition of finite curvature invariants is widely applied to determine~\cite{Dymnikova:1992ux,Ayon-Beato:1998hmi,Bronnikov:2000vy,Balart:2014cga,Fan:2016hvf} 
whether a BH is regular or not, 
which relates to the Markov limiting curvature conjecture \cite{Markov:1982ld,Frolov:1988vj,Frolov:2016pav,Frolov:2021afd}. 
Moreover, the other condition to ensure a non-singular spacetime is the completeness~\cite{Hawking:1973uf,Wald:1984rg} of timelike and null geodesics. 
These two conditions are {\em generally} not equivalent \cite{Geroch:1968ut,Olmo:2015bya,Carballo-Rubio:2019fnb,Carballo-Rubio:2021wjq,Zhou:2022yio}.
In this section, we shall show that the finiteness of curvature invariants and the completeness of geodesics are equivalent for 
 the spherically symmetric BHs with one shape function.

\subsection{Finite curvature invariants}

Let us begin with the metric with a spherical symmetry,
\begin{equation}
\label{eq:metric}
	\dif s^2=-f(r)\dif t^2+f^{-1}(r)\dif r^2+r^2\left(\dif \theta^2+\sin^2\theta\dif \varphi^2\right),
\end{equation}
where the shape function reads
\begin{equation}
	f(r)=1-\frac{2M \sigma(r)}{r}.
\end{equation}
Traditionally, the Kretschmann scalar $K\coloneqq R^{\mu\nu\alpha\beta}R_{\mu\nu\alpha\beta}$, Ricci scalar $R\coloneqq g^{\mu\nu}R_{\mu\nu}$, and contraction of two Ricci tensors $R_2\coloneqq R^{\mu\nu}R_{\mu\nu}$ are applied to check \cite{Fan:2016hvf,Balart:2014cga} the regularity of BHs. 
These three invariants are connected by a {\em syzegy} \cite{Zakhary:1997acs} (also known as {\rm Ricci decomposition}) as follows,
\begin{equation}
    C= K - 2 R_2+\frac{1}{3} R^2,
\end{equation}
where $C\coloneqq C^{\mu\nu\alpha\beta}C_{\mu\nu\alpha\beta}$ is contraction of two Weyl tensors. Alternatively, $R_2$ and $K$ can be replaced \cite{Frolov:2016pav} by $S$ and $C$, where $S$ is defined by $S:=S_{\mu\nu}S^{\mu\nu}$ and $S_{\mu\nu}$ by $ S_{\mu\nu}:=R_{\mu\nu}-\frac{1}{4}g_{\mu\nu} R$.

Fourteen independent curvature invariants can be established \cite{Weinberg:1972kfs} based on the Riemann curvature in the 4-dimensional spacetime, 
while a complete set of curvature invariants includes seventeen elements which are called Zakhary-Mcintosh (ZM) invariants \cite{Zakhary:1997acs,Pravda:2002us,Kraniotis:2021qah,Overduin:2022cir}. 
Therefore, it is natural to wonder whether the set of three invariants $\{K, R_2, R\}$ or $\{C, S, R\}$ can represent the finiteness of all curvature invariants. If not,  
 what elements should be included in the minimum set in order to determine an RBH? 
To answer this question for the metric Eq.\ \eqref{eq:metric},
we calculate the seventeen invariants, where six of them vanish. 
The non-vanishing eleven invariants can be separated into three groups.
\begin{enumerate}
\item  Ricci type constructed by Ricci tensors:
\begin{subequations}
\begin{equation}
\ii_5:=R=2\left(\ca_1+\ca_2\right),
\end{equation}  
\begin{equation}
\ii_6:=R_2=
2\left(\ca_1^2+\ca_2^2\right),
\end{equation}
\begin{equation}
     \ii_7\coloneqq R_{\mu}^{\;\;\nu}R_{\nu}^{\;\;\alpha}R_{\alpha}^{\;\;\mu}
     =2\left(\ca_1^3+\ca_2^3\right),
\end{equation}
\begin{equation}
    \ii_8\coloneqq
    R_{\mu}^{\;\;\nu}R_{\nu}^{\;\;\alpha}R_{\alpha}^{\;\;\beta}R_{\beta}^{\;\;\mu}
    =2\left(\ca_1^4+\ca_2^4\right);
\end{equation}
\end{subequations}
\item  Weyl type constructed by Weyl tensors:
\begin{subequations}
\begin{equation}
    \ii_1\coloneqq C^{\mu\nu}_{\quad \alpha\beta}C^{\alpha\beta}_{\quad \mu\nu}=\frac{4}{3} \ca_3^2,
\end{equation}
\begin{equation}
    \ii_3\coloneqq C^{\mu\nu}_{\quad \alpha\beta}C^{\alpha\beta}_{\quad \rho\sigma}C^{\rho\sigma}_{\quad \mu\nu}
    =\frac{4}{9} \ca_3^3;
\end{equation}
\end{subequations}
\item  Mixed type constructed by both Ricci and Weyl tensors:
\begin{subequations}
\begin{equation}
    \ii_9
    \coloneqq C_{\mu\alpha\beta\nu}R^{\alpha\beta}R^{\nu\mu}
    =-\frac{2}{3}\ca_3(\ca_2-\ca_1)^2,
\end{equation}
\begin{equation}
    \ii_{11}
    \coloneqq
    R^{\mu\nu} R^{\alpha\beta} \left(
    C_{\rho\mu\nu}^{\quad\; \sigma} C_{\sigma\alpha\beta}^{\quad\; \rho} 
    - C_{\rho\mu\nu}^{*\quad \sigma} C_{\sigma\alpha\beta}^{*\quad \rho}
    \right)
    =\frac{4}{9}\ca_3^2(\ca_2-\ca_1)^2,
\end{equation}
\begin{equation}
    \ii_{13}
    \coloneqq 
    R^{\mu \rho} 
    R_{\rho}^{\;\; \alpha}
    R^{\nu \sigma} R_{\sigma}^{\;\beta}
    C_{\mu\nu\alpha\beta}
    =\frac{2}{3}\ca_3(\ca_2-2\ca_1)^2,
\end{equation}
\begin{equation}
    \ii_{15}
    \coloneqq
    \frac{1}{16} R^{\mu\nu}R^{\alpha\beta}
    \left(
    C_{\rho\mu\nu\sigma}C_{\;\;\alpha\beta}^{\rho\quad\sigma}
    +
    C^*_{\rho\mu\nu\sigma}C_{\quad\alpha\beta}^{*\rho\quad\sigma}
    \right)
    =\frac{1}{36}\ca_3^2(\ca_2-\ca_1)^2,
\end{equation}
\begin{equation}
\begin{split}
    \ii_{16}
   & \coloneqq
   -\frac{1}{32} R^{\alpha\beta} R^{\mu\nu}
    \Big(
     C_{\rho\sigma\gamma\zeta}
    C^{\rho\quad\zeta}_{\;\;\alpha\beta}C^{\sigma\quad\zeta}_{\;\;\mu\nu}
    +C_{\rho\sigma\gamma\zeta}
    C^{*\rho\quad\zeta}_{\;\;\;\;\alpha\beta}C^{\sigma\quad\zeta}_{\;\;\;\;\mu\nu}\\
    &\qquad\qquad\qquad-C^*_{\rho\sigma\gamma\zeta}
    C^{*\rho\quad\zeta}_{\;\;\;\;\alpha\beta}C^{\sigma\quad\zeta}_{\;\;\mu\nu}
    +C^*_{\rho\sigma\gamma\zeta}
    C^{\rho\quad\zeta}_{\;\;\alpha\beta}C^{*\sigma\quad\zeta}_{\;\;\;\;\mu\nu}
    \Big)\\
    &=\frac{1}{108}\ca_3^3(\ca_2-\ca_1)^2,
\end{split}
\end{equation}
\end{subequations}
\end{enumerate}
where we have defined
\begin{equation}
\label{eq:3-term}
    \ca_1:= \frac{2 M \sigma'}{r^2},\qquad
    \ca_2:=\frac{ M \sigma''}{r},\qquad
    \ca_3:=-2\ca_1+\ca_2+\frac{6 M \sigma}{r^3}.
\end{equation}

The three quantities $\ca_1$, $\ca_2$ and $\ca_3$ in Eq.\ \eqref{eq:3-term} determine the finiteness of all non-vanishing invariants. 
Furthermore, $\ii_6$, as a specific representation, is necessary to determine if a BH is regular when $\sigma(r)$ is not a constant. The reason is that the squares of $\ca_1$ and $\ca_2$ in $\ii_6$ guarantee that no divergence terms appear. 
For more general cases, we can combine another invariant, for instance, $\ii_1$ with $\ii_6$.
Since $\ca_1$ and $\ca_2$ are finite, $\ca_3$ is also finite if $6M \sigma/r^3$ is finite.
In summary, $\ii_1$ and $\ii_6$, as representatives, can fully determine the regularity of the BHs with the spherical symmetry and one shape function.
We note that the number of elements in the complete set of spherically symmetric BHs is less than that of rotating RBHs \cite{Torres:2016pgk}.

A similar result can be obtained if the Kretschmann scalar $ K\coloneqq R^{\mu\nu\alpha\beta} R_{\mu\nu\alpha\beta}$ instead of $\ii_1$ is considered.
We rewrite $K$ in terms of $\ca_i$, $i=1, 2, 3$,
\begin{equation}
   K= \frac{4}{3}\left[
    2\left(\ca_1^2-\ca_1\ca_2+\ca_2^2\right)+\ca_3^2
    \right].
\end{equation}
If $\ii_6$ is finite, the regularity can be determined fully by $K$, i.e., $\{K, R_2\}$ is a complete set to determine whether a spherically symmetric BH with one shape function is regular. 

Furthermore, the requirement of a BH being regular at center $r=0$ can be obtained 
\begin{equation}
\label{eq:cond-finite}
    \sigma(r)\sim O(r^n),\quad
    n\ge 3.
\end{equation}
In other words, $\sigma(r)$ approaches zero not slower than $r^3$. Alternatively, a BH is regular at the center if we have
\begin{equation}
 \lim_{r\to 0} \sigma(r)/r^3=\const.<\infty.   
\end{equation}
This behavior of $\sigma$ around $r=0$ will help us to construct the RBHs in the next section.

\subsection{Complete geodesics}
\label{sec:geodesic}

In order to determine the condition that guarantees the geodesics being complete at $r=0$,\footnote{A geodesic is complete if its affine parameter can extend \cite{Hawking:1973uf} to infinity.} 
we start with the effective Lagrangian of null/timelike geodesics \cite{Chandrasekhar:1985kt},
\begin{equation}
2 \mathscr{L}=g_{\mu\nu} \frac{\dif  x^{\mu}}{\mathrm{~d} \tau} \frac{\dif  x^{\nu}}{\mathrm{~d} \tau},
\end{equation}
where $\tau$ is an affine parameter along a geodesic. 
For the spacetime given by metric Eq.\ \eqref{eq:metric}, the Lagrangian reads
\begin{equation}
\label{eq:lagrange}
\mathscr{L}=\frac{1}{2}
\left[-f(r)\dot{t}^{2}+\frac{\dot{r}^{2}}{f(r)}+r^{2} \dot{\theta}^{2}+\left(r^{2} \sin ^{2} \theta\right) \dot{\varphi}^{2}\right],
\end{equation}
which does not depend on $t$ and $ \varphi $ explicitly, 
i.e., $t$ and $\varphi$ are cyclic coordinates. 
Thus, the canonical momenta conjugate to $t$ and $\varphi$ are conserved quantities,
\begin{equation}
\label{eq:int-motion}
p_ {t}=-\frac{\partial \mathscr{L}}{\partial \dot{t}}=f(r)\dot{t}=E, \qquad p_ {\varphi}=\frac{\partial \mathscr{L}}{\partial \dot{\varphi}}=\left(r^{2} \sin ^{2} \theta\right) \dot{\varphi}=L,
\end{equation}
where $E$ and $L$ are the total energy and total angular momentum, respectively. To simplify the following discussion, we just concentrate on the motions on an equatorial plane, $\theta=\pp/2$, and set the initial angular momentum to be zero, $L=0$.

Next, we introduce the Painlevé–Gullstrand time \cite{Karl:2001rcs},
 \begin{equation}
 \label{eq:pg-time-dif}
 \dif t=\dif t_*-\frac{\sqrt{1-f(r)}}{f(r)} \dif r,
 \end{equation}
and recast the Lagrangian Eq.\ \eqref{eq:lagrange},
\begin{equation}
\label{eq:lag-pg}
    \mathscr{L}_*=\frac{1}{2}
    \left[
    -f(r) \dot t_*^2
    +2\sqrt{1-\frac{f(r)}{
    E^2 }} \dot r \dot t_*
    +\frac{\dot r^2}{E^2}
    \right].
\end{equation}
Moreover, the definition of the Painlevé–Gullstrand time provides 
\begin{equation}
\label{eq:pg-time-dot}
   \dot t_* =\frac{ \dot r}{f(r)}\sqrt{1-\frac{f(r)}{E^2}}+\frac{E}{f(r)},
\end{equation}
where we have replaced $\dot t$ by $E$ via Eq.\ \eqref{eq:int-motion}.
Thus, we can obtain the equation of radial motion by using Eqs.\ \eqref{eq:lag-pg} and \eqref{eq:pg-time-dot} together with $2 \mathscr{L}_*=k$, 
\begin{equation}
 \label{eq:radial-motion}
 \frac{\dif r}{\dif t_*}=\frac{f(r) \sqrt{1+\frac{k f(r)}{E^2}}}{\sqrt{1-\frac{f(r)}{E^2}} \sqrt{1+\frac{k f(r)}{E^2}}-1},
 \end{equation}
where $k=0, -1$ corresponds to null and timelike geodesics, respectively, and we consider only the ingoing motion. 

For timelike geodesics with $k=-1$, if we suppose that the test particle starts with $\dif r/ \dif t_*=0$, i.e., $E=1$, Eq.\ \eqref{eq:radial-motion} can be reduced \cite{Liang:2015gig} to
\begin{equation}
     \frac{\dif r}{\dif t_*}=-\sqrt{\frac{2M \sigma(r)}{r}},
\end{equation}
then the Painlevé–Gullstrand time can be calculated  
\begin{equation}
    t_*=-\int_{\infty}^0\dif r \sqrt{\frac{r}{2M \sigma(r)}},
\end{equation}
which is divergent at $r=0$ if $\sigma$ has the same asymptotic behavior around $r=0$ as Eq.~(\ref{eq:cond-finite}). 
In other words, the timelike geodesics are complete at $r=0$ if $\sigma \sim O(r^n)$ with $n\ge 3$, which is consistent with the requirement of finite curvatures.

For null geodesics with $k=0$, the initial velocity should be $\dif r/\dif r_*=-1$ at infinity, which also corresponds to $E=1$. Thus, Eq.\ \eqref{eq:radial-motion} becomes
\begin{equation}
    \frac{\dif r}{\dif t_*}=-1-\sqrt{\frac{2M \sigma(r)}{r}}.
\end{equation}
The Painlevé–Gullstrand time of null geodesics is divergent if $\sigma\sim 1$ as $r\to \infty$, which also meets the requirement of asymptotic flatness, i.e.\
\begin{equation}
    t_* =- \int_{\infty }^0 \frac{\dif r}{1+ \sqrt{\frac{2 M \sigma(r) }{r}}} \sim 
    - \int_{\infty }^0 \frac{\dif r}{1+ \sqrt{\frac{2 M}{r}}}\to -\infty.
\end{equation}
The null geodesics are complete for a massless test particle starting at infinity.
However, this cannot provide any constraints to $\sigma(r)$  in the vicinity of a BH center.
In summary, we have proved that the finiteness of curvature
invariants and the completeness of geodesics are equivalent for the spherically symmetric BHs with one shape function.


\section{Construction of our model}
\label{sec:construction}

We propose a shape function according to the asymptotic behavior given by Eq.\ \eqref{eq:cond-finite},
\begin{equation}
\label{eq:shape}
f(r,Q)=1-\frac{r^{2}+r \delta(Q)}{r^{3}+Q^{3}},
\end{equation}
where $\delta(Q)$ is a function of parameter $Q$ and vanishes when $Q$ reduces to an extreme value 
$Q=Q_{\rm ext}$ which corresponds to the extreme horizon $r_{\rm ext}$, 
meanwhile we have performed a transformation, $r\to r/(2M)$ and $Q\to Q/(2M)$, 
such that all variables are dimensionless.
The existence of an extreme horizon indicates that this BH model has two horizons at least.

Furthermore,  from $f(r_{\rm ext}, Q_{\rm ext})=0$ and $\delta(Q_{\rm ext})=0$, we derive the 
extreme horizon radius and charge,
\begin{equation}
    r_{\rm ext}=\frac{2}{3},\qquad
    Q_{\rm ext}=\frac{2^{2/3}}{3},
\end{equation}
respectively.
Thus, we select $\delta$ to be
\begin{equation}
\delta(Q)=(Q - Q_{\rm ext})^2.\label{seclectdelta}
\end{equation}
Actually, $\delta(Q)$ as a function of $Q$ is not unique,  and it is a relatively simple one we have selected. Alternatively, $\delta$ is a function depending on the difference between outer and inner horizons, $r_+-r_-$, and vanishes when the BH reduces to its extreme configuration, $r_+=r_-=r_{\rm ext}$, e.g., $\delta=r_+-r_-$.

\begin{figure}[!htb]
     \centering
 \includegraphics[width=0.6\textwidth]{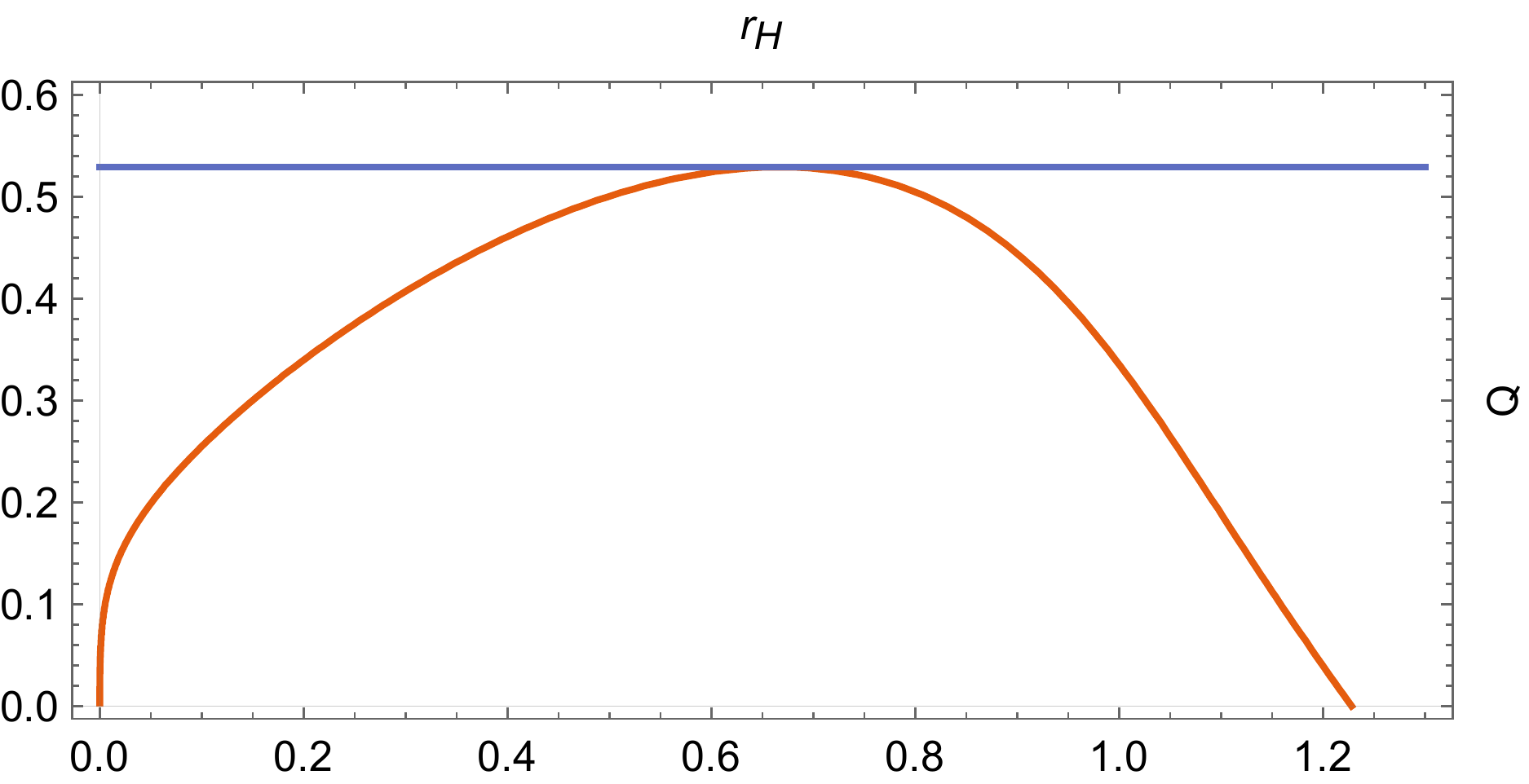}
     \captionsetup{width=.9\textwidth}
       \caption{The orange curve depicts the equation $f(r_{\rm H}, Q)=0$ and the blue line $Q=Q_{\rm ext}$.}
        \label{fig:hoirzon}
\end{figure}

The horizon curve $f(r_{\rm H}, Q)=0$ is shown in Fig.\ \ref{fig:hoirzon}, where it clearly shows that there are two horizons for a given value $Q\le Q_{\rm ext}$.
The maximum value of outer horizon $r_+$ is 
\begin{equation}
    \max\{r_+\}=\frac{1}{2} \left(1+\sqrt{1+4 Q_{\rm{ext}}^2}\right)
    \approx 1.228,
\end{equation}
which can be normalized to unity by a certain transformation. It is not necessary to make such a normalization because Eq.\ \eqref{eq:shape} does not reduce to the shape function of Schwarzschild BHs when $Q=0$.

Now let us analyze the change of divergence of curvature invariants in our model, see Eqs.(\ref{eq:shape})-(\ref{seclectdelta}). 
The corresponding curvature invariants around $r=0$ are
\begin{subequations}
\begin{equation}
K\sim\frac{8 \left(Q-Q_{\rm{ext}}\right)^4}{Q^6 r^2}
+\frac{24 \left(Q-Q_{\rm{ext}}\right)^2}{Q^6 r}
+\frac{24}{Q^6}
+O\left(r\right),
\end{equation}
\begin{equation}
\mathcal{I}_6\sim
\frac{10 \left(Q-Q_{\rm{ext}}\right){}^4}{Q^6 r^2}+\frac{36 \left(Q-Q_{\rm{ext}}\right){}^2}{Q^6 r}+\frac{36}{Q^6}+O\left(r\right),
\end{equation}
\end{subequations}
which are divergent if $Q\neq Q_{\rm ext}$.
Therefore, the BH that we constructed has curvature singularity at $r=0$, that is,
this BH is located at its {\em singular state} when $Q\neq Q_{\rm ext}$.
As $Q$ approaches the extreme value $Q_{\rm ext}$, the behaviors of curvature invariants around $r=0$ become
\begin{subequations}
\begin{equation}
K\sim\frac{24}{Q_{\rm{ext}}^6}-\frac{168 r^3}{Q_{\rm{ext}}^9}+O\left(r^4\right),
\end{equation}
\begin{equation}
\mathcal{I}_6\sim
\frac{36}{Q_{\rm{ext}}^6}-\frac{252 r^3}{Q_{\rm{ext}}^9}+O\left(r^4\right),
\end{equation}
\end{subequations}
which implies that the curvature invariants are finite and positive at $r=0$. Thus,
the BH is located at its {\em regular state} when $Q=Q_{\rm ext}$.
The transition of BHs due to the jump in the parameter $Q$ in our model is similar to the changes between BHs and wormholes discussed in Refs.~\cite{Simpson:2018tsi,Churilova:2019cyt}. In these two references, however, the change in the parameter $a$ or $m$ does not have to correspond to a physical process. In other words, the BHs and wormholes are more likely to be regarded as independent physical objects that are described by the same metric.
The following contexts are to investigate the changes of physical properties for our BH model as it evolves from its singular to regular state.

\section{Sources and energy conditions}
\label{sec:sources}

In this section, we discuss the matter sources that may generate the BH model Eqs.\ \eqref{eq:metric} and \eqref{eq:shape} in the framework of $F(R)$ gravity and give the energy conditions of the matters sources.
The motivation for interpreting our BH model in the context of $F(R)$ gravity is that the entropy of RBHs has a deviation from the entropy-area law 
of Einstein's gravity
if it is calculated by $\displaystyle S=\int \dif M/T$, where $T$ is BH temperature. 
In other words, either the entropy-area theorem or this thermodynamic formulas is no longer valid.
We anticipate that the modified gravity may shed some light on resolving this problem.

\subsection{Matter sources in our model}
\label{subsec:sources}

We suppose that the action is of the following form,
\begin{equation}
\label{eq:action}
I= \int \dif ^4x \sqrt{-g}\; F(R)+I_M,
\end{equation}
where $I_M$ represents the action of the matter sources that will be determined below.
The equation of motion with respect to metric reads \cite{Sotiriou:2008rp}
 \begin{equation}
 \label{eq:eq-of-motion}
   M_{\mu\nu}\coloneqq F'(R) R_{\mu\nu}-\frac{1}{2} F(R) g_{\mu\nu}-\left[\nabla_\mu \nabla_\nu-g_{\mu\nu} \square\right] F'(R)=- T_{\mu\nu},
\end{equation}
where we have defined
\begin{equation}
    T_{\mu\nu}=\frac{1}{\sqrt{-g}}\frac{\delta I_M}{\delta g^{\mu\nu}},
\end{equation}
which leads to a minus sign in the right hand side of Eq.\ \eqref{eq:eq-of-motion}

Now we analyze the algebraic structure of $M_{\mu\nu}$ in order
to establish the action of matter sources.
Since the metric Eq.\ \eqref{eq:metric} is spherically symmetric and has one shape function Eq.\ \eqref{eq:shape},
the Ricci tensor $R_{\mu\nu}$ has the algebraic structure $[(1,1)(11)]$, so does $F'(R) R_{\mu\nu}$. Here we adopt the Segr\'e notation \cite{Dolgachev:2012cag,Stephani:2003tm}, where symbol 1 corresponds to one component of Ricci tensors, all components in parentheses are equal, and a comma separates timelike and spacelike components.
The remaining part of $M_{\mu\nu}$ is of $[1,1(11)]$ because the structure of $g_{\mu\nu}$ is $[1,1(11)]$ and $R$ depends only on radial coordinate $r$.
As a result, the algebraic structure of $M_{\mu\nu}$ is $[1,1(11)]$.  
Therefore, the energy-momentum tensor $T_{\mu\nu}$ must have the same structure, otherwise, the equation of motion Eq.\ \eqref{eq:eq-of-motion} would not be satisfied.

Let us turn to the structure of the energy-momentum tensor of nonlinear electrodynamics. It is  $[(1,1)(11)]$, 
such that it can generate spherically symmetric RBHs with one shape function. 
However, only nonlinear electrodynamics is not enough \cite{Berej:2006cc} for its energy-momentum tensor to have the same algebraic structure as $M_{\mu\nu}$. 
To achieve our purpose, we introduce a scalar field following the method of Ref.\ 
\cite{Bronnikov:2021uta} because the energy-momentum tensor of a scalar field has the algebraic structure $[1(11,1)]$. 

The combination of nonlinear electromagnetic and scalar fields provides the algebraic structure $[1,1(11)]$. 
Thus, we write down the action for the matter source,
\begin{equation}
\label{eq:action-matter}
    I_M=\int \dif^4 x \sqrt{-g}\left[
    2 W(\phi) \partial^\mu\phi
    \partial_\mu\phi-2V(\phi)-\mathcal{L}(\mathcal{F})
    \right],
\end{equation}
where $V(\phi)$ is potential of scalar fields and $\mathcal{F}=F_{\mu\nu}F^{\mu\nu}$ is contraction of strength tensors. The introduction of $W(\phi)$ is to guarantee that the kinetic term $\partial^\mu\phi\partial_\mu\phi$ is positive definite, 
but this also leads to a redundant degree of freedom.  It seems a disadvantage but actually provides more possibilities for us to construct reasonable scalar fields.

The energy-momentum tensor of the scalar field reads
\begin{equation}
\label{eq:emt-phi}
    T_{\mu\nu}[\phi]  =
    2 W(\phi) \partial_\mu\phi
    \partial_\nu\phi- g_{\mu\nu}\left[
     W(\phi) \partial^\alpha\phi
    \partial_\alpha\phi-V(\phi)\right].
\end{equation}
After substituting the metric Eq.\ \eqref{eq:metric} into the above equation, we obtain
\begin{equation}
    T_{\mu}^{\nu}[\phi] =f(r) W(r) \phi'^2\diag\{-1,1,-1,-1\} +\delta_{\mu}^{\nu} V(r),
\end{equation}
which has the structure $[1(11,1)]$.
The energy-momentum tensor of nonlinear electrodynamics takes the form,
\begin{equation}
\label{eq:emt-a}
    T_{\mu\nu}[A]=-2\mathcal{L}_{\mathcal{F}} F_{\mu\alpha}F_\nu^{\;\;\alpha}+
    \frac{1}{2}g_{\mu\nu} \mathcal{L}(\mathcal{F}),
\end{equation}
where $\mathcal{L}_{\mathcal{F}}$ denotes the derivative of $\mathcal{L}$ with respect to $\mathcal{F}$. The explicit form of $T^{\nu}_{\mu}[A]$ is 
\begin{equation}
    T^\nu_\mu[A]=\frac{1}{2} \diag\left\{
    \mathcal{L}, 
    \mathcal{L}, 
    \mathcal{L}-\frac{4Q^2}{r^4}\mathcal{L}_{\mathcal{F}},
    \mathcal{L}-\frac{4Q^2}{r^4}\mathcal{L}_{\mathcal{F}}, 
    \right\},
\end{equation}
where the contraction of strength tensors reads
\begin{equation}
\mathcal{F}=2Q^2/r^4,\label{confieldten}
\end{equation}
and $Q$ is magnetic charge~\cite{Fan:2016hvf,Bronnikov:2021uta}.
The algebraic structure of $T^\nu_\mu[A]$ is
$[(1,1)(11)]$,
therefore the combination of two matters 
$T_{\mu}^{\nu}[\phi]+T^\nu_\mu[A]$ has $[1,1(11)]$ as we expected.

\begin{figure}[!htb]
     \centering
         \includegraphics[width=\textwidth]{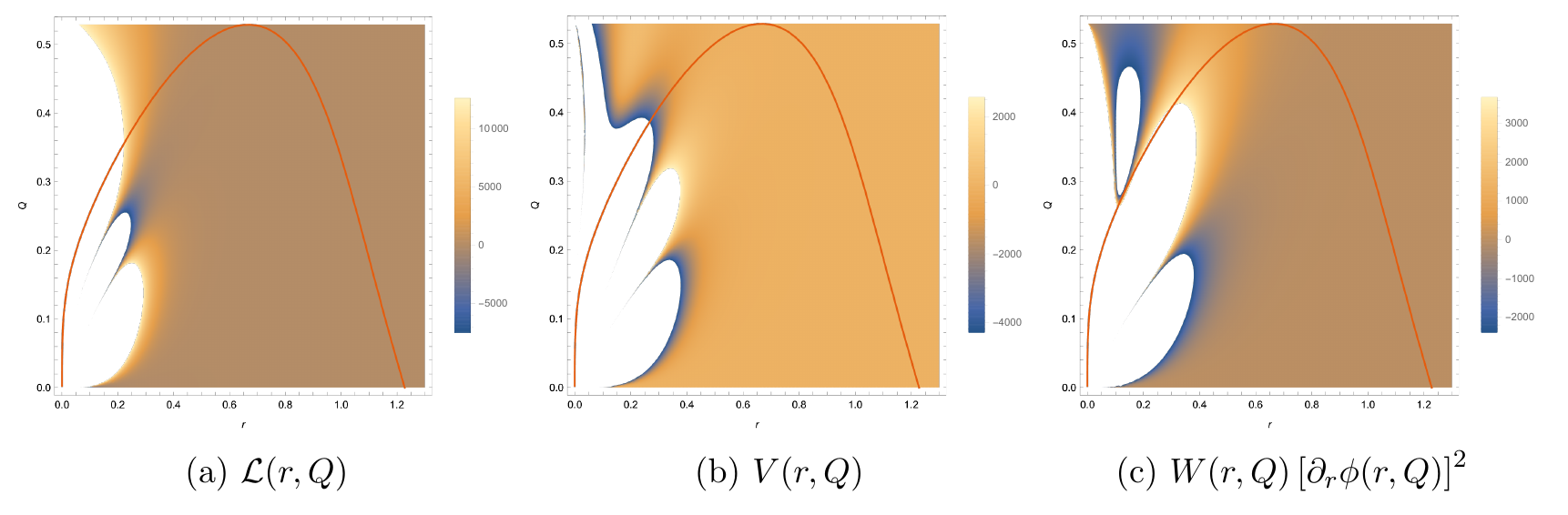}
      \captionsetup{width=.9\textwidth}
       \caption{Density plots of matter sources. The orange curves correspond to the horizons, $f(r_{\rm H}, Q)=0$, where $\alpha=1$ is set.}
        \label{fig:matter-source}
\end{figure}

\begin{figure}[!htb]
     \centering
         \includegraphics[width=\textwidth]{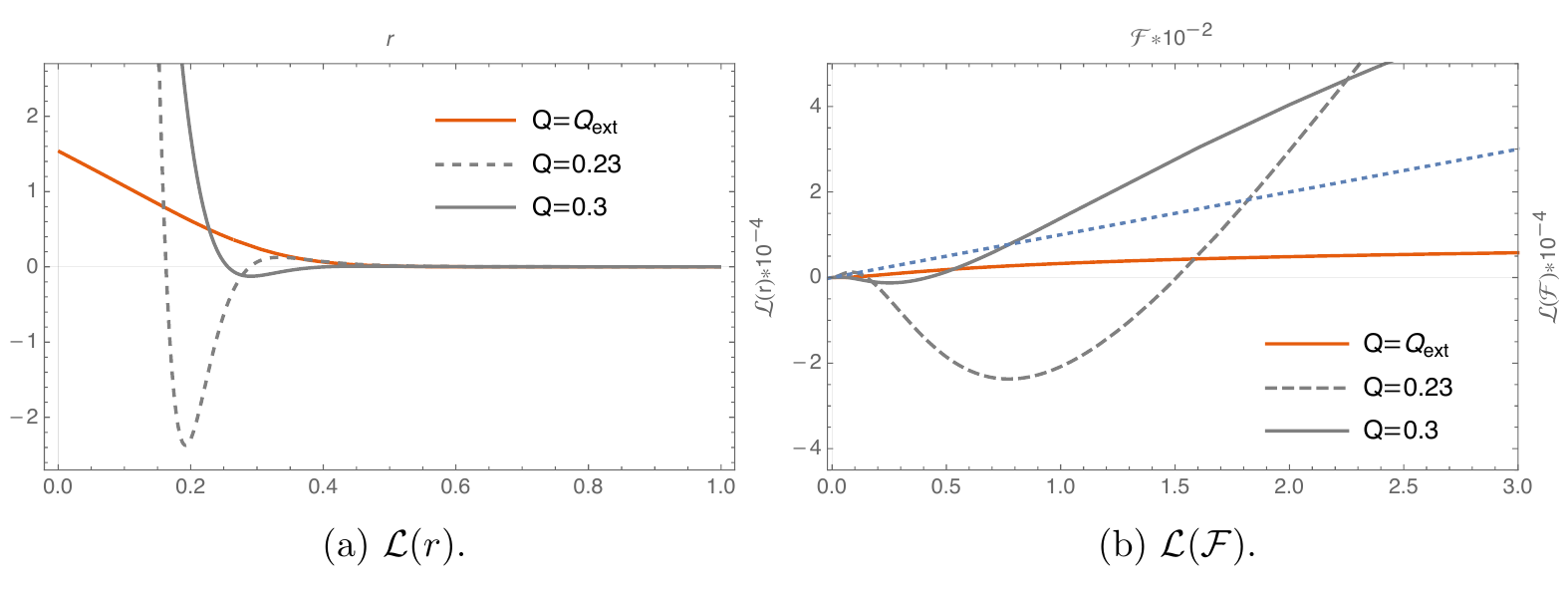}
      \captionsetup{width=.9\textwidth}
       \caption{Nonlinear electrodynamic source. The orange curves correspond to a regular state, while the gray curves are a singular state. In addition, the blue dotted line depicts Maxwell's asymptotic, $\mathcal{L}(\mathcal{F})=\mathcal{F}$. $\alpha=1$ is set.}
        \label{fig:nonl-elect}
\end{figure}

\begin{figure}[!htb]
     \centering
         \includegraphics[width=\textwidth]{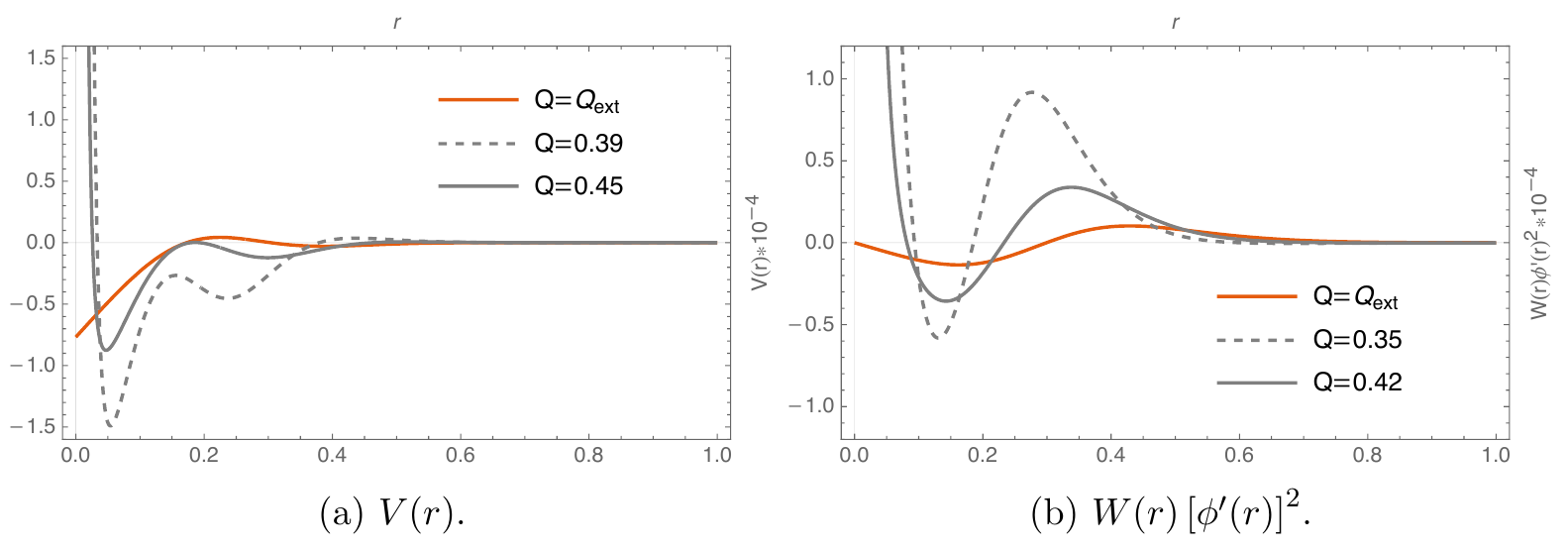}
      \captionsetup{width=.9\textwidth}
       \caption{Scalar field source. The orange curves correspond to a regular state, while the gray curves a singular state, where $\alpha=1$ is set.}
        \label{fig:scalar-funs}
\end{figure}

To give $T_{\mu}^{\nu}[\phi]+T^\nu_\mu[A]$ explicitly, we suppose that $F(R)$ has the same form with the Starobinsky inflation model \cite{Starobinsky:1980te}, 
\begin{equation}
    F(R)=R+\alpha R^2,\label{FRform1}
\end{equation}
but with parameter $\alpha$ that will be determined via the formula of entropy below.
Then,  
we solve $\mathcal{L}$, $V$, and $W\phi'^2$ as functions of radial coordinate $r$ by substituting Eqs.\ \eqref{eq:shape}, \eqref{eq:emt-phi}, \eqref{eq:emt-a} and \eqref{FRform1} into Eq.\ \eqref{eq:eq-of-motion}, see App.\ \ref{sec:Analytical-matter} for the derivations and results.\footnote{By further considering Eq.\ \eqref{confieldten}, we give $\mathcal{L(\mathcal{F})}$. In addition, we can separate $W$ from $W\phi'^2$ by using the ansatz Eq.~(\ref{eq:scalar-field}). See App.\ \ref{sec:Analytical-matter} for the details.}
We verify that the following Klein-Gordon equation holds automatically, 
\begin{equation}
    \frac{\partial W}{\partial \phi}\partial_\mu\phi\partial^\mu\phi+2W(\phi)\Box\phi+\frac{\partial V}{\partial \phi}=0,
\end{equation}
or
\begin{equation}
\label{eq:KG-eq}
    2 W \left[\left(r f'+2 f\right) \phi '+r f \phi ''\right]+r f W' \phi '+\frac{r V'}{\phi'}=0,
\end{equation}
when the metric Eq.~(\ref{eq:metric}) with the shape function Eq.\ \eqref{eq:shape}, together with the relevant $\mathcal{L}$, $V$, $W$ and $\phi$, is substituted into it, 
which means that the metric Eq.~(\ref{eq:metric}) with the shape function Eq.\ \eqref{eq:shape} is indeed the solution of $F(R)$ gravity with the matter source described by Eqs.~(\ref{eq:action}) and (\ref{eq:action-matter}).
We then draw $\mathcal{L}$, $V$, and $W\phi'^2$ in the density plots of full two-dimensional parameter space, $\{r, Q\}$, see Fig.\ \ref{fig:matter-source}. 
We make two comments. The first is that $\mathcal{L}$, $V$ and $W \phi'^2$ may change their signs along the $r$-axis. This signifies that it is appropriate and necessary to introduce $W$, otherwise a negative $\phi'^2$, $\phi'^2<0$, would lead to a contradiction. The second comment is that these solutions may diverge at the BH center.
To exhibit this divergent nature clearly, 
we plot the solutions with specific values of $Q$ in Figs.\ \ref{fig:nonl-elect} and \ref{fig:scalar-funs}, where $\mathcal{L}$, $V$ and $W \phi'^2$ are divergent at $r=0$ for a singular state but approach finite values for a regular state.

We further list the asymptotic behaviors of $\mathcal{L}$, $V$, and $W \phi'^2$ for both singular and regular states in order to illustrate the properties of the two states clearly. 

We consider a singular state at first. 
As $r\to\infty$, we have 
\begin{equation}
    \mathcal{L} \sim \mathcal{L}_0-
    \frac{40 \pp  \alpha  \left(-7 \delta ^2+18 Q^3-7 \delta  Q\right)}{3 \sqrt{3} Q^8}-\frac{2 \delta }{r^4}+O\left(r^{-5}\right),\label{lagsing}
\end{equation}
where $\mathcal{L}_0$ is an integration constant 
which can be fixed by  $\mathcal{L}\to 0$ as $r$ approaches infinity
\cite{Bronnikov:2017tnz}, 
\begin{equation}
    \mathcal{L}_0=
    \frac{40 \pp  \alpha  \left(-7 \delta ^2+18 Q^3-7 \delta  Q\right)}{3 \sqrt{3} Q^8}.
\end{equation}
Correspondingly, we obtain 
\begin{equation}
    V\sim \frac{126 \alpha  Q^3}{r^8}+O\left(r^{-9}\right),\label{vsing}
\end{equation}
and
\begin{equation}
    W\phi'^2\sim -\frac{252\alpha Q^3}{r^8}+O\left(r^{-9}\right).\label{wsing}
\end{equation} 
As $r\to 0$, we have
\begin{equation}
   \mathcal{L}\sim \frac{16 \alpha  \delta }{Q^3 r^3}+\frac{12 \alpha  \delta ^2}{Q^6 r^2}+\frac{4 \left(24 \alpha +Q^3\right) \delta }{Q^6 r}+O\left(r^0\right),
\end{equation}
\begin{equation}
    V\sim\frac{4\alpha \delta }{Q^3 r^3}-\frac{18\alpha \delta ^2}{Q^6 r^2}-\frac{60\alpha \delta }{Q^6 r}+O\left(r^0\right),
\end{equation}
\begin{equation}
    W\phi'^2\sim \frac{12 \alpha  \delta }{Q^3 r^3}-\frac{60 \alpha  \delta }{Q^6}-\frac{252 \alpha  r}{Q^6}+O\left(r^2\right),
\end{equation}
which are divergent at $r=0$ in the form of $r^{-3}$.

Now we turn to the asymptotic behaviors of a regular state. As $r\to\infty$, we derive
\begin{equation}
    \mathcal{L}\sim \frac{6 Q_{\rm ext}^3}{r^6}+O\left(r^{-8}\right),
\end{equation}
which is different from Eq.~(\ref{lagsing}) when compared with the Lagrangian of a singular state. But the asymptotic behaviors of $V$ and $W \phi'^2$ remain unchanged when compared with those of a singular state, see Eqs.~(\ref{vsing}) and (\ref{wsing}), just with the replacement of $Q$ by $Q_{\rm ext}$. 
On the other hand, as $r\to 0$, the obvious differences emerge between singular and regular states, i.e., 
$\mathcal{L}$, $V$ and $W\phi'^2$ do not diverge any longer for a regular state,
\begin{equation}
    \mathcal{L}\sim \frac{10 \alpha  \left(32 \sqrt{3} \pp  Q_{\rm ext}+9\right)}{3 Q_{\rm ext}^6}+\frac{6}{Q_{\rm ext}^3}+O\left(r^1\right),
\end{equation}
\begin{equation}
    V\sim -\frac{5 \alpha  \left(32 \sqrt{3} \pp  Q_{\rm ext}+9\right)}{3 Q_{\rm ext}^6}+O\left(r^1\right),
\end{equation}
\begin{equation}
  W \phi'^2  \sim -\frac{252 \alpha  r}{Q_{\rm ext}^6}+O\left(r^2\right).
\end{equation}

\begin{figure}[!htb]
     \centering
     \begin{subfigure}[b]{0.45\textwidth}
         \centering
         \includegraphics[width=\textwidth]{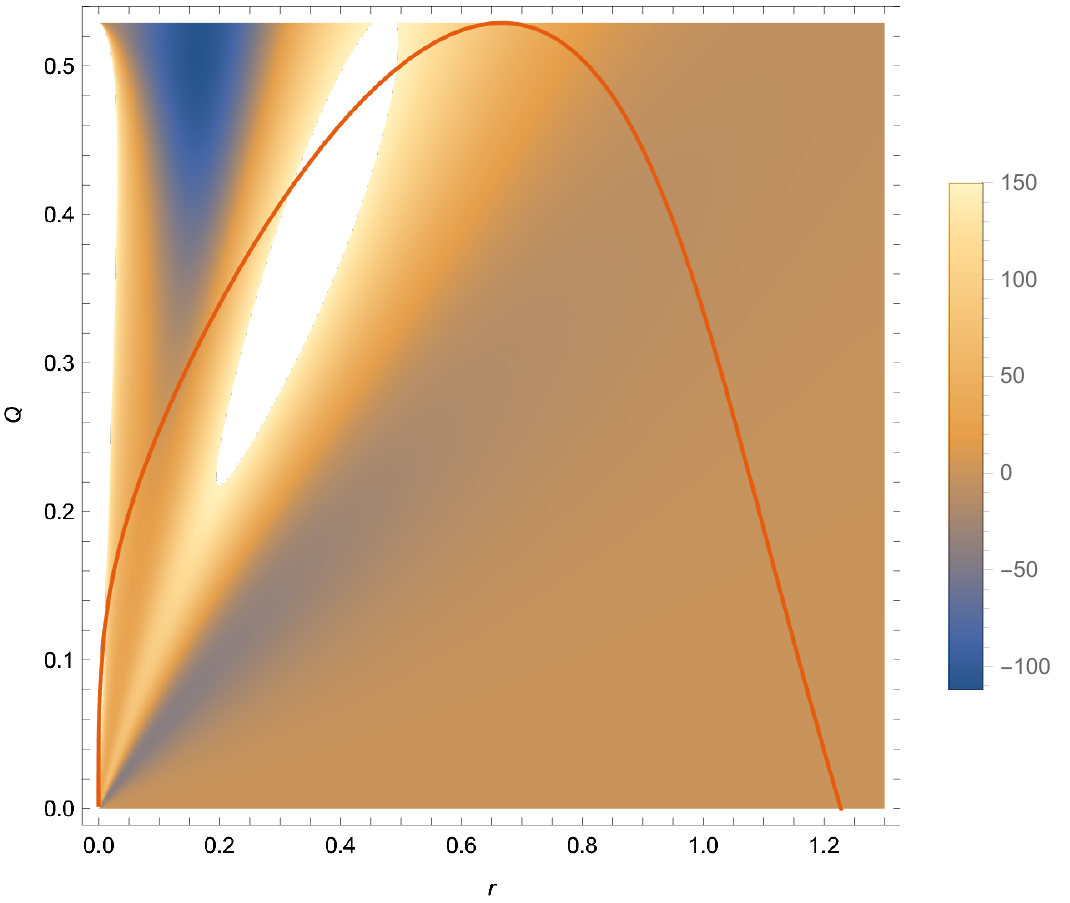}
         \caption{$W(r,Q)$}
         \label{fig:sep-w}
     \end{subfigure}
     \begin{subfigure}[b]{0.45\textwidth}
         \centering
         \includegraphics[width=\textwidth]{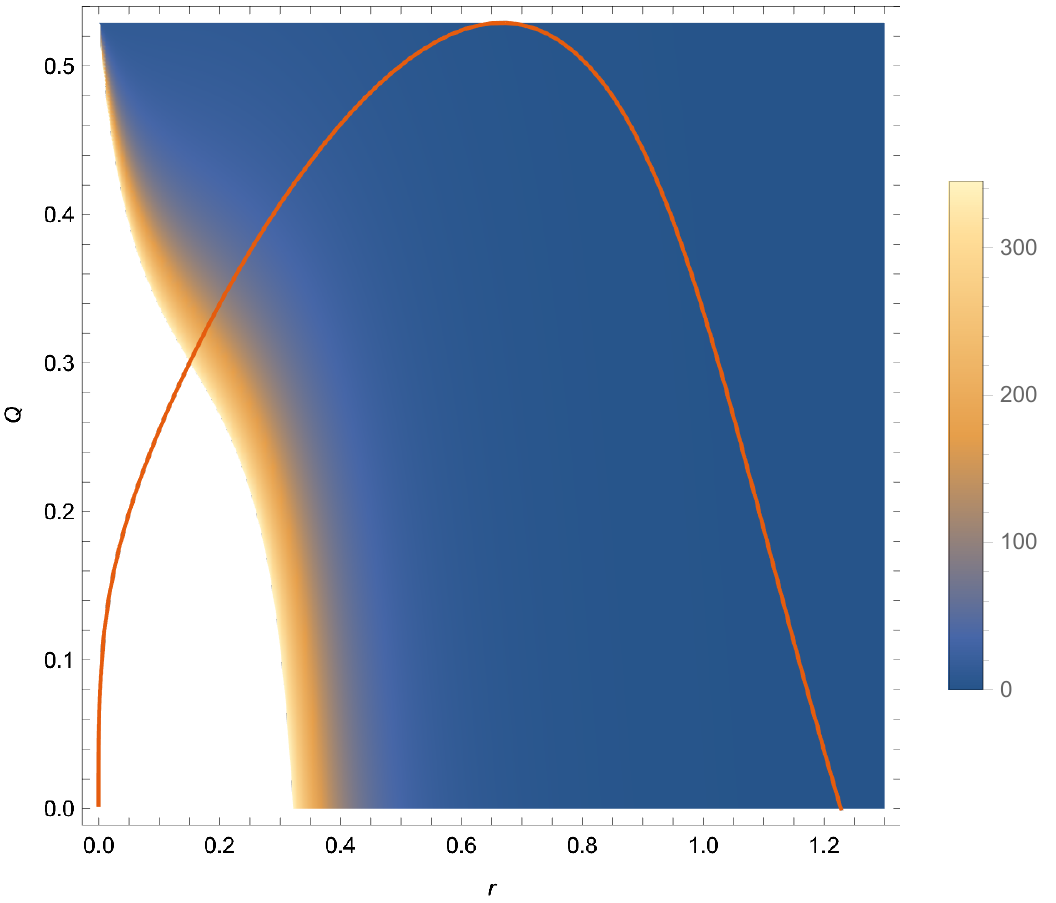}
         \caption{$\left[\partial_r\phi(r,Q)\right]^2$}
         \label{fig:sep-phi}
     \end{subfigure}
      \captionsetup{width=.9\textwidth}
       \caption{Separation of $W(r,Q)$ and $\left[\partial_r\phi(r,Q)\right]^2$ from $W\phi'^2$. $\alpha=1$ is set.}
        \label{fig:sep-scalar}
\end{figure}

In order to determine $W$ and $\phi$ separately, we introduce the following ansatz for a scalar field,
\begin{equation}
\label{eq:scalar-field}
   \left[\phi '(r)\right]^2= \frac{\delta +r^2}{r^2 \left(r^3+Q^3\right)^{4/3}}.
\end{equation}
Thus, $W$ and $\phi'^2$ can be separated as shown in Fig.\ \ref{fig:sep-scalar}, where $\phi'^2$ is positive definite although it diverges in a certain region of the parameter space, $\{r, Q\}$, see Fig.\ \ref{fig:sep-phi}.

\begin{figure}[!htb]
     \centering
         \includegraphics[width=.6\textwidth]{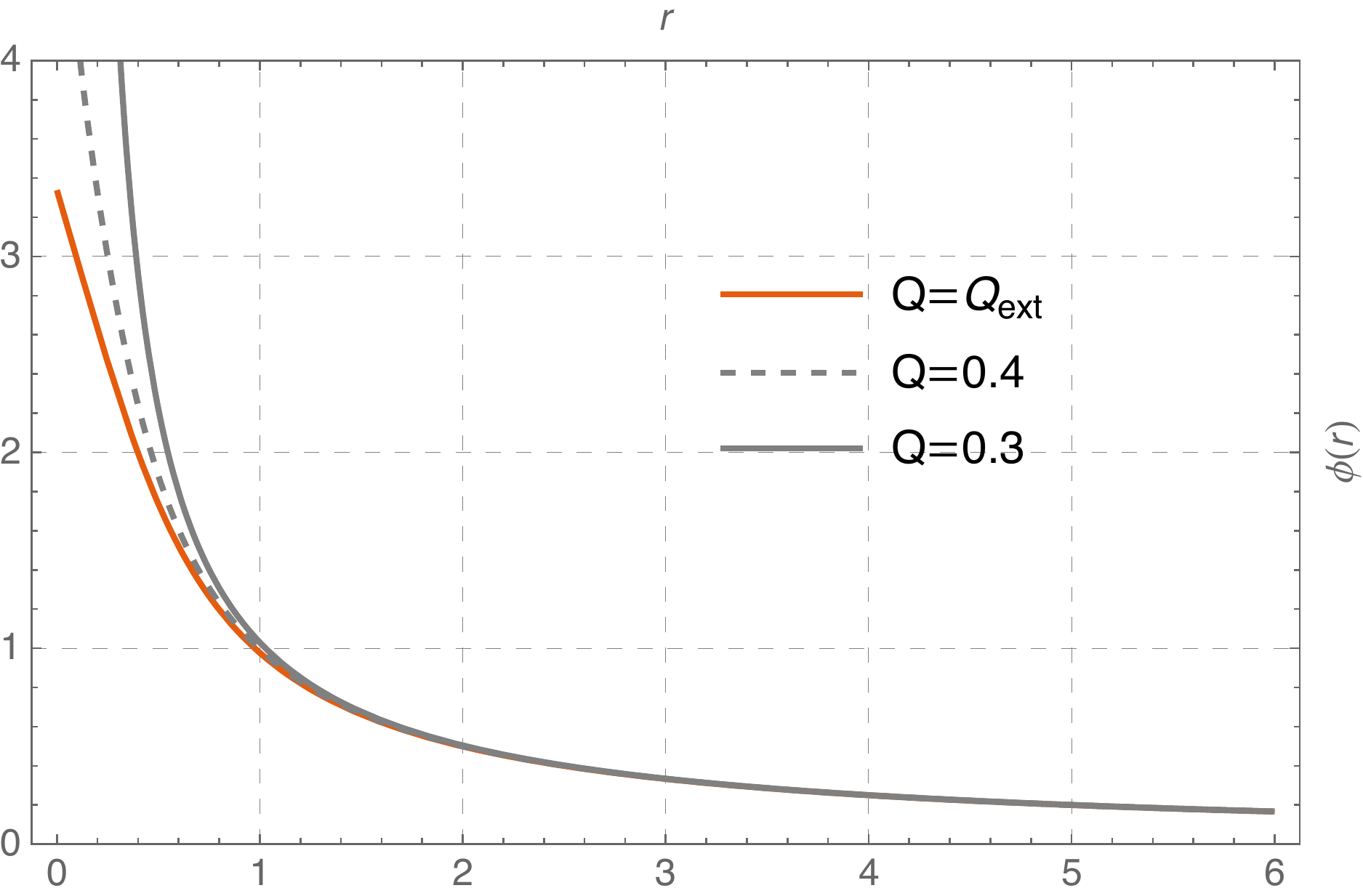}
      \captionsetup{width=.9\textwidth}
       \caption{$\phi(r)$. $\alpha=1$ is set.}
        \label{fig:phi-sol}
\end{figure}

Furthermore, we provide the numeric solution of $\phi$ in Fig.\ \ref{fig:phi-sol}.
We note that $\phi$ is a monotone decreasing function of radial coordinate $r$ 
and vanishes as $r$ approaches infinity. 
Moreover, $\phi$ diverges at the BH center except for $Q=Q_{\rm ext}$, i.e., when the BH stays at its regular state, the scalar field is bounded,
\begin{equation}
\label{eq:max-phi}
    \max\{\phi(r, Q_{\rm ext})\} \approx 3.3386.
\end{equation}

\begin{figure}[!htb]
     \centering
     \begin{subfigure}[b]{0.45\textwidth}
         \centering
         \includegraphics[width=\textwidth]{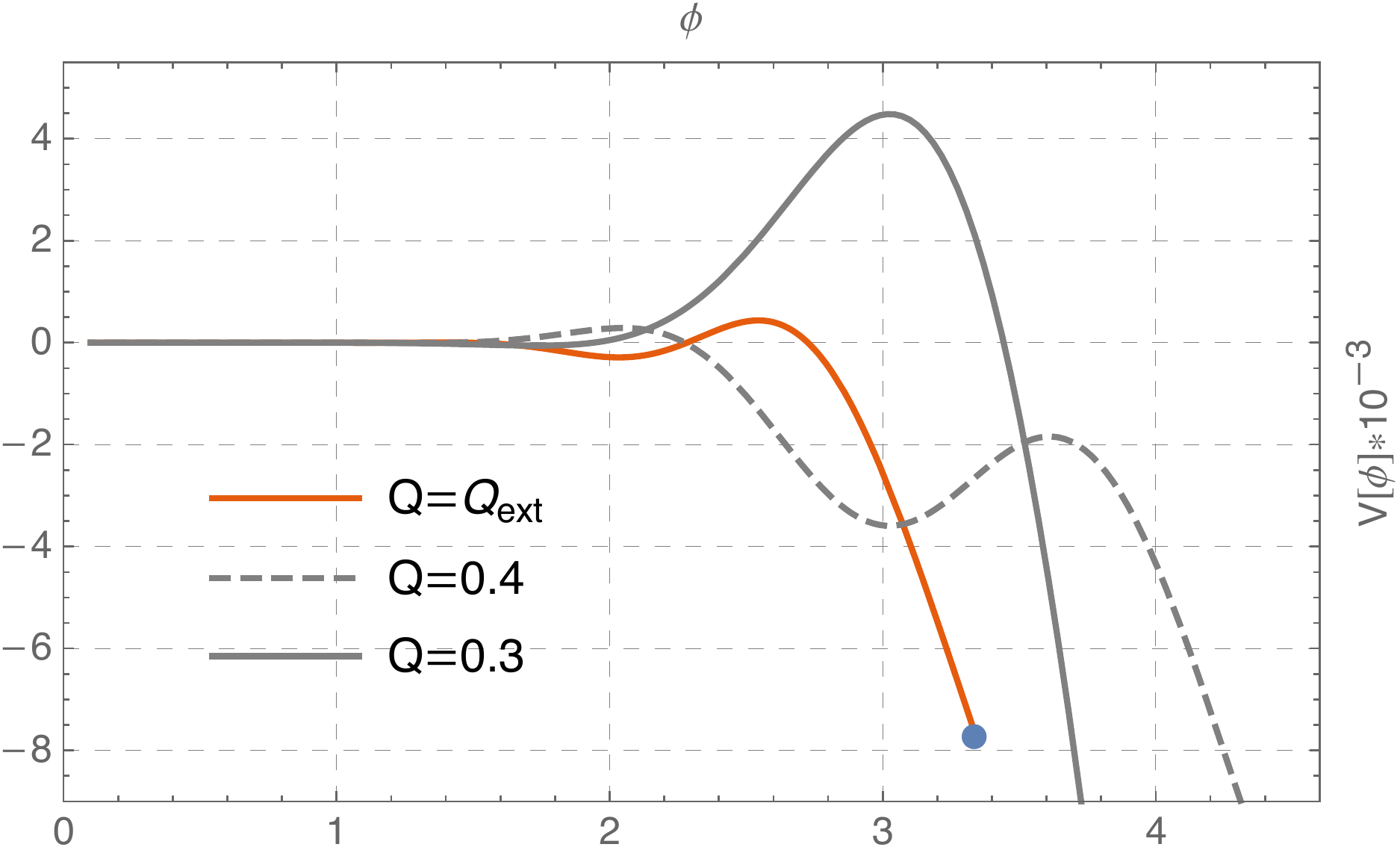}
         \caption{$V(\phi)$}
         \label{fig:V-phi}
     \end{subfigure}
     \begin{subfigure}[b]{0.45\textwidth}
         \centering
         \includegraphics[width=\textwidth]{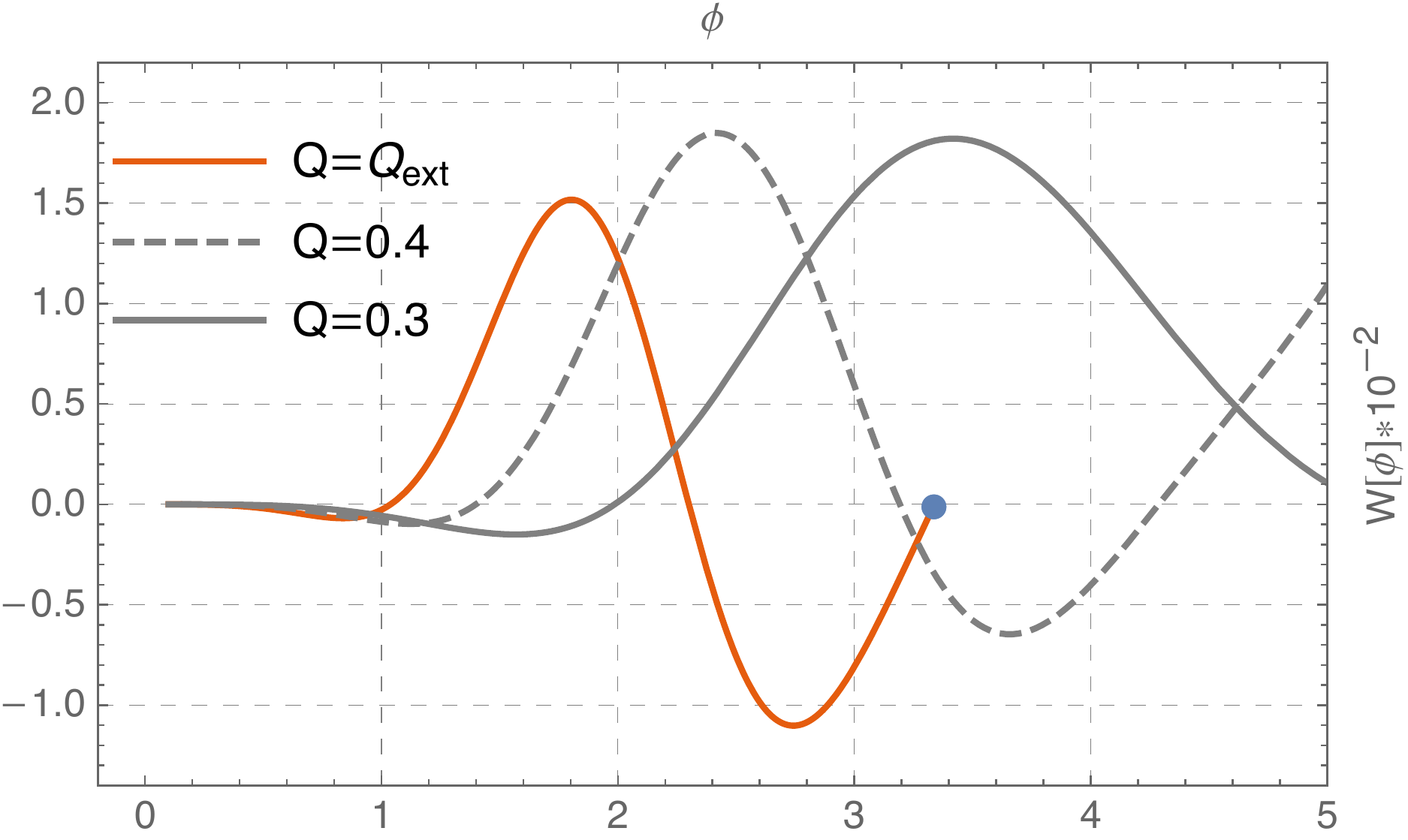}
         \caption{$W(\phi)$}
         \label{fig:W-phi}
     \end{subfigure}
      \captionsetup{width=.9\textwidth}
       \caption{Dependence of $V$ and $W$ on $\phi$. The orange curves (regular states) end up at blue points, while the gray curves (singular states) do not.  $\alpha=1$ is  set.}
        \label{fig:sep-scalar-2}
\end{figure}

The dependence of $V$ and $W$ on $\phi$ is exhibited in Fig.\ \ref{fig:sep-scalar-2}. 
It is worth emphasizing that $V$ and $W$ will end up in the extreme value of $\phi$, see Eq.\ \eqref{eq:max-phi},  when the BH is in its regular state (orange curves in Fig.\ \ref{fig:sep-scalar-2}).
Further, we note that $W(\phi)$ vanishes at $\phi_{\rm max} \approx 3.3386$. This implies $W\phi'^2=0$ at the BH center. In other words, the scalar field does not have kinetic energy at $r=0$ when the BH is in its regular state.

\subsection{Energy conditions of matter sources}

It is known that RBHs can bypass \cite{Ansoldi:2008jw,Zaslavskii:2010qz} the Penrose singularity
theorem because they violate the strong energy condition (SEC). 
In addition, the other energy conditions, like the null energy condition (NEC), the weak energy condition (WEC), and the dominant energy condition (DEC), are applied to examine the pathological behaviors of spacetime, see, e.g., Ref.\  \cite{Maeda:2021jdc} and the references within. Therefore,
the energy conditions play an important role in the study of RBHs \cite{Toshmatov:2017kmw,Simpson:2018tsi,Lan:2021ayk,Lan:2022bld}.

In the current subsection, we investigate the energy conditions in our model Eq. \eqref{eq:shape} from two aspects: One is the effective energy conditions of spacetime and the other aspect is the energy conditions of each matter source.

Given a diagonalized energy-momentum tensor, $T^\nu_\mu =\diag\{-\rho_0, p_1, p_2, p_3\}$,
the energy conditions can be expressed in terms of the diagonal components of $T^\nu_\mu$ as follows:
\begin{equation}
    \begin{split}
        \text{NEC:}& \quad \rho_0+p_i \geq 0, \\
        \text{WEC:}& \quad \rho_0 \geq 0 \bigcap \rho_0+p_i \geq 0, \\
        \text{SEC:}& \quad \rho_0+\sum_{i=1}^3 p_i \geq 0 \bigcap \rho_0+p_i \geq 0,\\
        \text{DEC:}& \quad \rho_0 \geq 0 \bigcap \rho_0-\left|p_i\right| \geq 0,
    \end{split}
\end{equation}
where $i=1,2,3$. 

\begin{figure}[!htb]
     \centering
     \begin{subfigure}[b]{0.2425\textwidth}
         \centering
         \includegraphics[width=\textwidth]{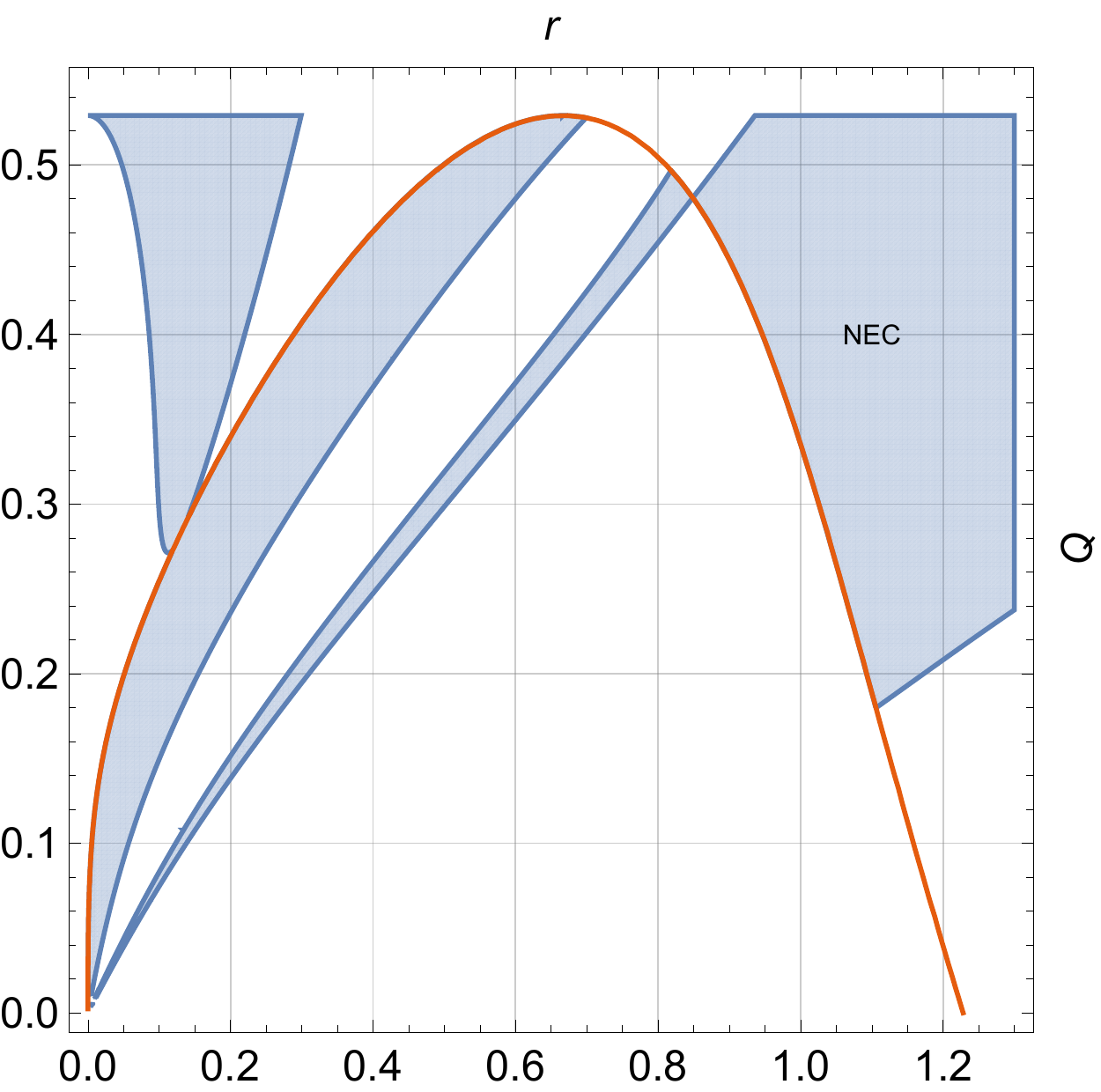}
         \caption{NEC}
         \label{fig:nec-plot}
     \end{subfigure}
     \begin{subfigure}[b]{0.2425\textwidth}
         \centering
         \includegraphics[width=\textwidth]{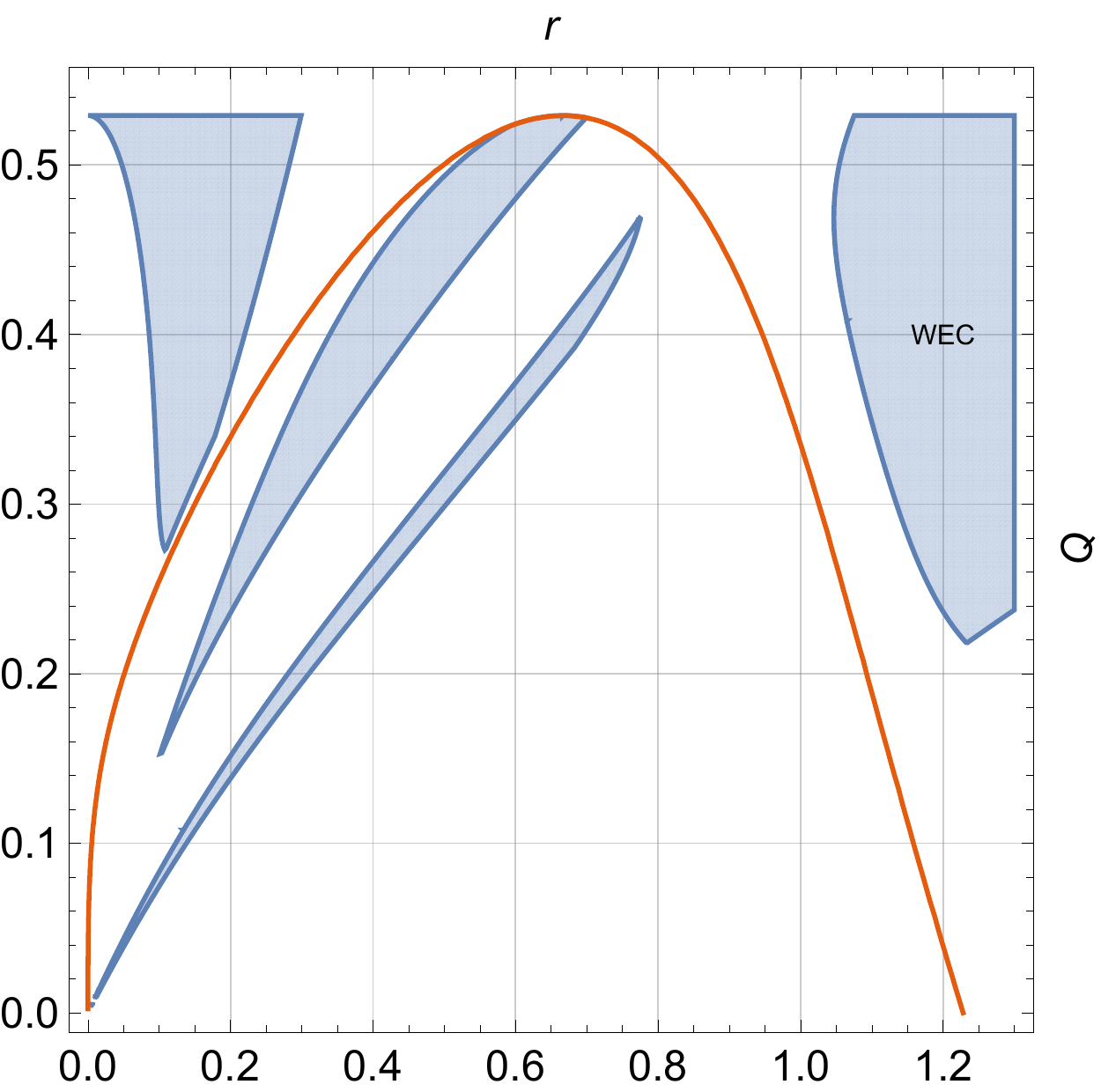}
         \caption{WEC}
         \label{fig:wec-plot}
     \end{subfigure}
     \begin{subfigure}[b]{0.2425\textwidth}
         \centering
         \includegraphics[width=\textwidth]{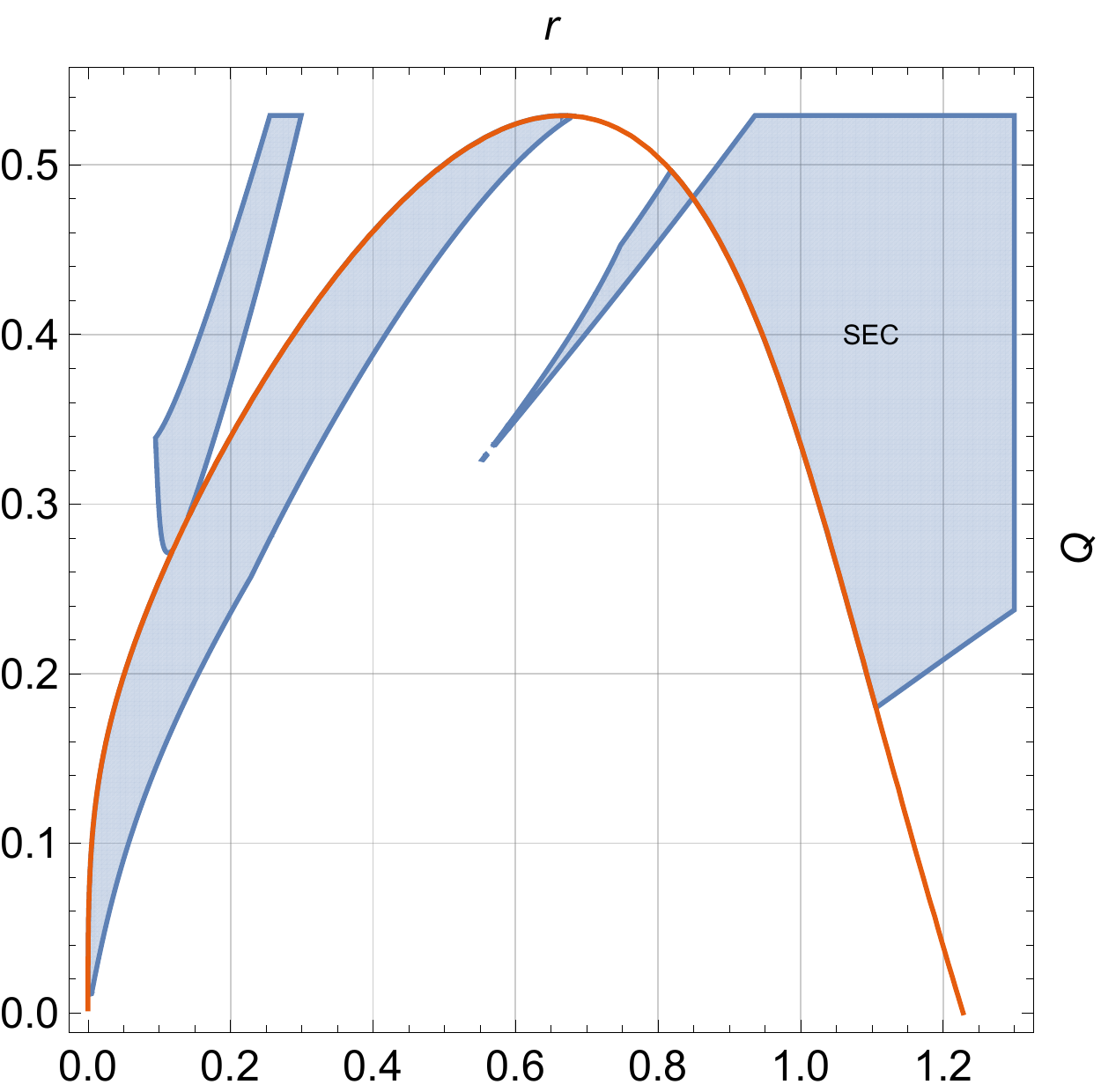}
         \caption{SEC}
         \label{fig:sec-plot}
     \end{subfigure}
     \begin{subfigure}[b]{0.2425\textwidth}
         \centering
         \includegraphics[width=\textwidth]{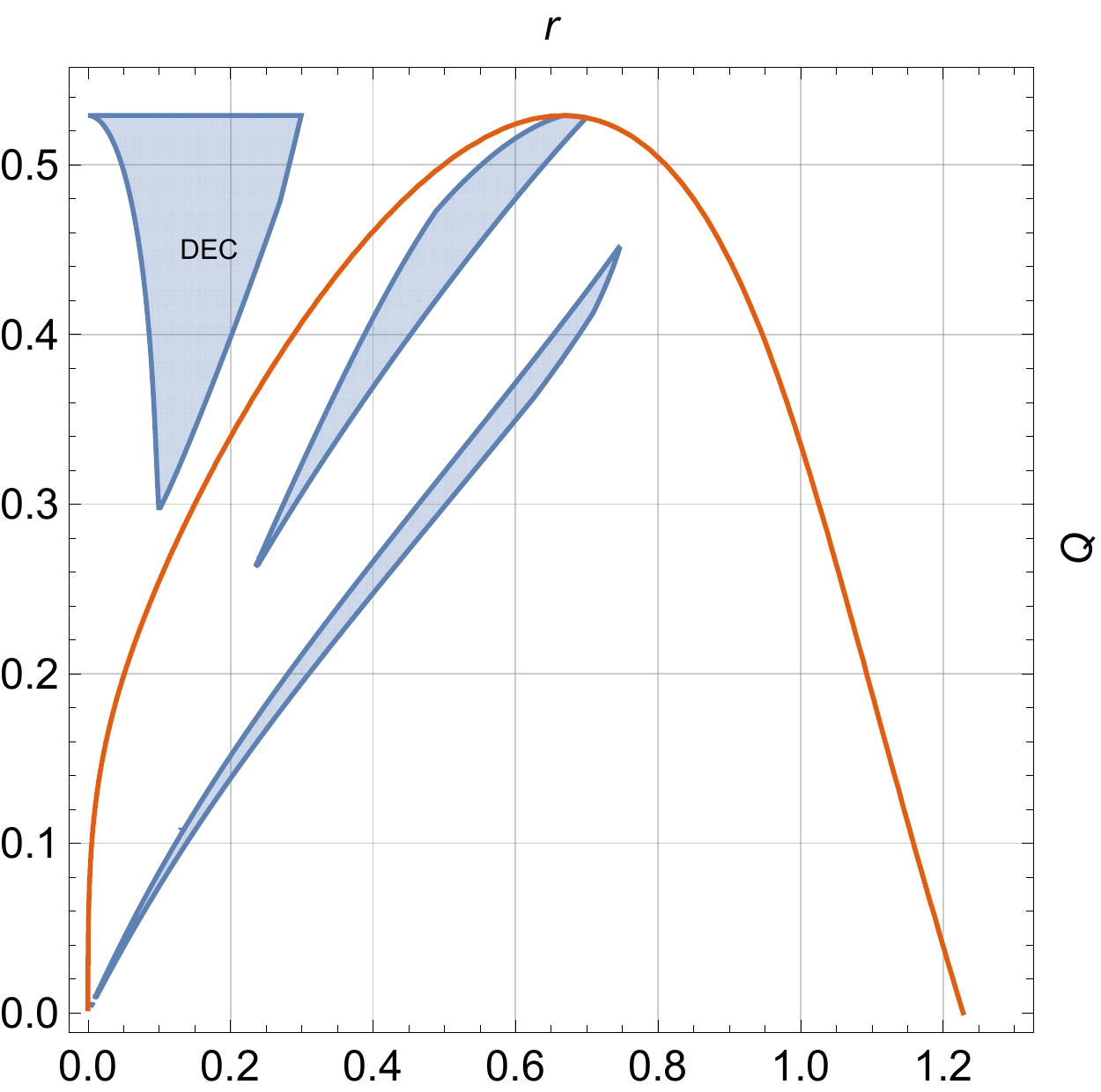}
         \caption{DEC}
         \label{fig:dec-plot}
     \end{subfigure}
      \captionsetup{width=.9\textwidth}
       \caption{Effective energy conditions of spacetime in $F(R)$ gravity. $\alpha=1$ is set.}
        \label{fig:en-cond}
\end{figure}

For the matter sources Eq.\ \eqref{eq:action-matter}, we define the energy densities and pressures of scalar and monopole fields, respectively,  using our conventions in Sec.\ \ref{subsec:sources}, 
\begin{subequations}
\begin{equation}
\label{eq:therm-val-scal}
    \rho_0[\phi] =T^0_0[\phi],\qquad
    p_i[\phi] =-T^i_i[\phi],
\end{equation}  
\begin{equation}
\label{eq:therm-val-a}
    \rho_0[A] =T^0_0[A],\qquad
    p_i[A] =-T^i_i[A].
\end{equation} 
\end{subequations}
Thus, the effective energy density and pressure of spacetime in $F(R)$ gravity can be introduced directly by the contributions of scalar field Eq.\ \eqref{eq:therm-val-scal} and monopole field Eq.\ \eqref{eq:therm-val-a},
\begin{equation}
    \rho_0[M]=T^0_0[\phi]+T^0_0[A],\qquad
    p_i[M] =-T^i_i[\phi]-T^i_i[A],
\end{equation}
where $M$ within a square bracket denotes the matter consisting of $\phi$ and $A$.
The effective energy conditions of spacetime are shown in Fig.\ \ref{fig:en-cond}, where we use the blue area to mark the validity of energy conditions and the orange curve to mark the horizon, $f(r_{\rm H}, Q)=0$.

\begin{figure}[!htb]
     \centering
     \begin{subfigure}[b]{0.2425\textwidth}
         \centering
         \includegraphics[width=\textwidth]{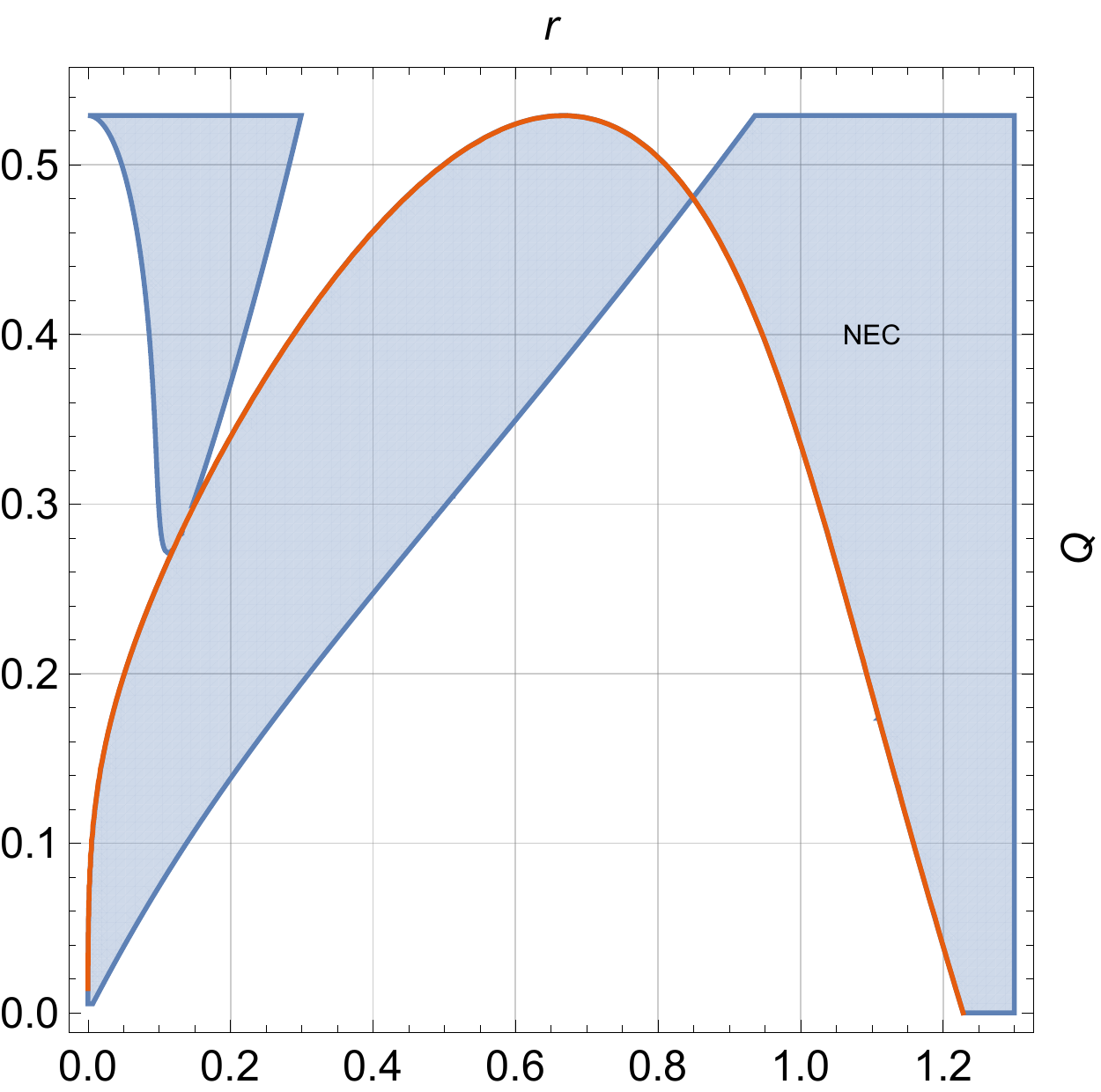}
         \caption{NEC}
         \label{fig:nec-scalar}
     \end{subfigure}
     \begin{subfigure}[b]{0.2425\textwidth}
         \centering
         \includegraphics[width=\textwidth]{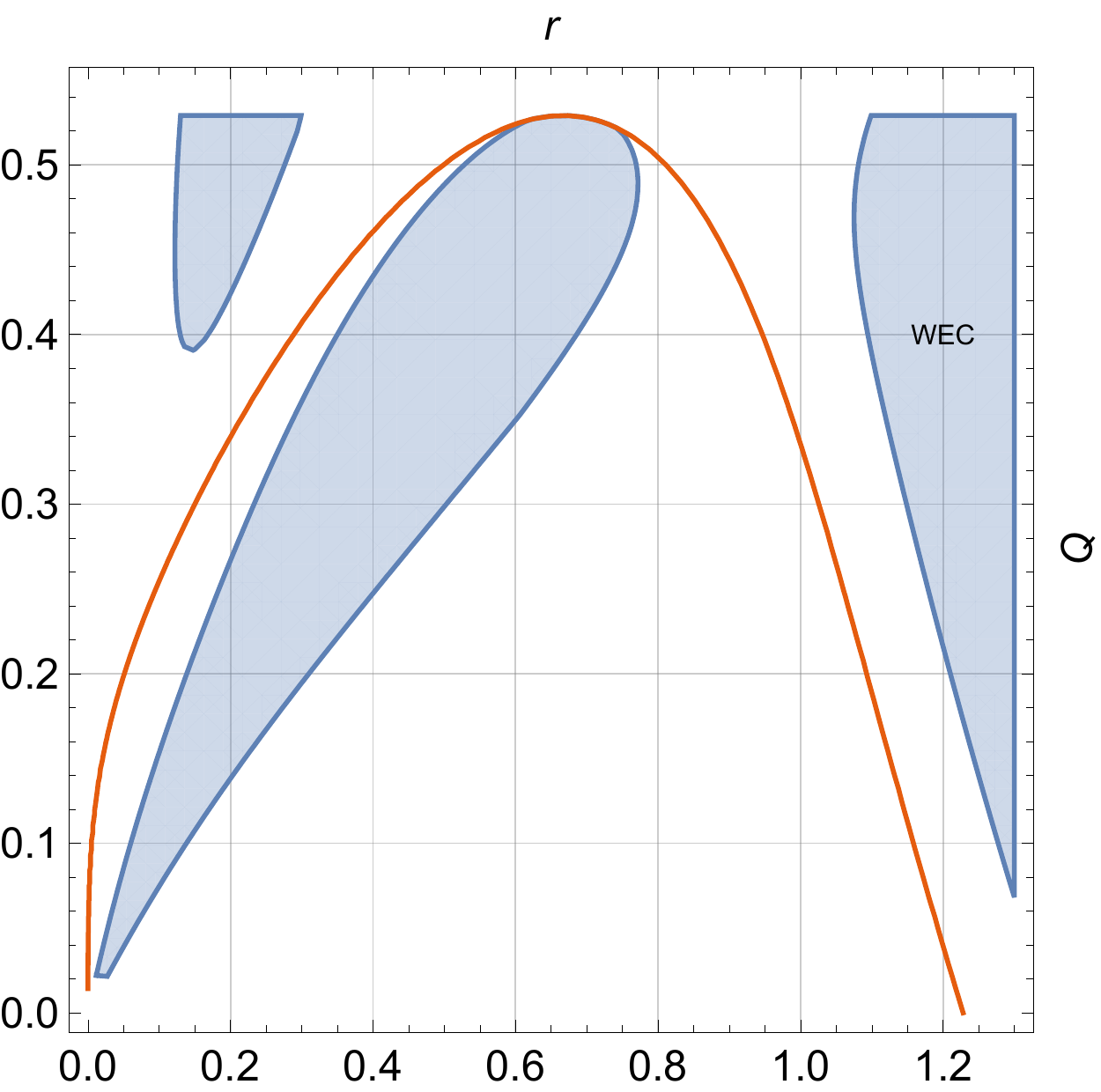}
         \caption{WEC}
         \label{fig:wec-scalar}
     \end{subfigure}
     \begin{subfigure}[b]{0.2425\textwidth}
         \centering
         \includegraphics[width=\textwidth]{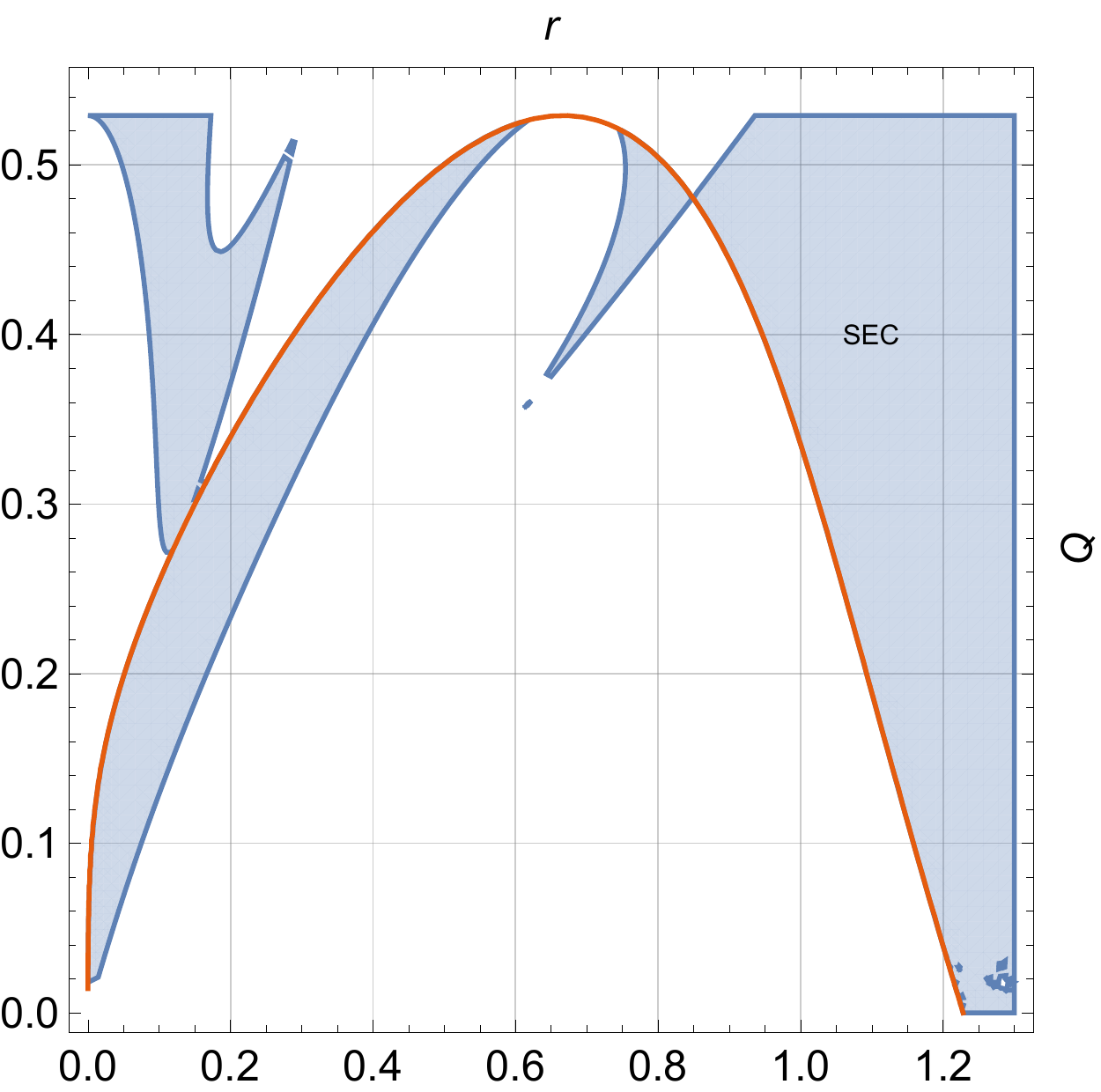}
         \caption{SEC}
         \label{fig:sec-scalar}
     \end{subfigure}
     \begin{subfigure}[b]{0.2425\textwidth}
         \centering
         \includegraphics[width=\textwidth]{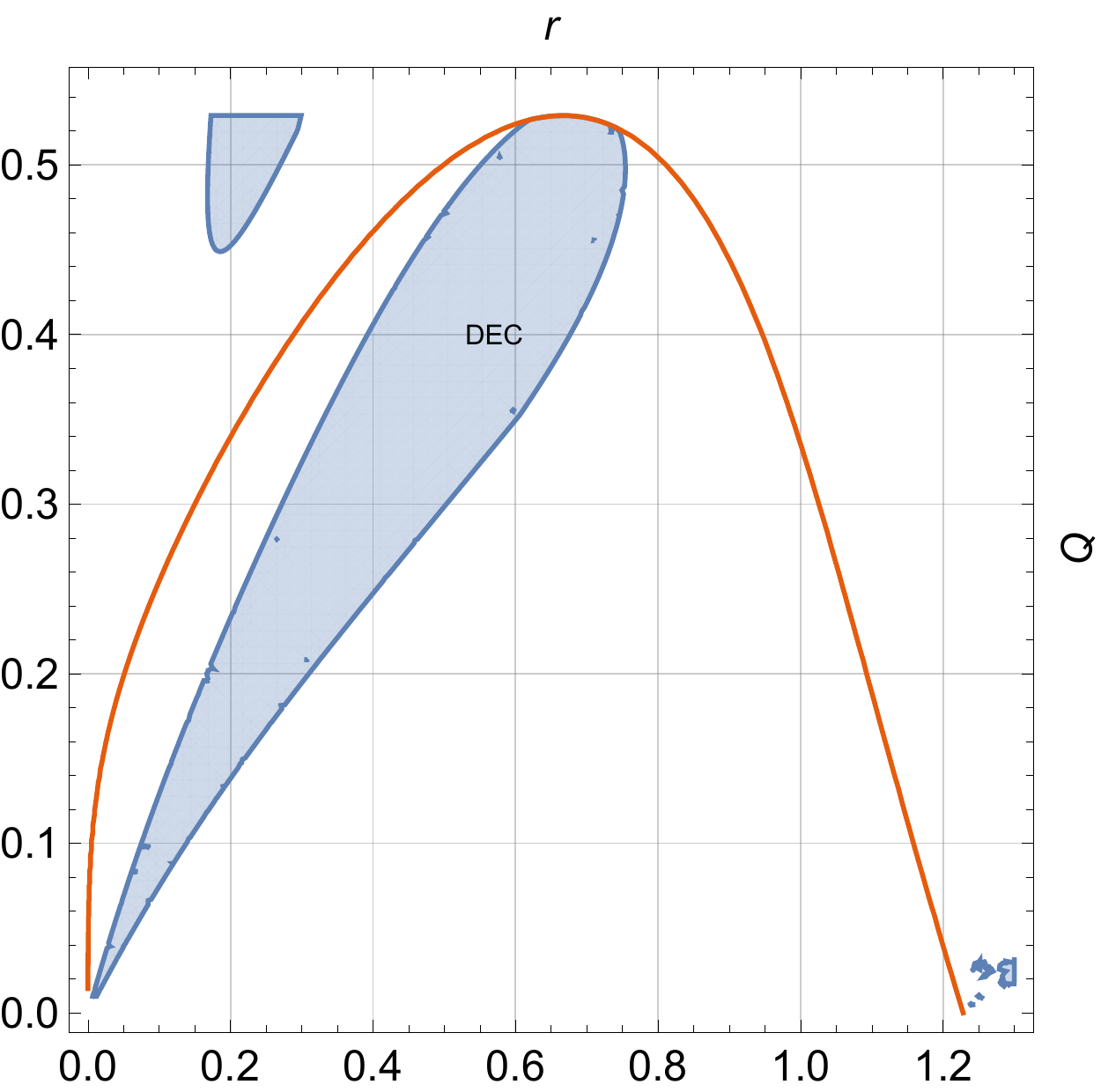}
         \caption{DEC}
         \label{fig:dec-scalar}
     \end{subfigure}
      \captionsetup{width=.9\textwidth}
       \caption{Energy conditions of the scalar field}
        \label{fig:en-cond-scalar}
\end{figure}

\begin{figure}[!htb]
     \centering
     \begin{subfigure}[b]{0.2425\textwidth}
         \centering
         \includegraphics[width=\textwidth]{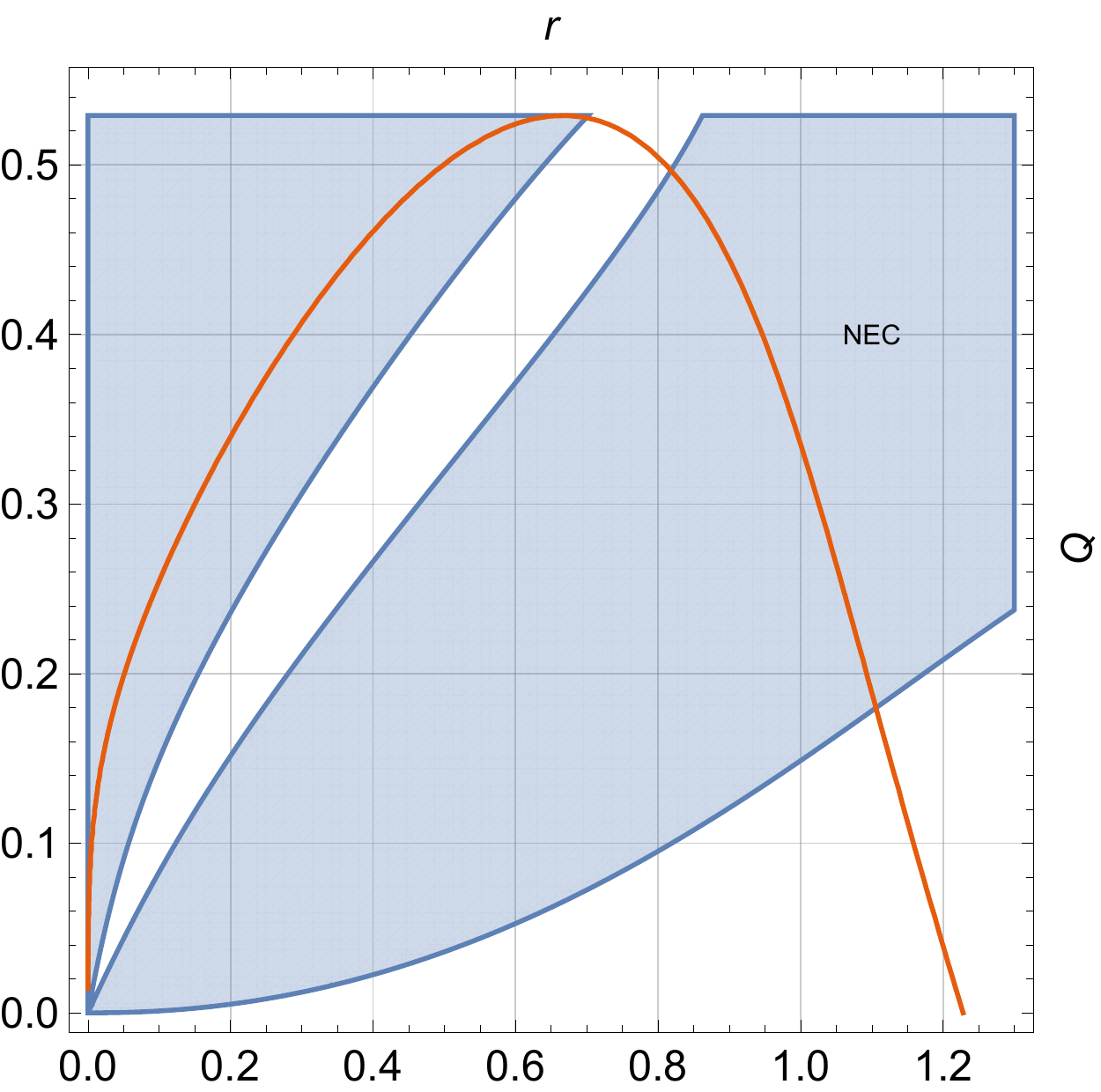}
         \caption{NEC}
         \label{fig:nec-vector}
     \end{subfigure}
     \begin{subfigure}[b]{0.2425\textwidth}
         \centering
         \includegraphics[width=\textwidth]{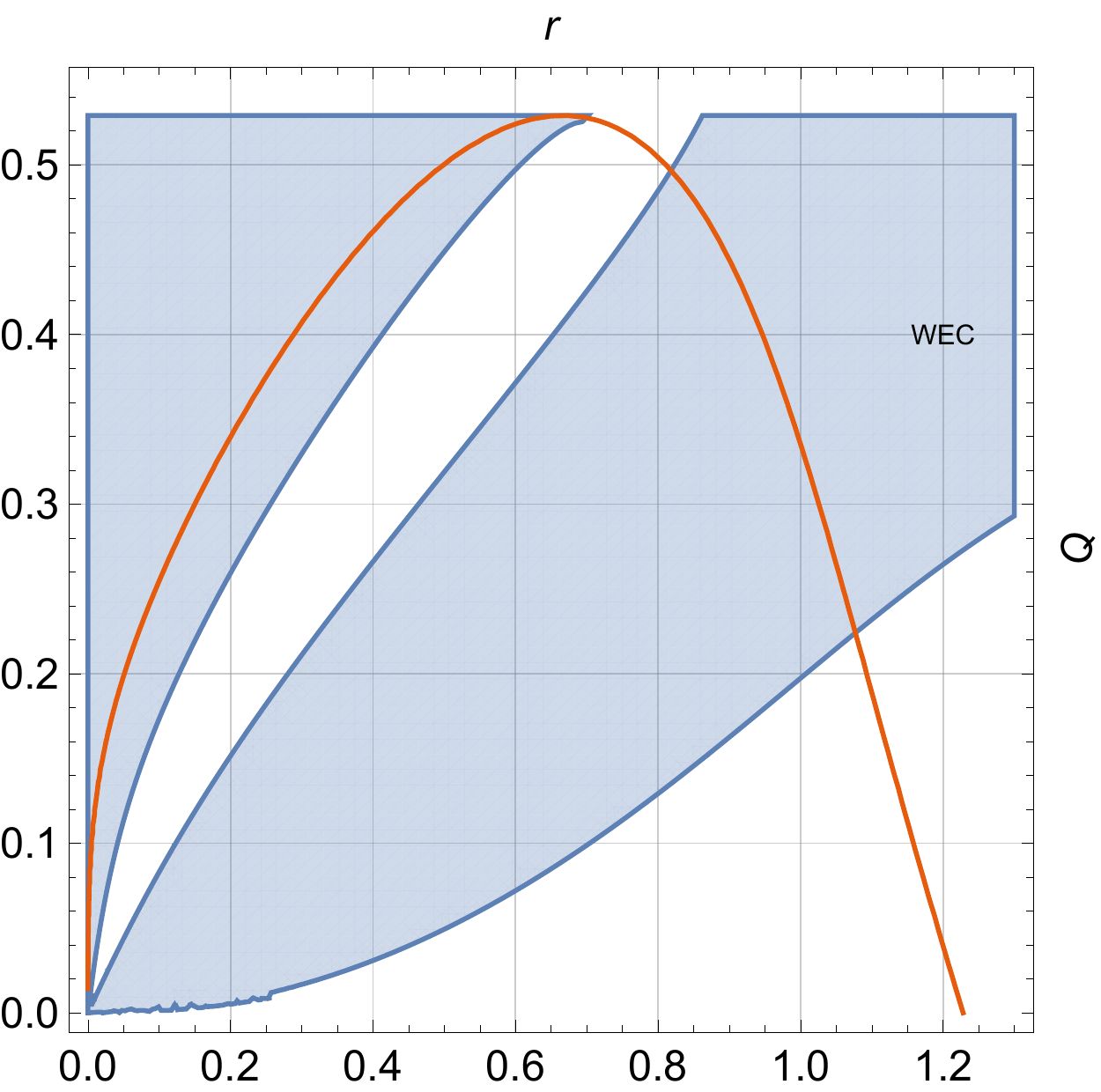}
         \caption{WEC}
         \label{fig:wec-vector}
     \end{subfigure}
     \begin{subfigure}[b]{0.2425\textwidth}
         \centering
         \includegraphics[width=\textwidth]{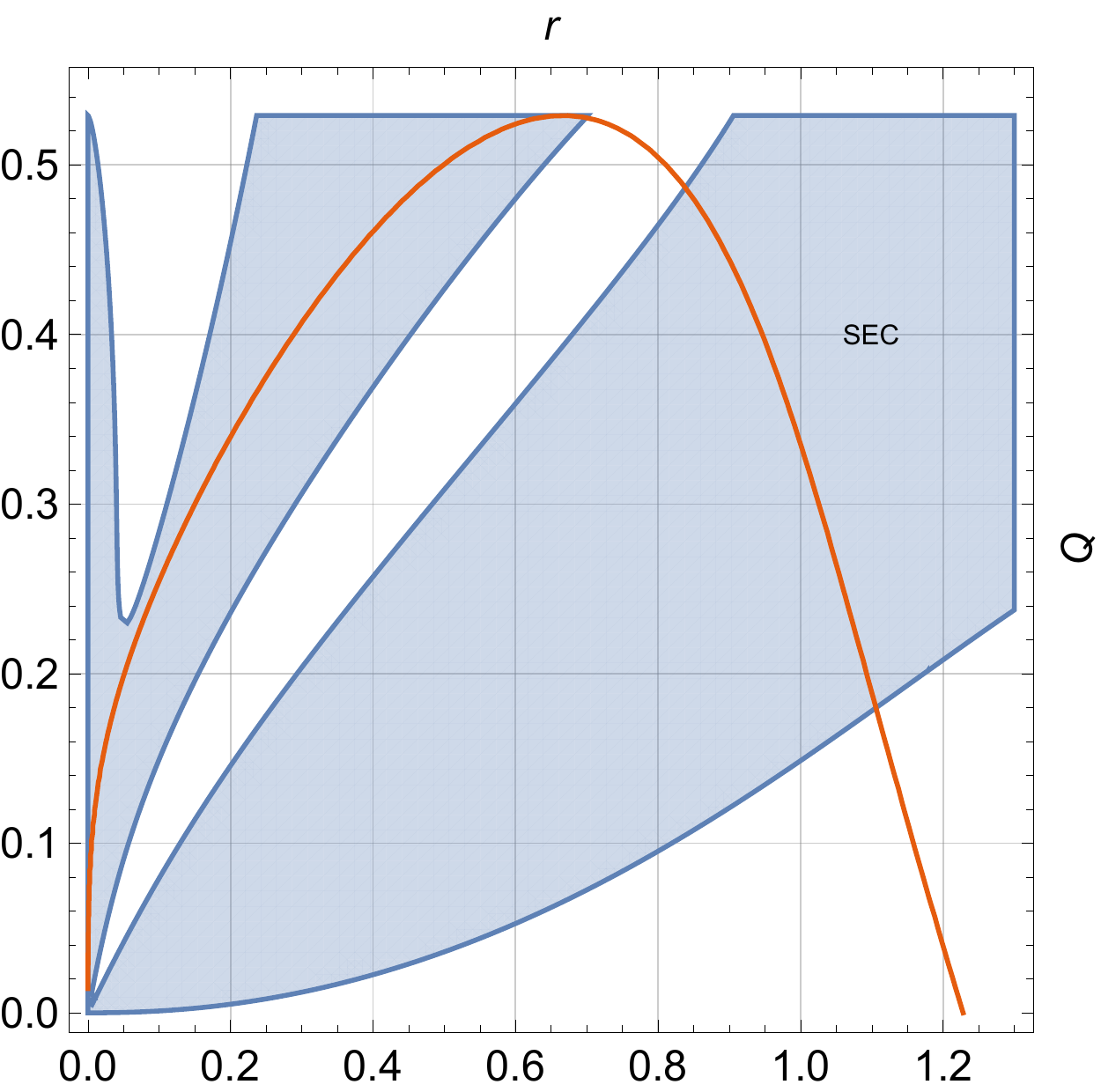}
         \caption{SEC}
         \label{fig:sec-vector}
     \end{subfigure}
     \begin{subfigure}[b]{0.2425\textwidth}
         \centering
         \includegraphics[width=\textwidth]{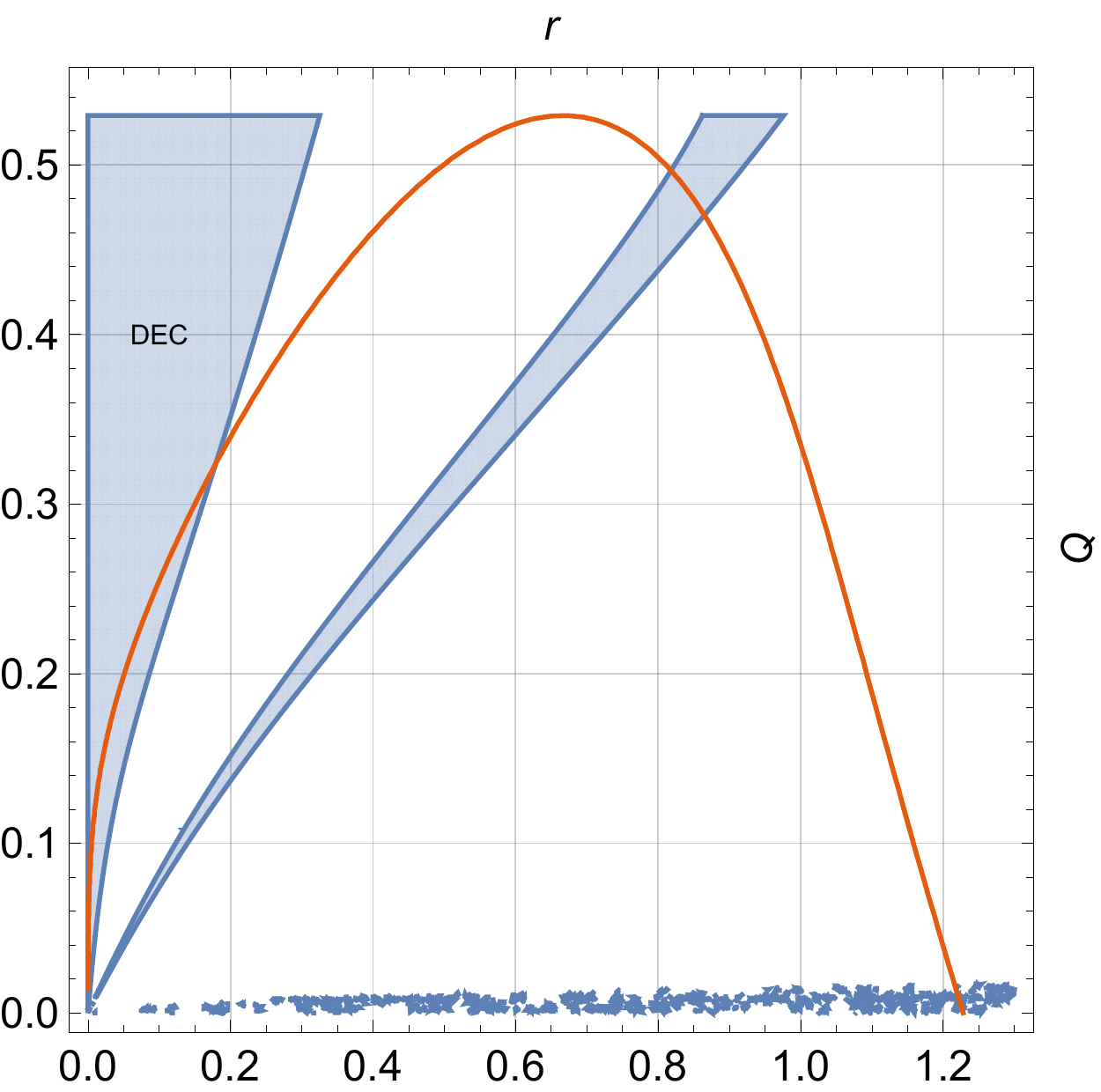}
         \caption{DEC}
         \label{fig:dec-vector}
     \end{subfigure}
      \captionsetup{width=.9\textwidth}
       \caption{Energy conditions of the monopole field}
        \label{fig:en-cond-vector}
\end{figure}

We note that the effective energy conditions of spacetime are badly damaged, especially the DEC. 
In addition, the NEC, the WEC, and the SEC are disrupted outside the horizon when $Q$ becomes small. 
To figure out which ingredient of matter is responsible for such damage, we draw the energy conditions associated with the scalar and monopole fields in Figs.\ \ref{fig:en-cond-scalar} and \ref{fig:en-cond-vector}, respectively.

It can be seen that the DEC of scalar and monopole fields is destroyed outside the horizon, whereas the NEC, WEC, and SEC are damaged outside the horizon by the monopole field when the charge is small.
In addition, 
Fig.\ \ref{fig:sec-plot}
 demonstrates that the SEC is broken near $r=0$ even in the singular state,
and such a violation is caused by the scalar field, which can be deduced when we  compare 
Fig.\ \ref{fig:sec-scalar} with Fig.\ \ref{fig:sec-vector}.
As a result, it is not possible to give the reason for the existence of singularity just from the point of view of energy conditions.
Perhaps we need to start with the complete theory of geodesic convergence and then analyze Raychaudhuri's equation \cite{Lan:2021ngq}.

\section{Quasinormal frequencies in our model}
\label{sec:qnf}

In this section, we consider the dynamic properties of our model by 
investigating the frequencies of quasinormal modes (QNMs) and their damping limit for a massless scalar field perturbation, i.e., the so-called asymptotic quasinormal frequencies (QNFs). 
The calculations we shall make are based on the 13th-order WKB approach improved by the Pad\' e approximation \cite{Matyjasek:2017psv,Konoplya:2019hlu}, the light ring/QNMs correspondence \cite{Cardoso:2008bp,Wei:2019jve}, and the monodromy method \cite{Motl:2003cd,Natario:2004jd,Lan:2022qbb}.

\subsection{Scalar field perturbation}

\begin{figure}[!htb]
     \centering
     \begin{subfigure}[b]{0.45\textwidth}
         \centering
         \includegraphics[width=\textwidth]{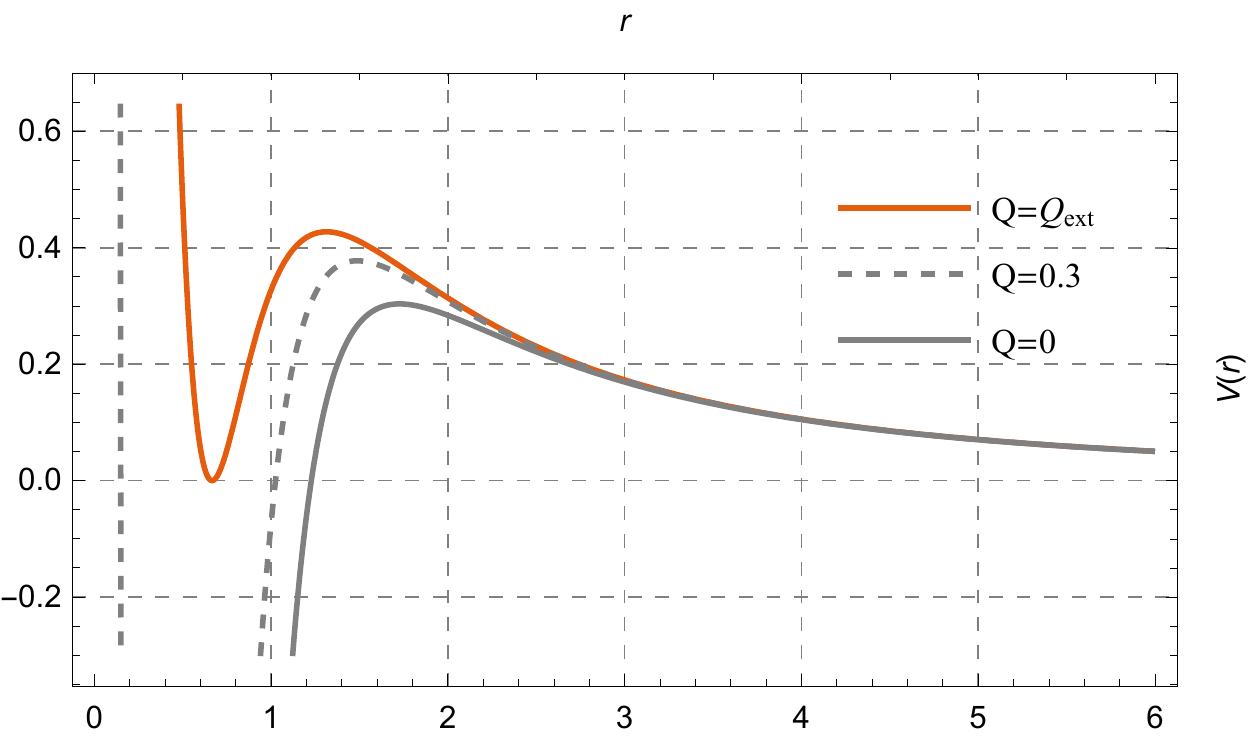}
         \caption{$V(r)$}
         \label{fig:Vr-plot}
     \end{subfigure}
     \begin{subfigure}[b]{0.45\textwidth}
         \centering
         \includegraphics[width=\textwidth]{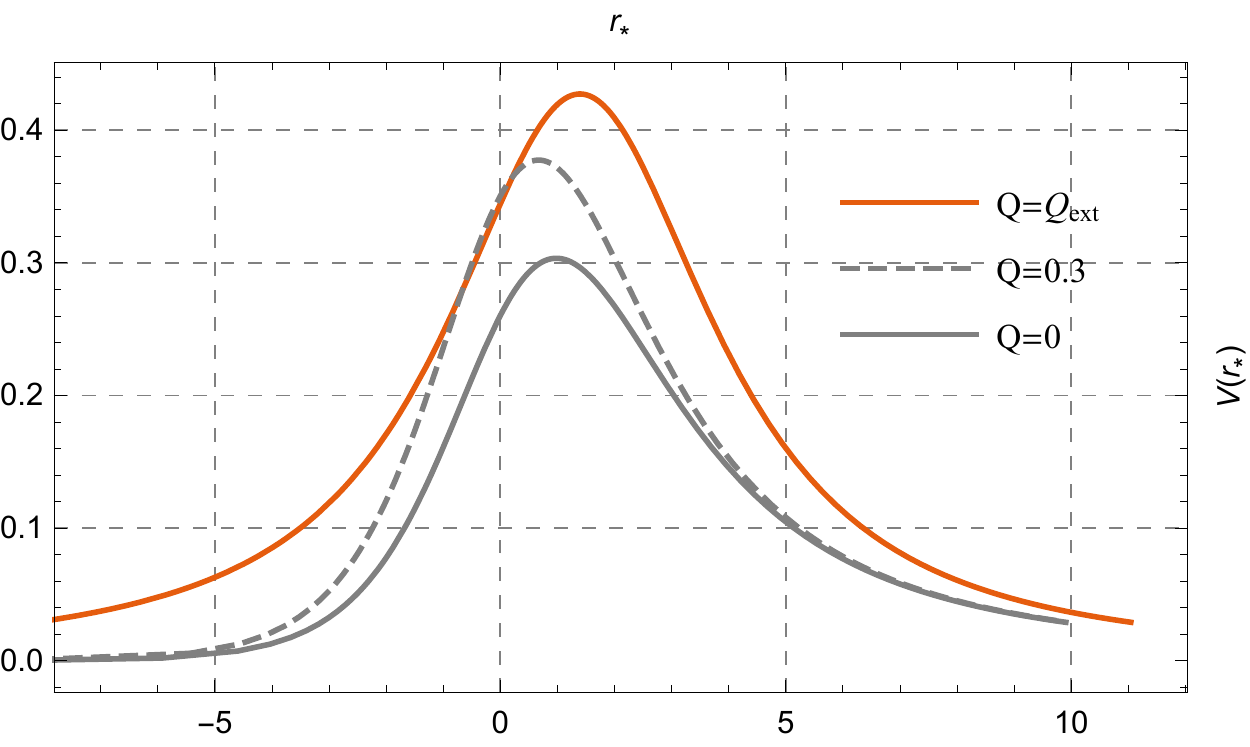}
         \caption{$V(r_*)$}
         \label{fig:Vrt-plot}
     \end{subfigure}
      \captionsetup{width=.9\textwidth}
       \caption{Effective potential with respect to $r$ and $r_*$, respectively. $\alpha=l=1$ is set.}
        \label{fig:veff}
\end{figure}

We write down the equation of a massless scalar field \cite{Kokkotas:1999bd,Berti:2009kk,Konoplya:2011qq},
\begin{equation}
	\frac{1}{\sqrt{- g}} \partial_{\mu}  \left( \sqrt{- g} g^{\mu \nu}
	\partial_{\nu} \Phi \right) = 0. 
\end{equation}
To separate the variables, we take the ansatz,
\begin{equation}
	\Phi = \frac{\Psi (r)}{r} Y_{l}^m (\theta, \phi) \me^{- \mi
		\omega t},
\end{equation}
and then obtain the equation satisfied by the radial component,
\begin{equation}
\label{eq:master}
	\frac{\dif^2 \Psi}{\dif r_*^2} + \omega^2 \Psi = V_{\rm eff}(r) \Psi, 
\end{equation}
where $\displaystyle r_*\coloneqq \int \dif r/f(r)$ denotes the tortoise coordinate
and the effective potential reads
\begin{equation}
\label{eq:eff-pont}
	V _{\rm eff}(r) = f (r)  \left[ \frac{l (l + 1)}{r^2} +   \frac{f'(r)}{r}
	 \right].
\end{equation}
Substituting the shape function Eq.\ \eqref{eq:shape} into the above equation, we plot the effective potential for three $Q$ values in Fig.\ \ref{fig:veff}, where Fig.\ \ref{fig:Vr-plot} is for $V(r)$ and Fig.\ \ref{fig:Vrt-plot} for $V(r_*)$. 

The shape of the effective potential implies that we are dealing with a scattering problem. The so-called quasinormal modes are defined as the solution of Eq.\ \eqref{eq:master} satisfying  the particular boundary conditions,
\begin{equation}
\label{eq:boundary}
    \Psi \sim \me^{\pm \mi \omega r_*}, \qquad r_*\to \pm \infty.
\end{equation}
To compute the QNFs, we use the 13th-order WKB approach improved by the Pad\'e approximation
\cite{Matyjasek:2017psv,Konoplya:2019hlu}.

\begin{figure}[!htb]
     \centering
     \begin{subfigure}[b]{0.45\textwidth}
         \centering
         \includegraphics[width=\textwidth]{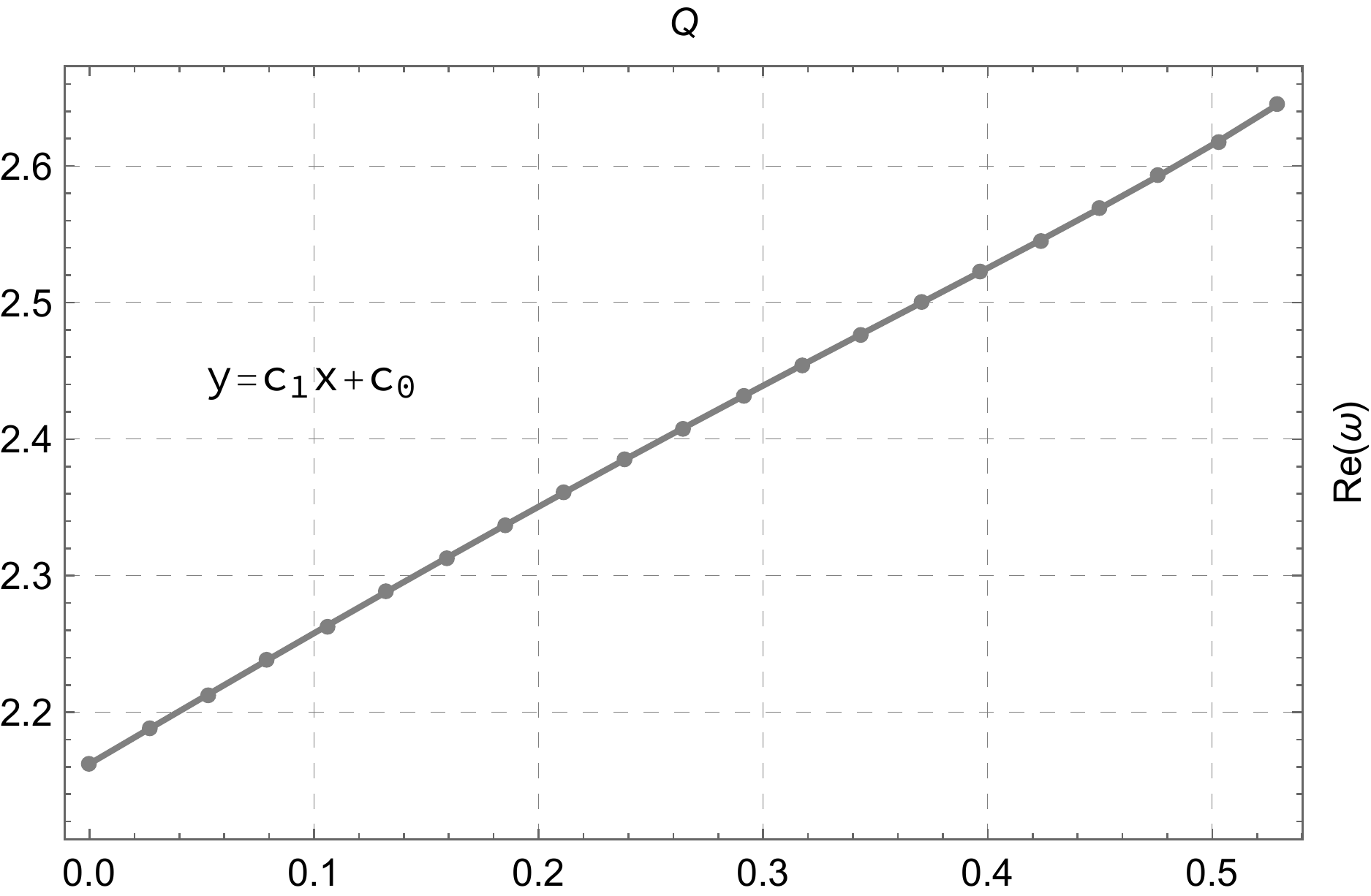}
         \caption{$\Re(\omega)$ with respect to $Q$}
         \label{fig:qnf-re-Q}
     \end{subfigure}
     \begin{subfigure}[b]{0.45\textwidth}
         \centering
         \includegraphics[width=\textwidth]{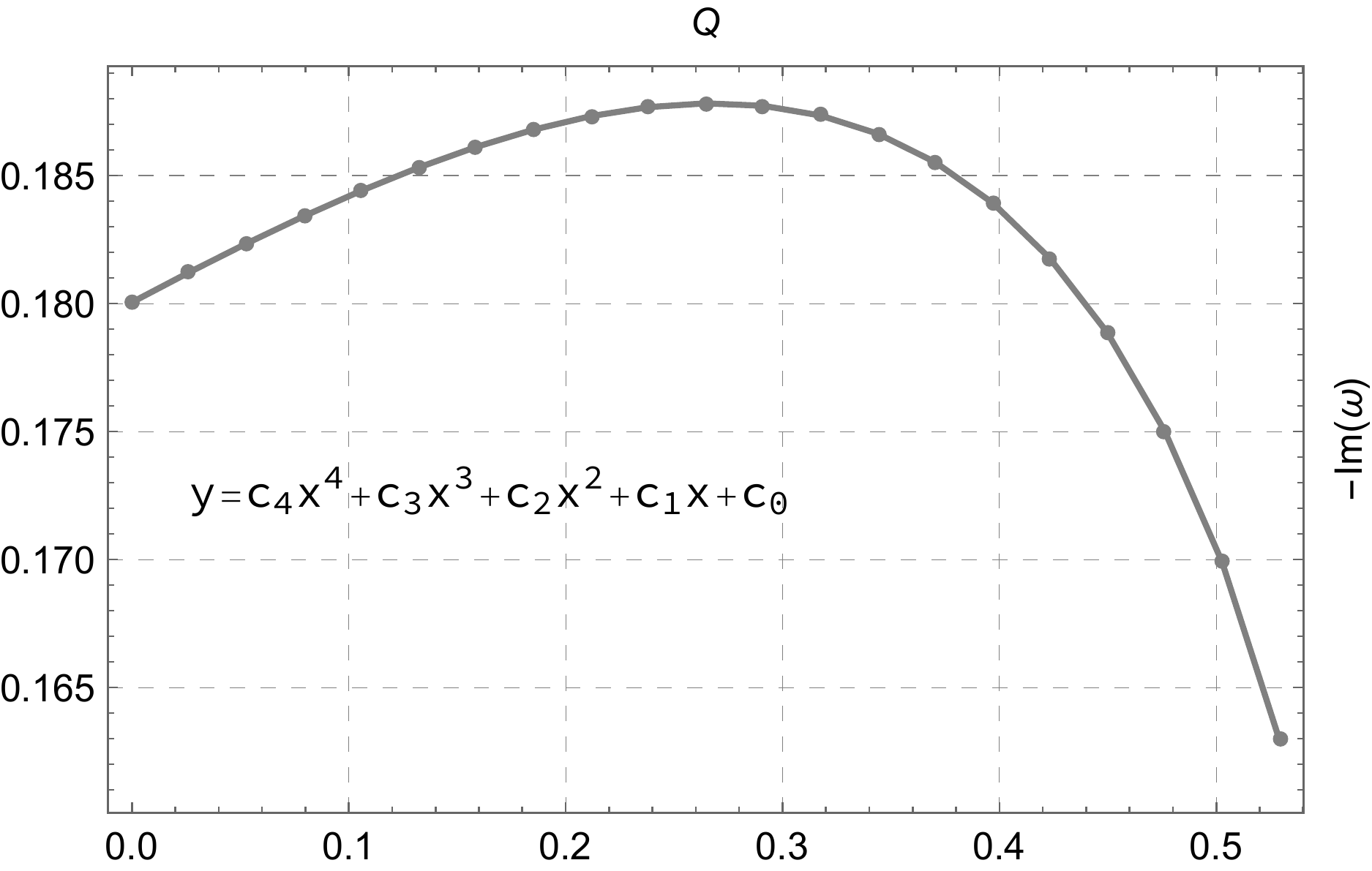}
         \caption{$-\Im(\omega)$ with respect to $Q$}
         \label{fig:qnf-im-Q}
     \end{subfigure}
      \captionsetup{width=.9\textwidth}
       \caption{Dependence of QNFs on the charge $Q$ with $l=6$.}
        \label{fig:qnf-6}
\end{figure}

For a specific value of $l=6$, we compute the QNFs with different values of $Q$. 
The relation is shown in Fig.\ \ref{fig:qnf-6}, where we find the curves that fit the data best via the polynomials of $Q$.
The linear relation of the real part of the QNFs with charge is obvious in Fig.\ \ref{fig:qnf-re-Q}, 
where the fit line is of the form,
\begin{equation}
\label{eq:linear-fit}
    y=c_1 x +c_0,
\end{equation}
with $c_1\approx0.8989$ and $c_0\approx2.1675$. However, the imaginary part shows the nonlinear relation with respect to the charge,
\begin{equation}
    y=c_4 x^4 + c_3 x^3 + c_2 x^2+c_1 x +c_0,
\end{equation}
where the fit coefficients take the values as follows:
\begin{equation}
    c_0\approx 0.1798,\quad
    c_1\approx 0.0594,\quad
    c_2\approx -0.2010,\quad
    c_3\approx 0.6202,\quad
    c_4\approx -1.0660.
\end{equation}

\begin{figure}[!htb]
     \centering
    \includegraphics[width=0.7\textwidth]{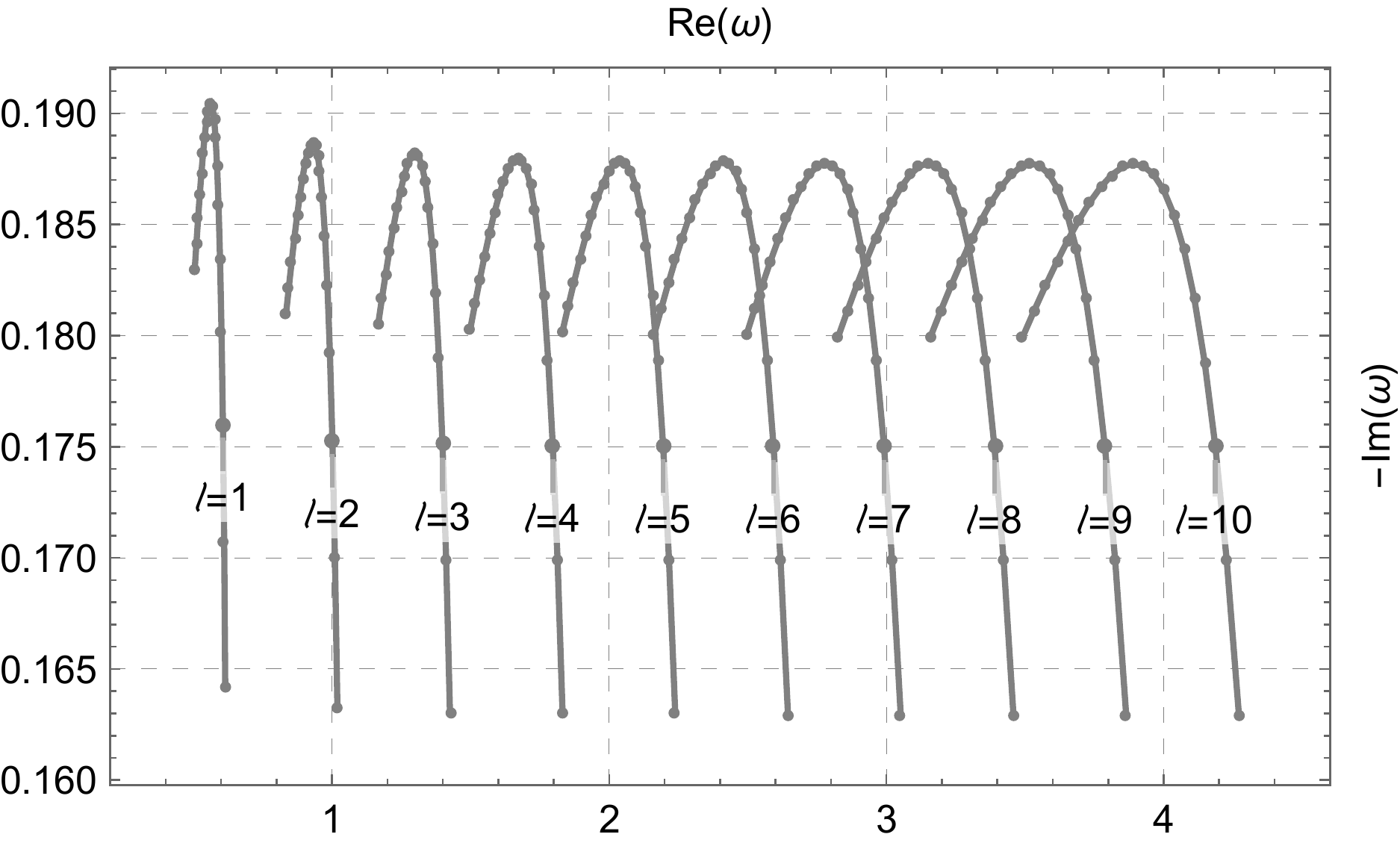}
      \captionsetup{width=.9\textwidth}
       \caption{Quasinormal frequencies in the complex frequency plane}
        \label{fig:qnf}
\end{figure}

Fig.\ \ref{fig:qnf} depicts the relationship between $\Re(\omega)$ and $-\Im(\omega)$  with more values of charge $Q\in [0, Q_{\rm ext}]$ and multipole number $l\in[1,10]$, 
where the frequencies with the same multipole number $l$ are joined into a curve.
Meanwhile, all curves have a similar shape:
One global minimum is located at $Q_{\rm ext}$ and one global maximum at a critical value $Q_{\rm C}$.
We calculate these critical values $Q_{\rm C}$ for a varying multipole number $l\in[1,10]$ and their corresponding QNFs, see Tab.\ \ref{tab:maximum}.
\begin{table}[ht!]
\begin{center}
\begin{tabular}{l*{9}{c}r}
\hline\hline
              & $l=1$ & $l=2$ & $l=3$ & $l=4$ & $l=5$  & $l=6$ & $l=7$ & $l=8$ & $l=9$ & $l=10$\\
\hline
$Q_{\rm C}$ & 0.2861 & 0.2910 & 0.2929 & 0.2938 & 0.2943 & 0.2946 & 0.2947 & 0.2949 & \boxit{1.1in} 0.2950 & 0.2950 \\
$\Re(\omega)$  & 0.5681 & 0.9394 & 1.3126 & 1.6864 &  2.0603 & 2.4345 &  2.8086 & 3.1829 & 3.5571 & 3.9314\\
$-\Im(\omega)$ & 0.1903 & 0.1886 & 0.1881 & 0.1879 &  0.1878 & 0.1877 &  0.1877 & \boxit{1.7in} 0.1876 & 0.1876 & 0.1876 \\
\hline
\end{tabular}
\caption{$Q_{\rm C}$, the real part, and the maximum of the minus imaginary part with respect to the multipole number, $l\in[1,10]$.\label{tab:maximum}}
\end{center}
\end{table}
The data circled by the red boxes in Tab.\ \ref{tab:maximum} show a possible phenomenon that $Q_{\rm C}$ and $-\Im(\omega)$ converge to constants in the eikonal limit ($l\gg 1$).

\begin{figure}[!htb]
     \centering
     \begin{subfigure}[b]{0.4\textwidth}
         \centering
         \includegraphics[width=\textwidth]{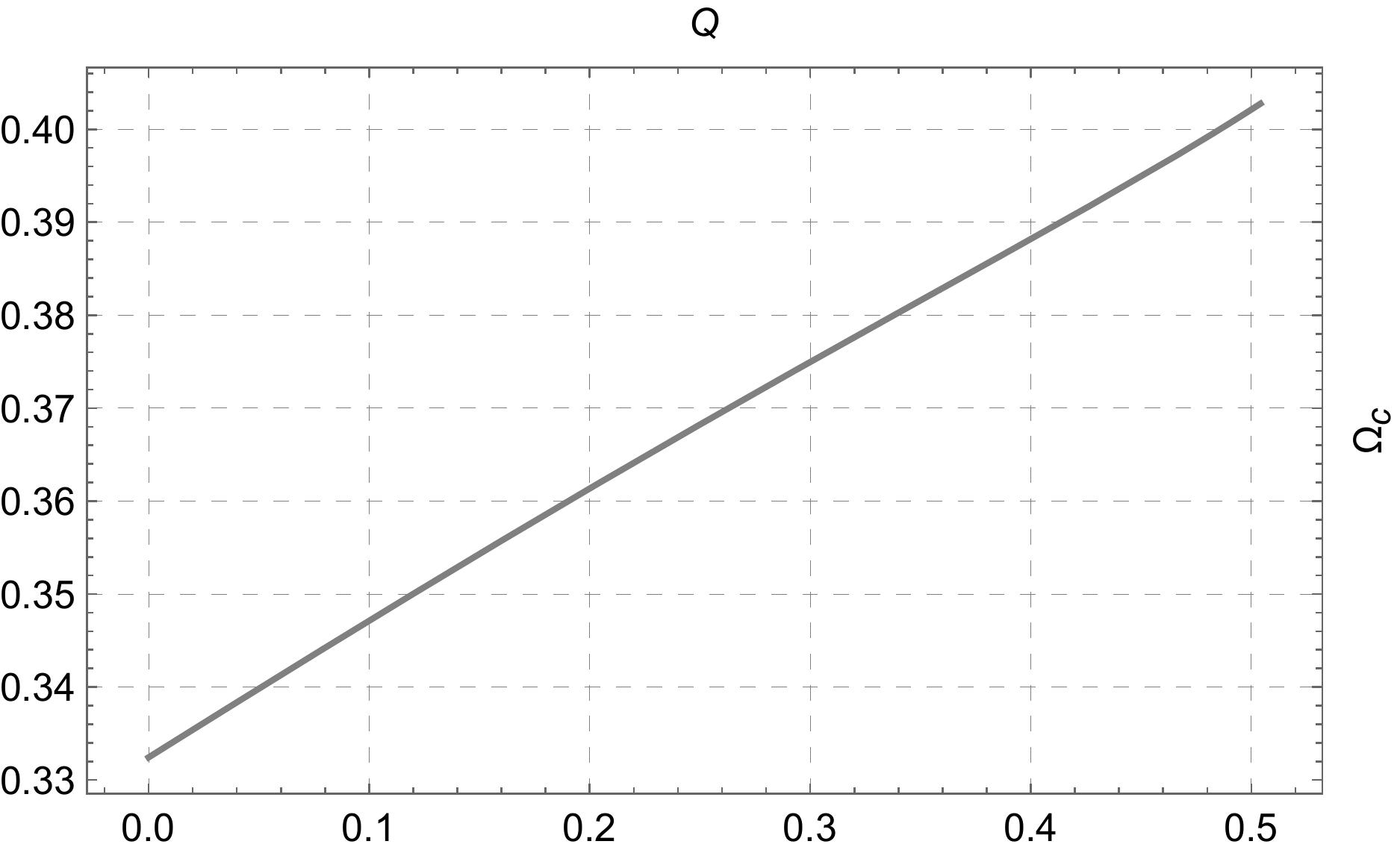}
         \caption{$\Omega_c$ with respect to $Q$}
         \label{fig:ang-vel}
     \end{subfigure}
     \begin{subfigure}[b]{0.4\textwidth}
         \centering
         \includegraphics[width=\textwidth]{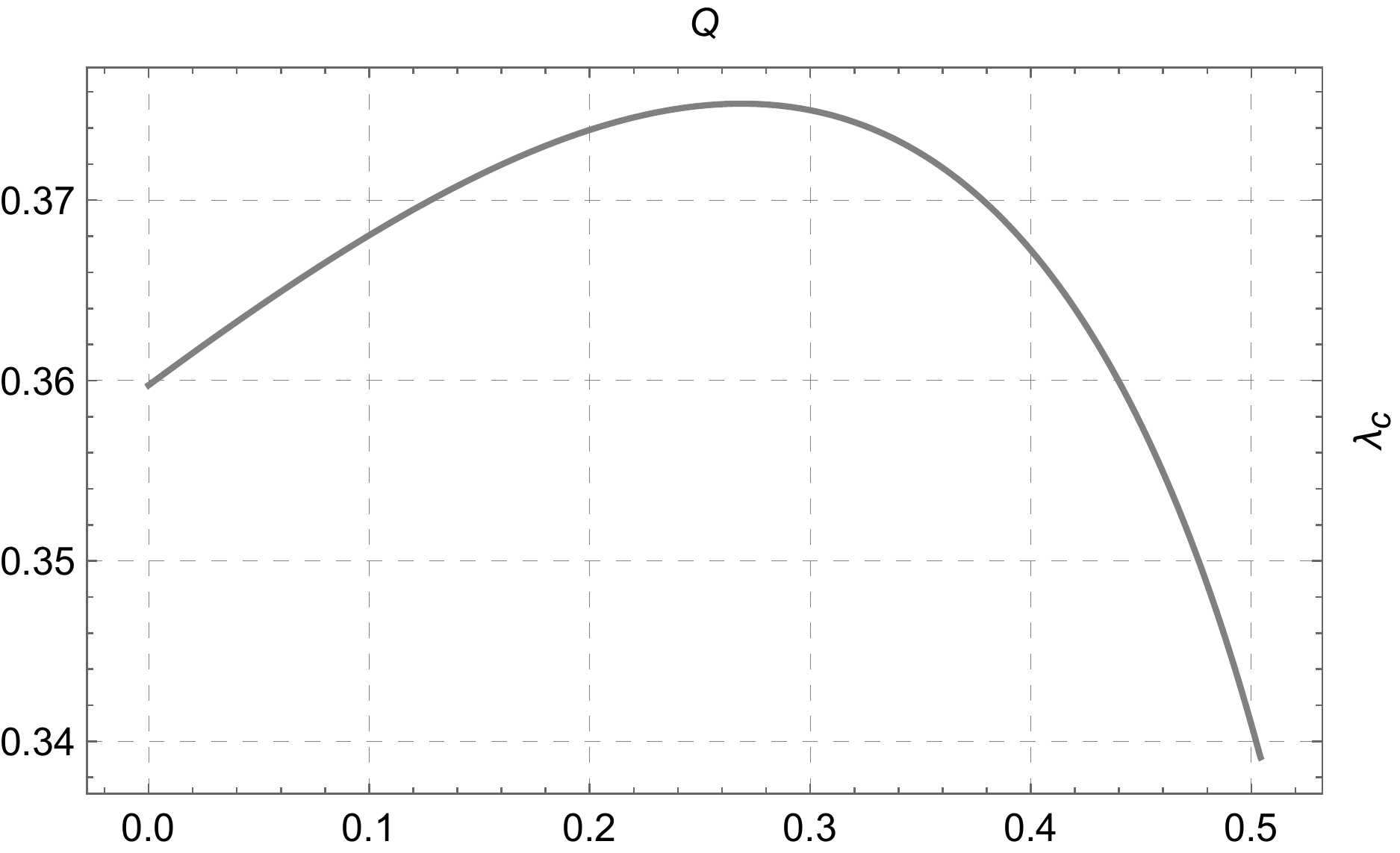}
         \caption{$\lambda_c$ with respect to $Q$}
         \label{fig:lya-exp}
     \end{subfigure}
      \captionsetup{width=.9\textwidth}
       \caption{Angular velocity and Lyapunov exponent}
        \label{fig:photon-sph}
\end{figure}

To provide an analytic result for the QNFs, we apply the {\em light ring/QNMs correspondence} \cite{Cardoso:2008bp}, 
according to which the QNFs can be cast in the following form,
\begin{equation}
    \omega = \Omega_c \,l -\mi \left(n+\frac{1}{2}\right)|\lambda_c|,
\end{equation}
where 
$\Omega_c$ denotes the angular velocity of a test scalar particle moving along an unstable null geodesic and $\lambda_c$ is the Lyapunov exponent. 
The applicability of this correspondence is limited, as evidenced by the case of the Einstein-Lovelock theory~\cite{Konoplya:2017wot}. Furthermore, if the magnetic field’s influence on photon spheres is disregarded in our model, the correspondence is valid. However, if the magnetic field’s influence on photon spheres is considered, the correspondence is invalid, as in the cases discussed in Refs.~\cite{Stuchlik:2019uvf,Toshmatov:2019gxg}.
Note that
the subscript $c$ means that the angular velocity and Lyapunov exponent are calculated at the radius of a circular null geodesic, $r_c$, determined by 
\begin{equation}
    2 f_c =  r_c f'_c,
\end{equation}
where $f_c:=f(r_c)$ and a prime means the derivative with respect to $r$. Thus we have \cite{Wei:2019jve}
\begin{equation}
    \Omega_c^2=\frac{f_c}{r_c^2},\qquad
    \lambda_c^2 =\frac{f_c}{2r_c^2}\left(
    2f_c- r_c^2 f''_c
    \right).
\end{equation}
For the shape function Eq.\ \eqref{eq:shape}, we obtain the equation of a photon sphere,
\begin{equation}
    Q^3 r_c \left(Q-Q_{\rm{ext}}\right)^2+4 r_c^4 \left(Q-Q_{\rm{ext}}\right)^2-2 \left(r_c^3+Q^3\right)^2+3 r_c^5=0,
\end{equation}
the angular velocity, 
\begin{subequations}
    \begin{equation}
        \Omega_c^2 =\frac{-r_c \left(Q-Q_{\rm ext}\right){}^2+r_c^3-r_c^2+Q^3}{r_c^2 \left(r_c^3+Q^3\right)},
    \end{equation}
and the Lyapunov exponent,
    \begin{equation}
    \begin{split}
        \lambda_c^2 r_c^2 \left(r_c^3+Q^3\right)^4=
        &\left[-r_c \left(Q-Q_{\rm{ext}}\right){}^2+r_c^3-r_c^2+Q^3\right]\\
        & \times\Big[-Q^6 r_c \left(Q-Q_{\rm{ext}}\right)^2-8 Q^3 r_c^4 \left(Q-Q_{\rm{ext}}\right)^2\\
        &\;\;\;\;\;\;+2 r_c^7 \left(Q-Q_{\rm{ext}}\right)^2+3 Q^6 r_c^3+3 Q^3 r_c^6-9 Q^3 r_c^5+r_c^9+Q^9\Big],           
    \end{split}
    \end{equation}
\end{subequations}
respectively.
The dependencies of $\Omega_c$ and $\lambda_c$ on $Q$ are shown in Fig.\ \ref{fig:photon-sph}.
The linear relation of  $\Omega_c$ and $Q$ can be found approximately,
\begin{equation}
    \Omega_c\approx k_1 Q+ k_0,
\end{equation}
where $k_0=0.3339$ and $k_1=0.1360$. When the above equation is compared with Eq.~(\ref{eq:linear-fit}), the former differs from the latter by a factor of $l=6$ due to $\Re(\omega)=\Omega_c\,l$. 
We can see from Fig.\ \ref{fig:lya-exp} that the Lyapunov exponent takes its maximum value, $\lambda_c^{\rm max}\approx 0.3753$, when  $Q_{\rm C}\approx 0.2689$, which corresponds to the maximum of the minus imaginary part due to $-\Im(\omega)= \left(n+\frac{1}{2}\right)|\lambda_c|$. These data are consistent with those in Tab.\ \ref{tab:maximum} for the case of $l=6$.

We end this subsection by summarizing the dynamic evolution in our model: 
As the charge $Q$ goes to its extreme value $Q_{\rm ext}$, 
the minus imaginary part approaches to its minimum. 
In other words, 
the decay amplitude becomes minimum. As a result, the regular state corresponds to the most stable state of our BH model.

\subsection{Asymptotic quasinormal frequencies}
\label{sec:aqnf}

In this subsection, we calculate asymptotic quasinormal frequencies using the monodromy method \cite{Motl:2003cd,Natario:2004jd}. This method is based on the Stokes  portrait \cite{Lan:2022qbb} 
which consists of Stokes lines $\Re[r_*]=0$, Stokes fields ${\bm V}_{\rm st}=\left\{
\Im\left[1/f\right],\;
\Re\left[1/f\right]
\right\}$, and complex horizons, i.e., the complex roots of the equation, $f(r_{\rm H})=0$. 
In  fact, the Stokes portrait shows a branch of Riemann surfaces described by $r_*(r)$.
The operations, $\Re$ and $\Im$, appear because we analytically continue the radial coordinate into the complex plane, $r\to x+\mi y$ or $r\to \rho\, \me^{\mi\varphi}$, where $x, y, \rho, \varphi \in \mathbb{R}$.

\begin{figure}[!htb]
     \centering
     \begin{subfigure}[b]{0.32\textwidth}
         \centering
         \includegraphics[width=\textwidth]{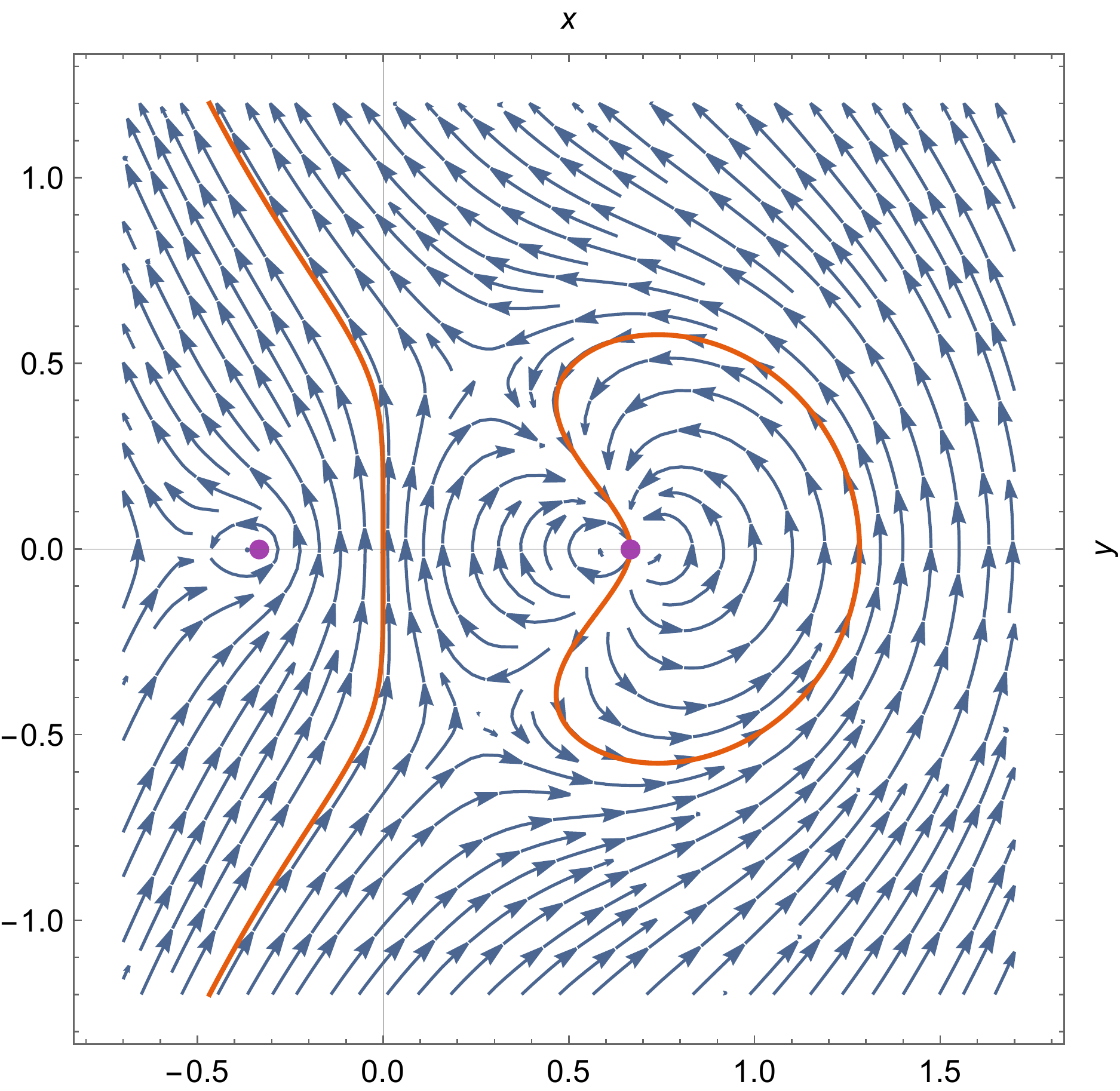}
         \caption{$Q=Q_{\rm ext}$}
         \label{fig:stokes-Q-Qe}
     \end{subfigure}
     \begin{subfigure}[b]{0.32\textwidth}
         \centering
         \includegraphics[width=\textwidth]{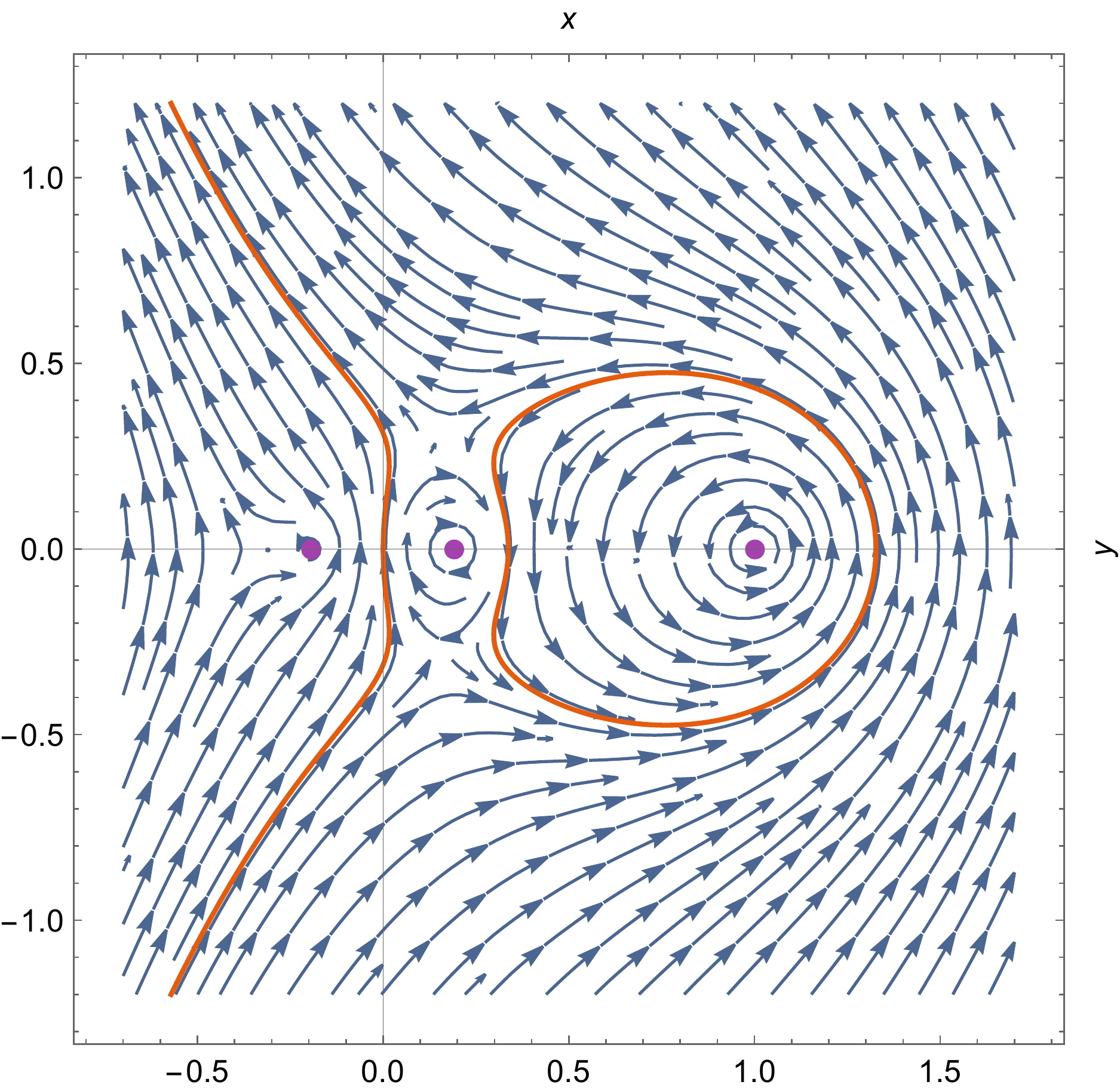}
         \caption{$Q=1/3$}
         \label{fig:stokes-Q-0.3}
     \end{subfigure}
     \begin{subfigure}[b]{0.32\textwidth}
         \centering
         \includegraphics[width=\textwidth]{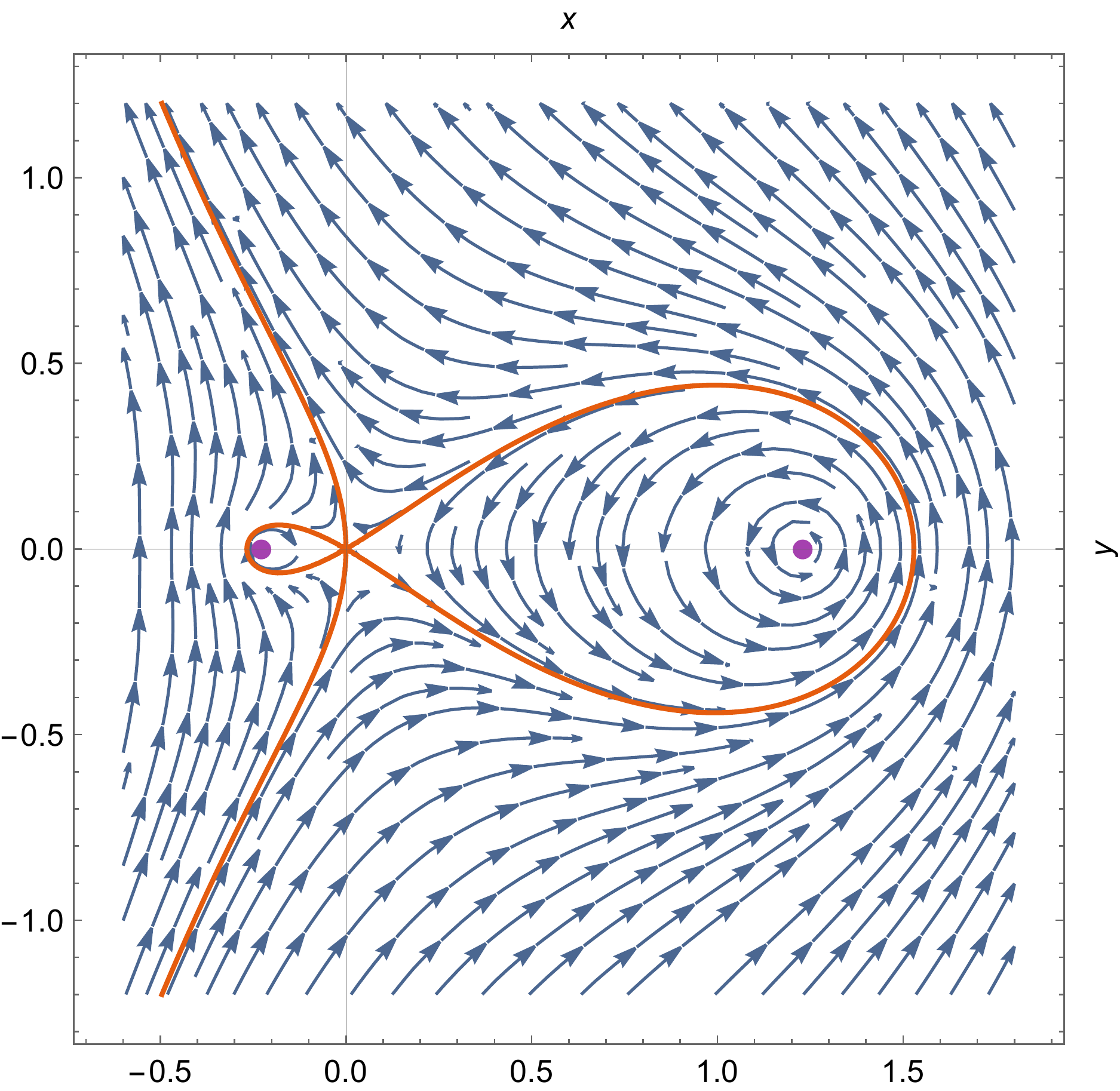}
         \caption{$Q=0$}
         \label{fig:stokes-Q-0}
     \end{subfigure}
      \captionsetup{width=.9\textwidth}
       \caption{Stokes portraits with $r_0=0$ as the central point over which the Stokes lines (orange curves) must cross.
       The blue vector field corresponds to the Stokes field and the purple points to the complex horizons.}
        \label{fig:stokes-xy}
\end{figure}

\begin{figure}[!htb]
     \centering
     \begin{subfigure}[b]{0.4\textwidth}
         \centering
         \includegraphics[width=\textwidth]{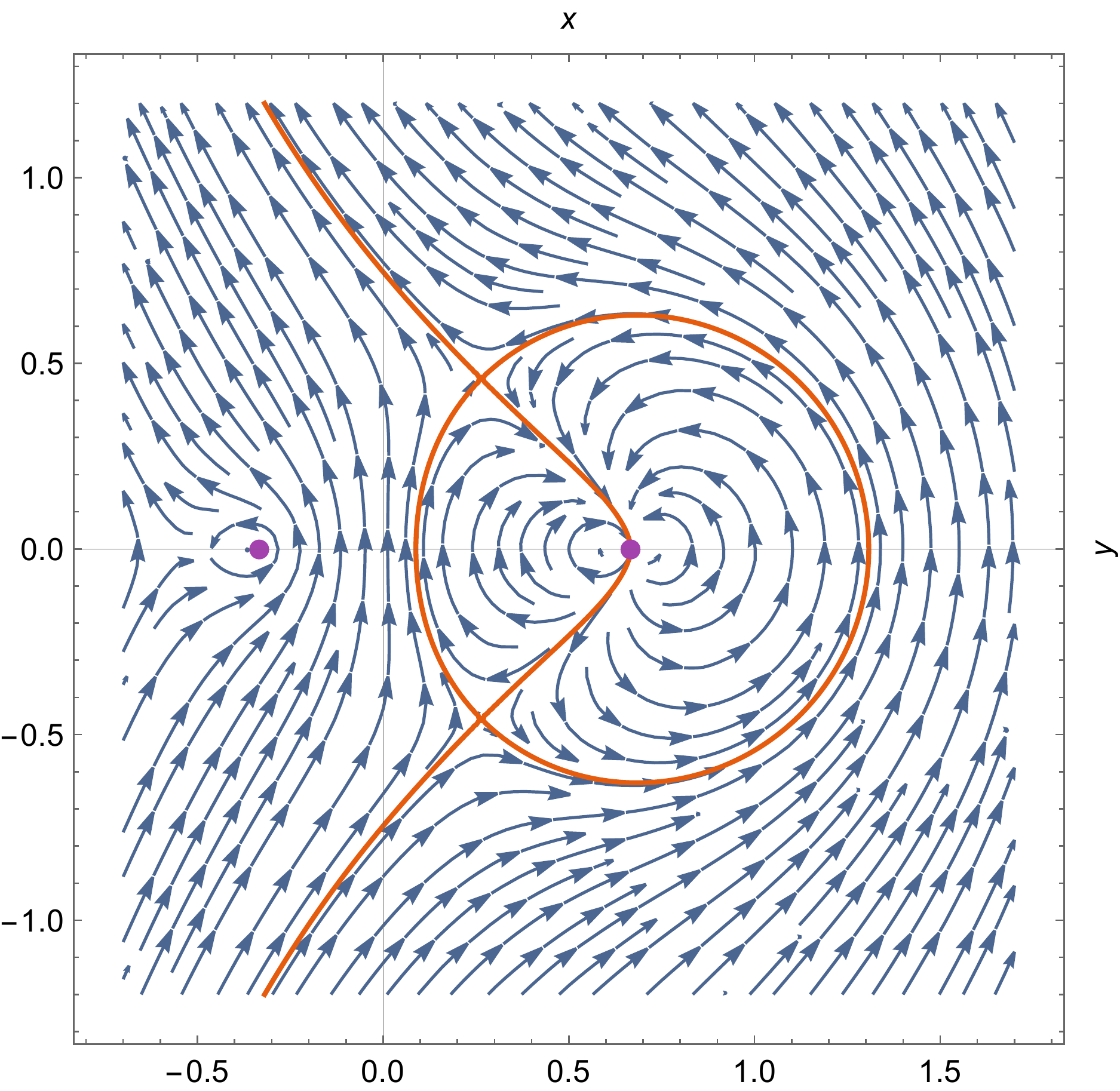}
         \caption{$Q=Q_{\rm ext}$}
         \label{fig:stokes-Q-Qe-2}
     \end{subfigure}
     \begin{subfigure}[b]{0.4\textwidth}
         \centering
         \includegraphics[width=\textwidth]{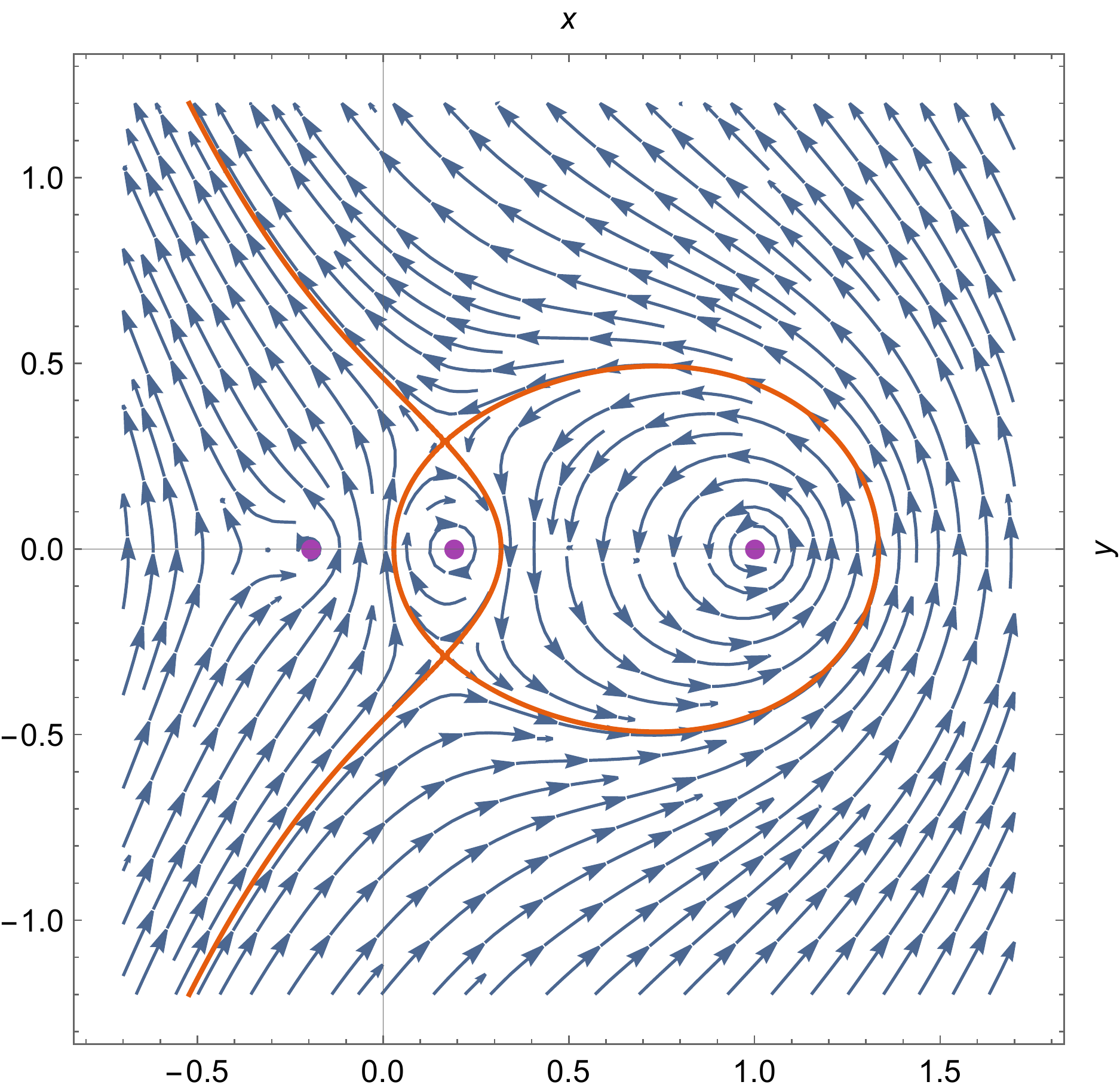}
         \caption{$Q=1/3$}
         \label{fig:stokes-Q-0.3-2}
     \end{subfigure}
      \captionsetup{width=.9\textwidth}
       \caption{Stokes portraits, where the Stokes lines cross over $r_2=(-1)^{1/3} Q$ or $r_3=-(-1)^{2/3} Q$.}
        \label{fig:stokes-xy-2}
\end{figure}

\begin{figure}[!htb]
     \centering
     \begin{subfigure}[b]{0.4\textwidth}
         \centering
         \includegraphics[width=\textwidth]{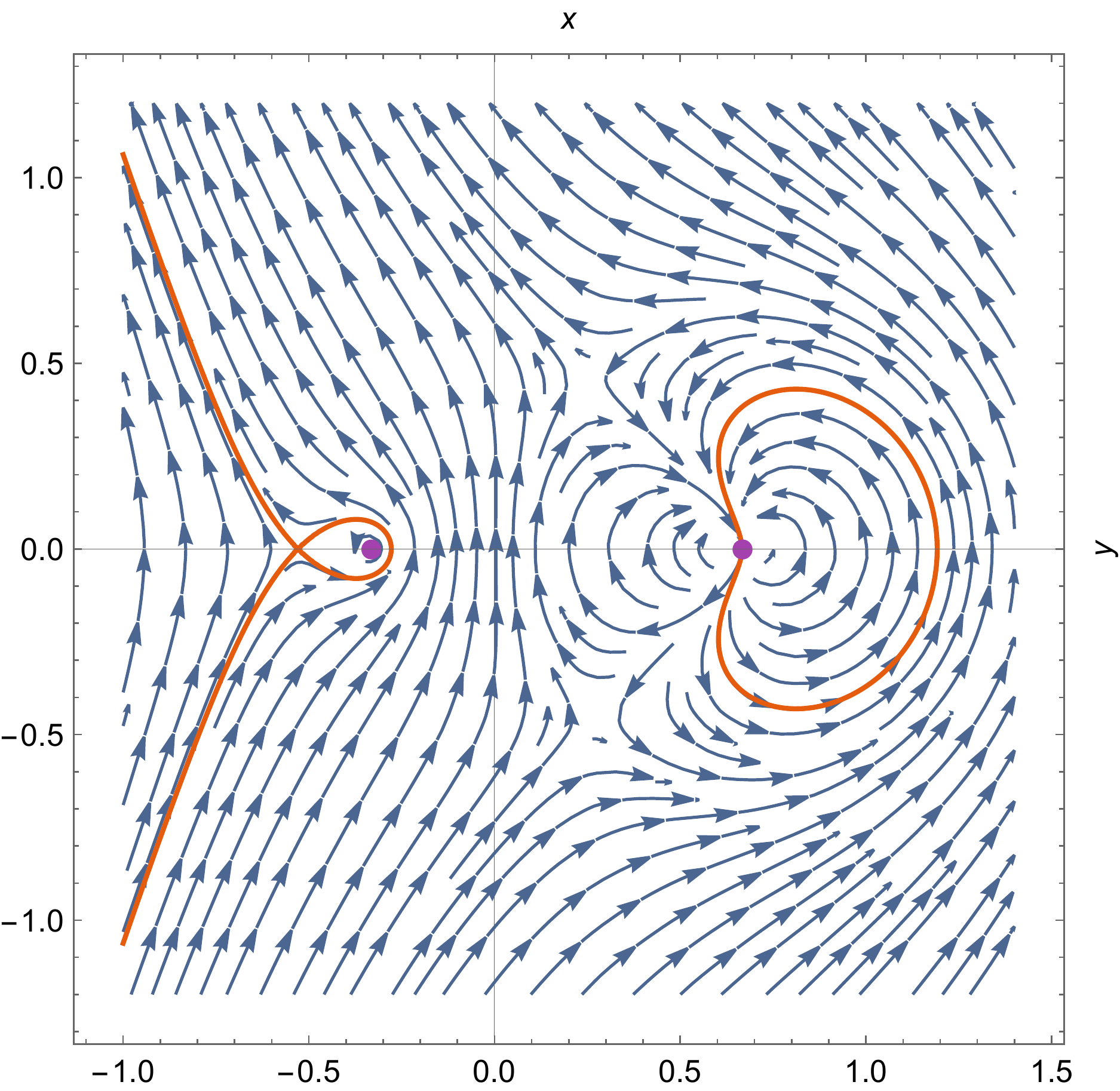}
         \caption{$Q=Q_{\rm ext}$}
         \label{fig:stokes-Q-Qe-3}
     \end{subfigure}
     \begin{subfigure}[b]{0.4\textwidth}
         \centering
         \includegraphics[width=\textwidth]{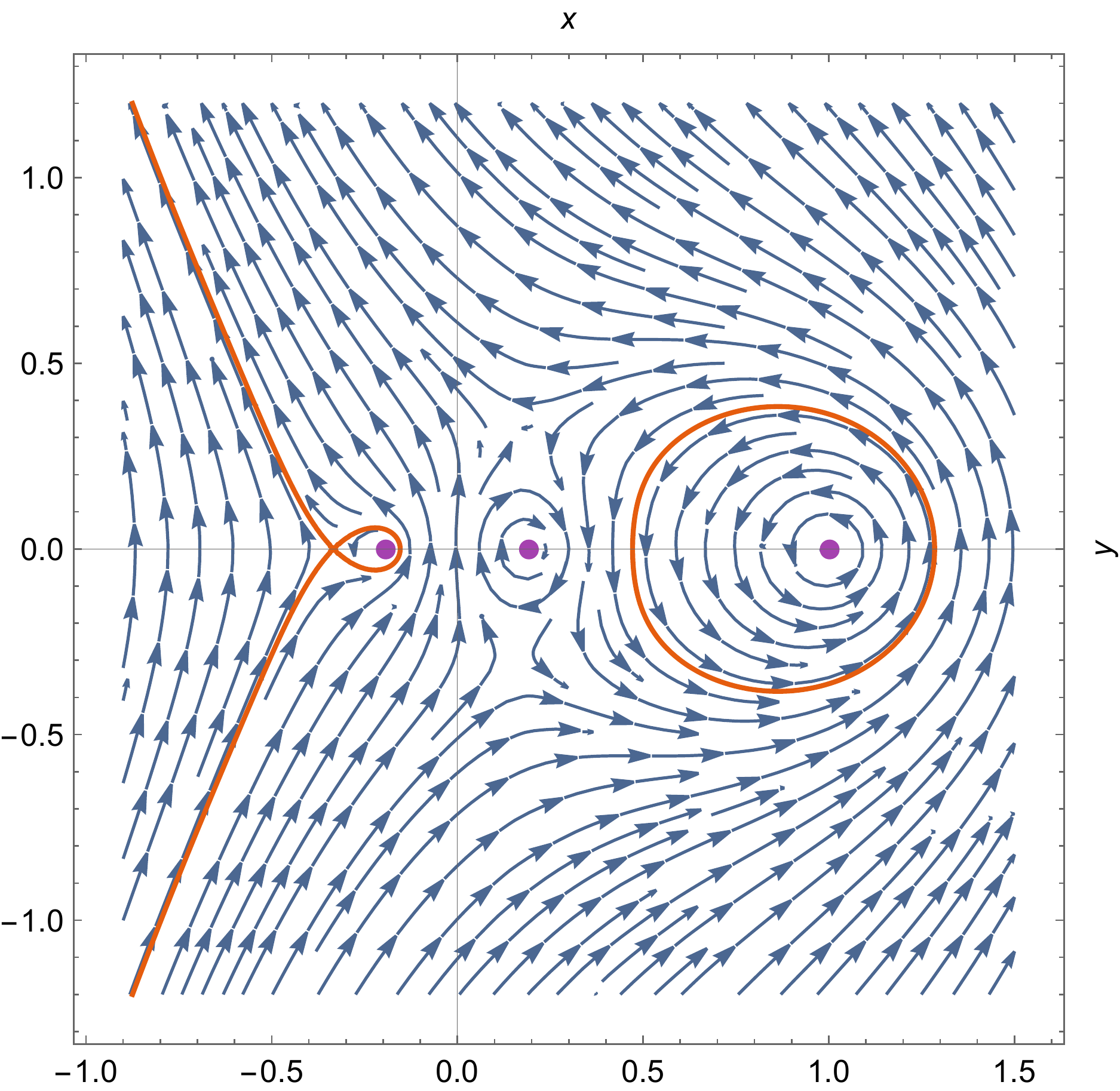}
         \caption{$Q=1/3$}
         \label{fig:stokes-Q-0.3-3}
     \end{subfigure}
      \captionsetup{width=.9\textwidth}
       \caption{Stokes portraits, where the Stokes lines cross over $r_1=-Q$.}
        \label{fig:stokes-xy-3}
\end{figure}

We start with the analysis of singularities of differential equations. 
The master equation, Eq.\ \eqref{eq:master}, has four {\em regular singular points}\footnote{For a given differential equation, $y''(x) + P(x) y'(x) +Q(x) y(x)=0$, if either $P(x)$ or $Q(x)$ diverges at $x=x_0$, but $(x-x_0) P(x)$ and $(x-x_0)^2 Q(x)$ remain finite, then $x_0$ is called a regular singular point.} due to the behavior of effective potential Eq.\ \eqref{eq:eff-pont}, 
\begin{equation}
\label{eq:regsig-ps}
    r_0=0, \qquad 
    r_1=-Q,\qquad
    r_2=(-1)^{1/3} Q,\qquad
    r_3 = -(-1)^{2/3} Q,
\end{equation}
where only $r_1$, $r_2$ and $r_3$ are ``singular points'' of Stokes lines if $Q\neq 0$, see Figs.\ \ref{fig:stokes-xy},  \ref{fig:stokes-xy-2} and \ref{fig:stokes-xy-3}.
Here, the ``singular point'' refers to the point where the curve (Stokes line) has self-intersection in the sense of algebraic geometry \cite{Shafarevich2013}.
Thus, when using the monodromy method, $r=r_0$ should be treated as an ordinary point except $Q=0$.
If $Q$ equals zero, the four regular singular points, see Eq.\ \eqref{eq:regsig-ps}, merge into one. In this case, $r_0$ reduces to one singular point of Stokes lines, see Fig.\ \ref{fig:stokes-Q-0}.

\begin{figure}[!htb]
     \centering
     \begin{subfigure}[b]{0.32\textwidth}
         \centering
         \includegraphics[width=\textwidth]{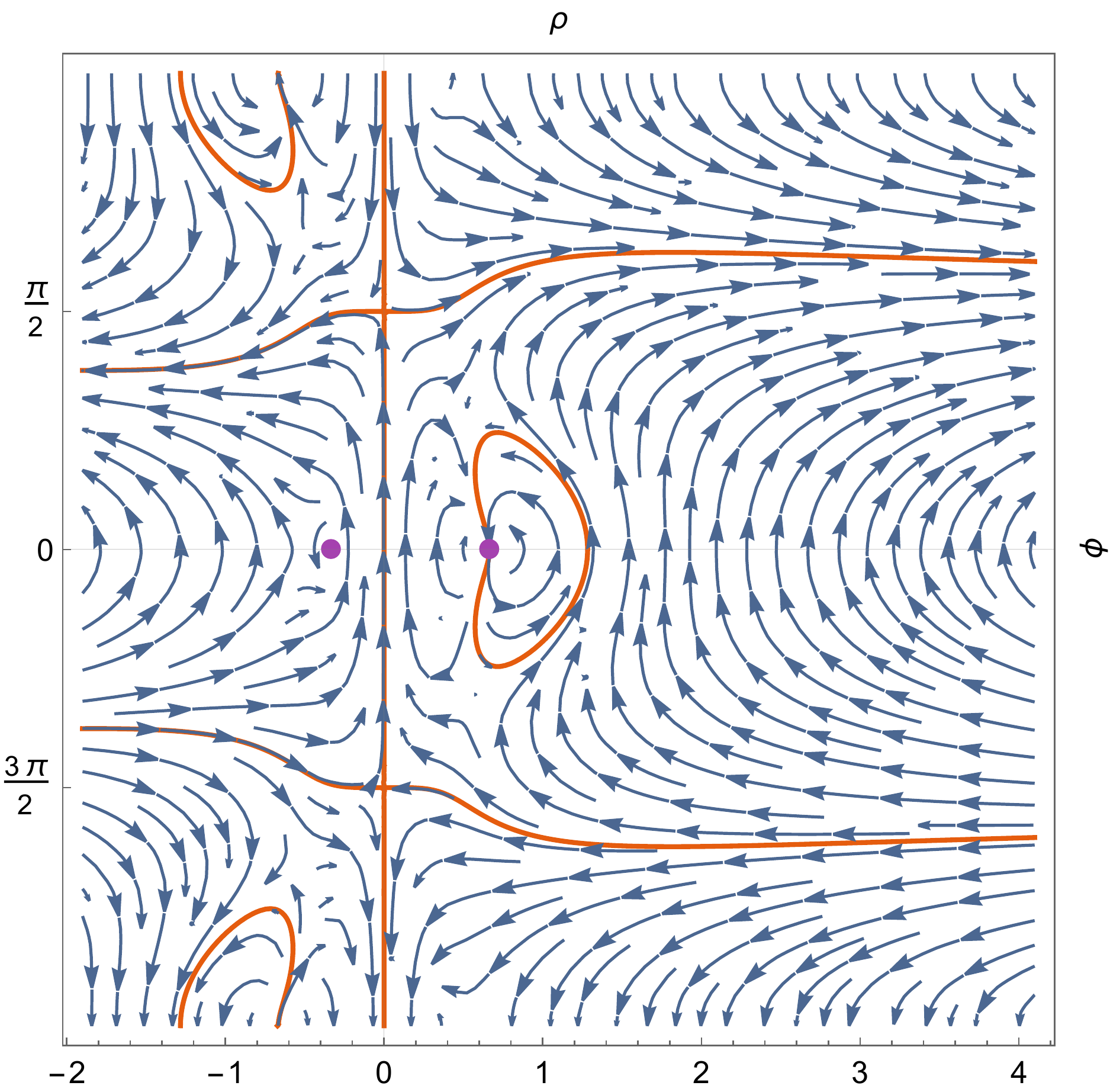}
         \caption{$Q=Q_{\rm ext}$}
         \label{fig:stokes-ag-Qe}
     \end{subfigure}
     \begin{subfigure}[b]{0.32\textwidth}
         \centering
         \includegraphics[width=\textwidth]{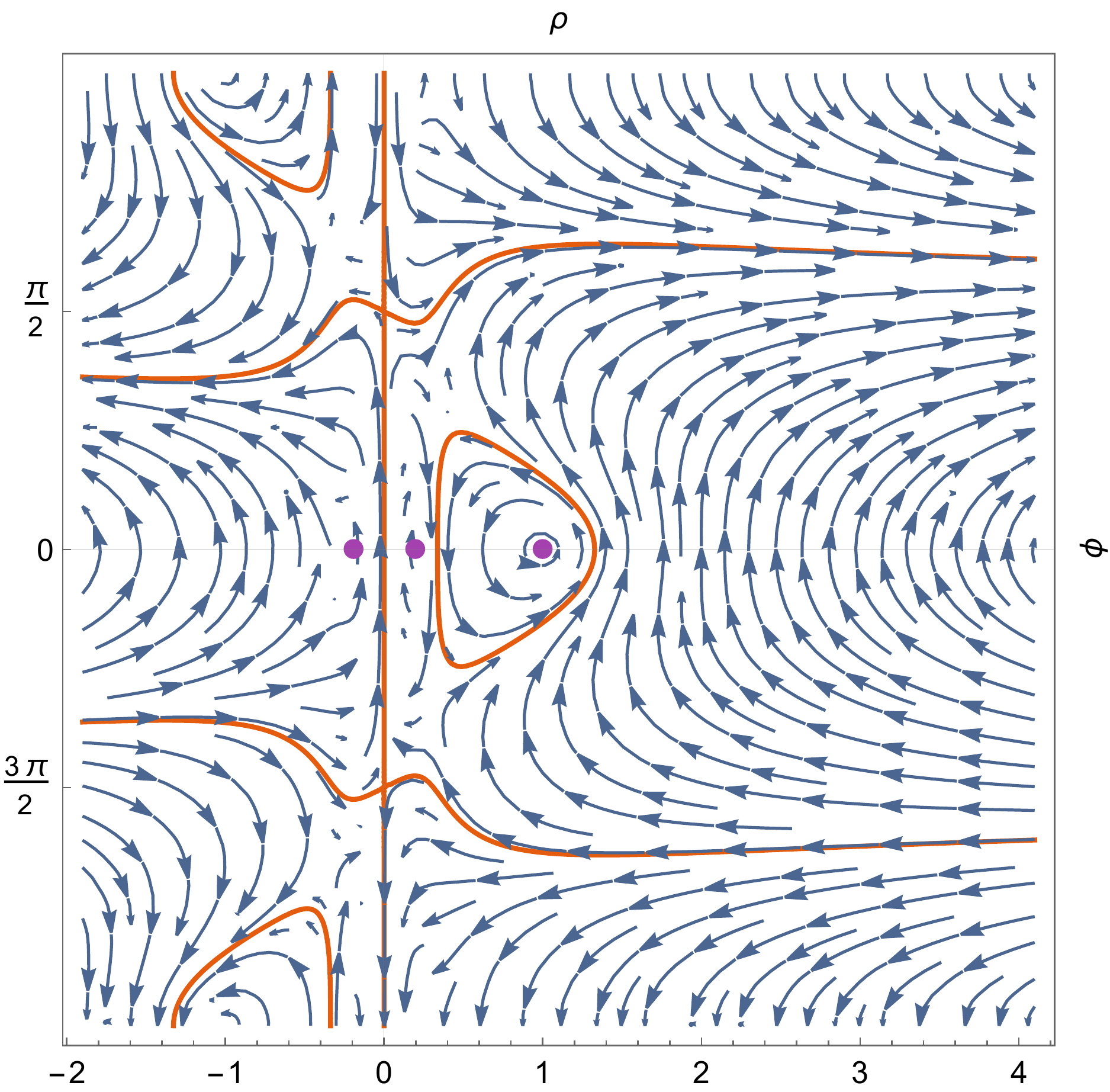}
         \caption{$Q=1/3$}
         \label{fig:stokes-ag-Q0.3}
     \end{subfigure}
     \begin{subfigure}[b]{0.32\textwidth}
         \centering
         \includegraphics[width=\textwidth]{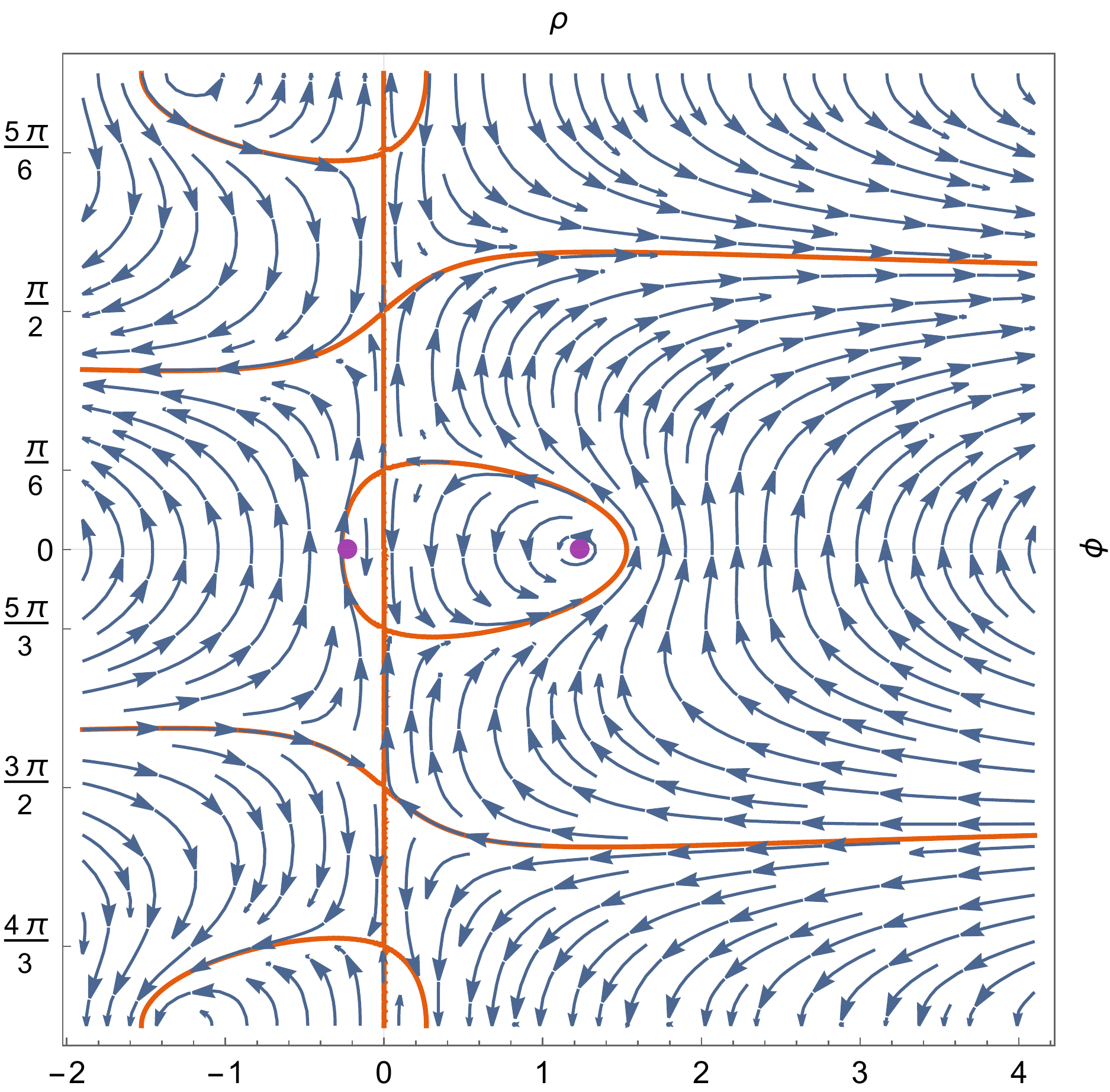}
         \caption{$Q=0$}
         \label{fig:stokes-ag-Q0}
     \end{subfigure}
      \captionsetup{width=.9\textwidth}
       \caption{Polar Stokes portraits corresponding to Fig.\ \ref{fig:stokes-xy}.}
        \label{fig:stokes-phi}
\end{figure}

\begin{figure}[!htb]
     \centering
     \begin{subfigure}[b]{0.4\textwidth}
         \centering
         \includegraphics[width=\textwidth]{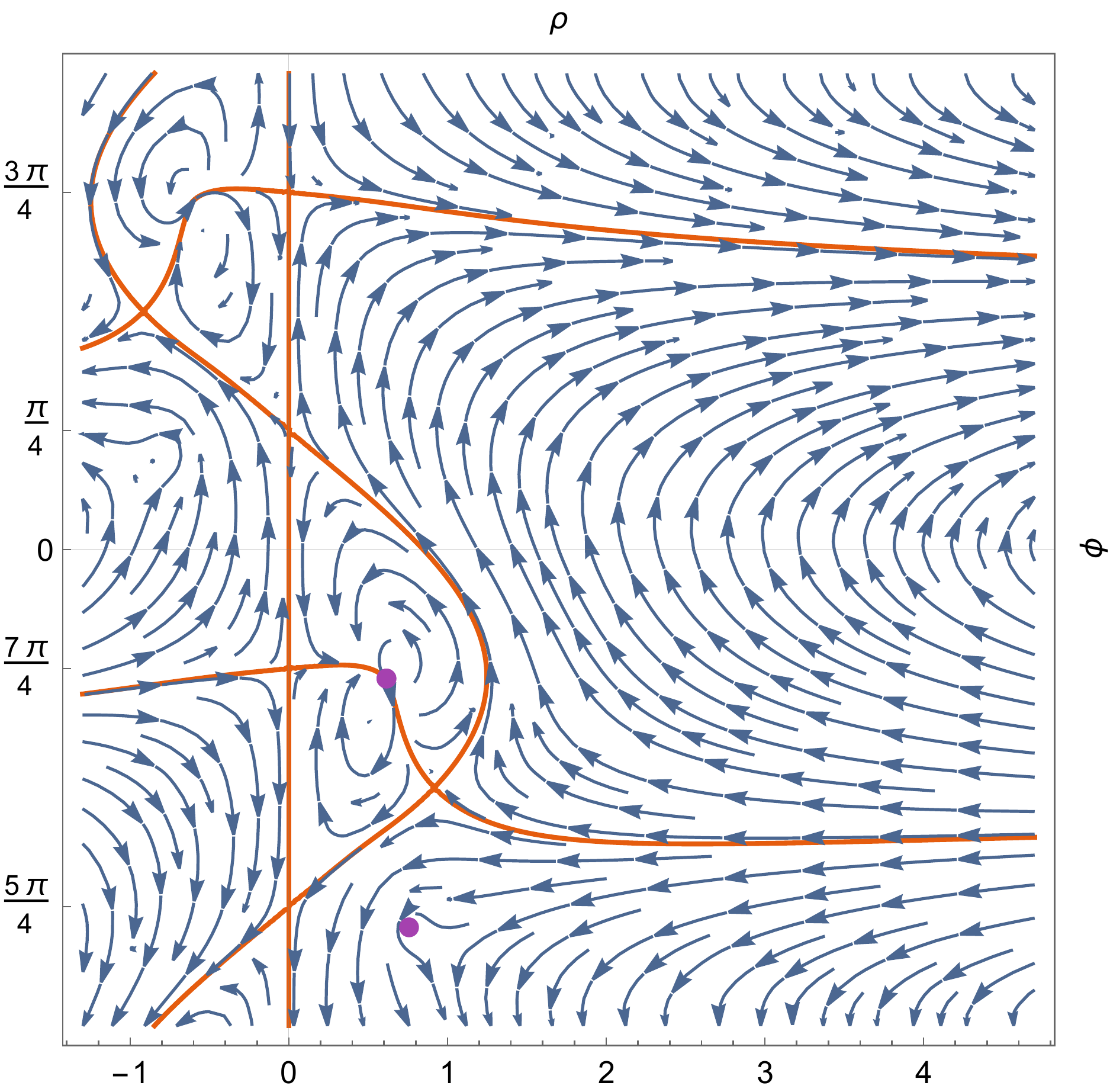}
         \caption{$Q=Q_{\rm ext}$}
         \label{fig:stokesphi-Q-Qe-2}
     \end{subfigure}
     \begin{subfigure}[b]{0.4\textwidth}
         \centering
         \includegraphics[width=\textwidth]{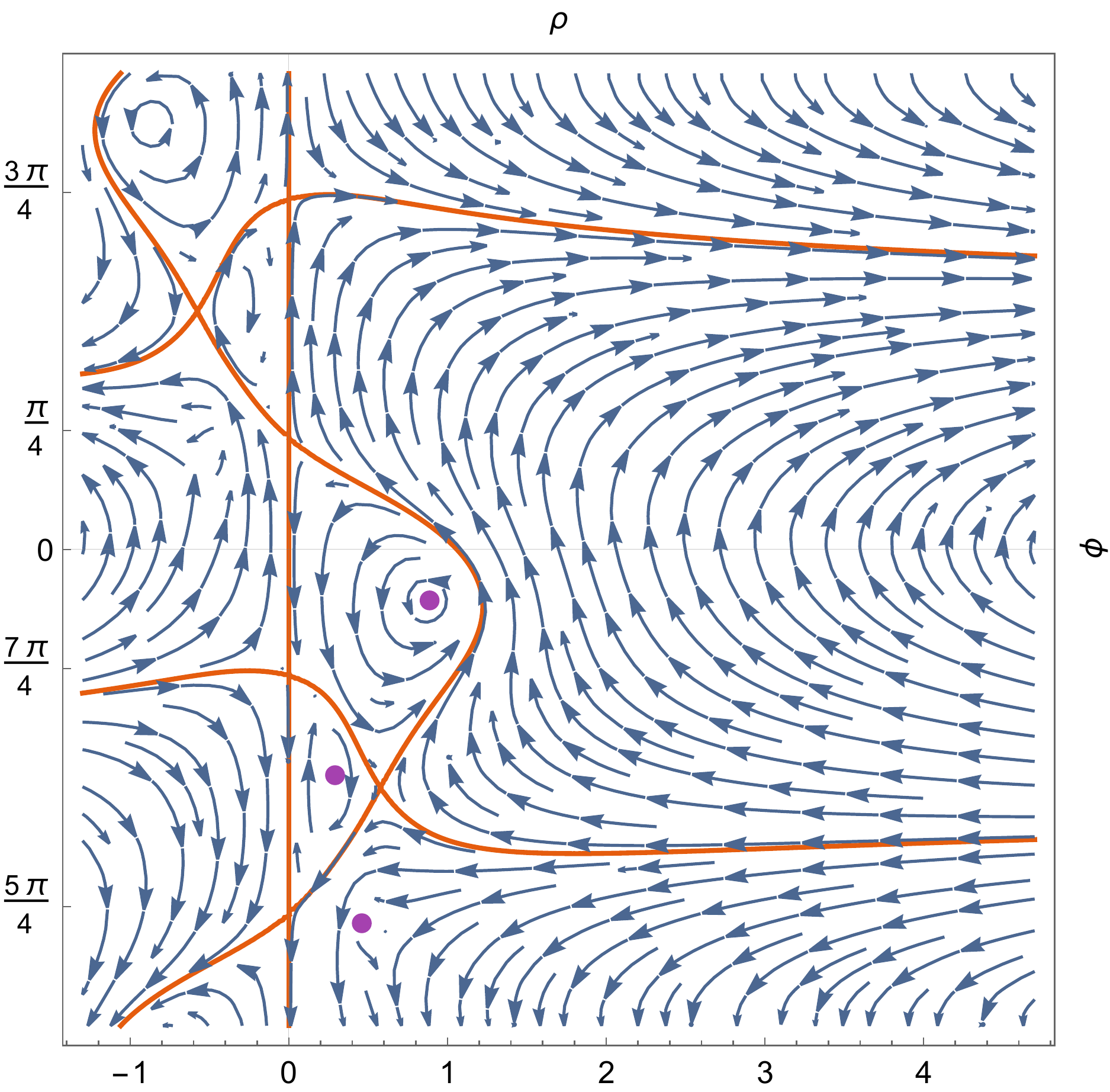}
         \caption{$Q=1/3$}
         \label{fig:stokesphi-Q-0.3-2}
     \end{subfigure}
      \captionsetup{width=.9\textwidth}
       \caption{Polar Stokes portraits corresponding to $r_2=(-1)^{1/3} Q$.}
        \label{fig:stokes-phi-2}
\end{figure}

\begin{figure}[!htb]
     \centering
     \begin{subfigure}[b]{0.4\textwidth}
         \centering
         \includegraphics[width=\textwidth]{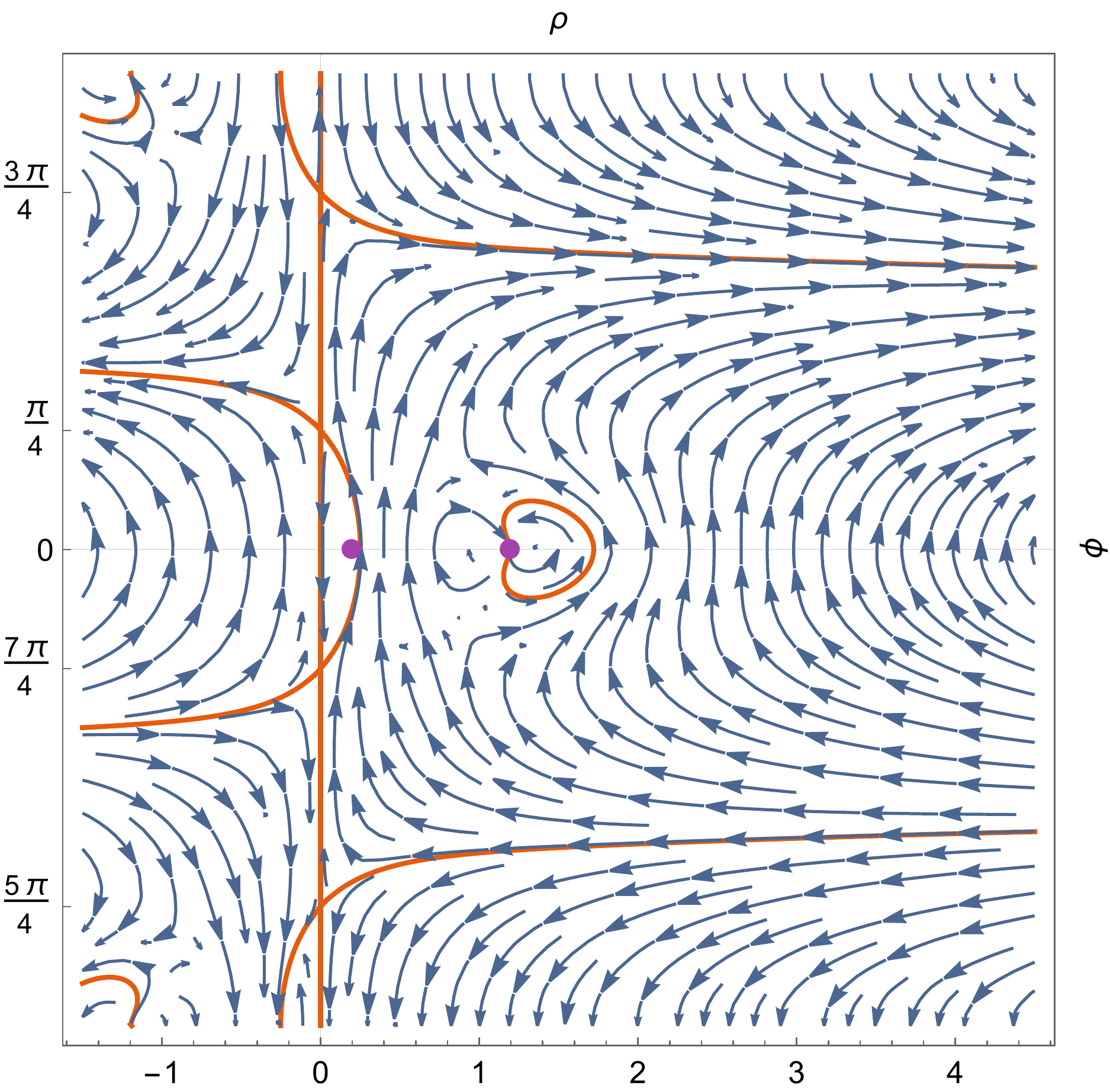}
         \caption{$Q=Q_{\rm ext}$}
         \label{fig:stokesphi-Q-Qe-3}
     \end{subfigure}
     \begin{subfigure}[b]{0.4\textwidth}
         \centering
         \includegraphics[width=\textwidth]{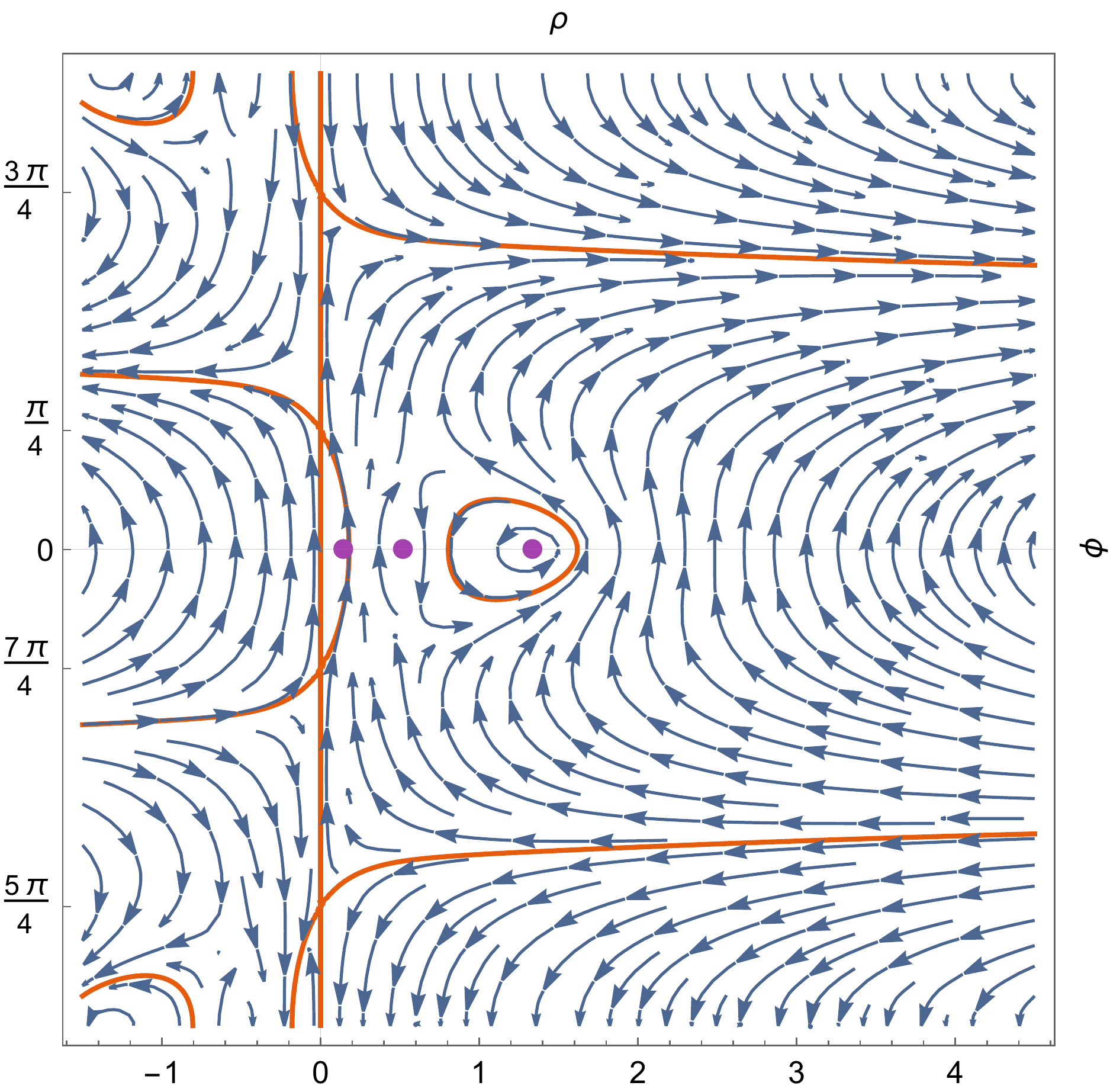}
         \caption{$Q=1/3$}
         \label{fig:stokesphi-Q-0.3-3}
     \end{subfigure}
      \captionsetup{width=.9\textwidth}
       \caption{Polar Stokes portraits corresponding to $r_1=-Q$.}
        \label{fig:stokes-phi-3}
\end{figure}

The Stokes portraits of the model Eq.\ \eqref{eq:shape} can be separated into three groups according to the ways how the Stokes lines cross over a regular singular point.
\begin{enumerate}
    \item Stokes lines cross over the point $r = r_0$, see   Figs.\ \ref{fig:stokes-xy} and \ref{fig:stokes-phi}, where the Stokes lines in Fig.\ \ref{fig:stokes-xy} are built in the Cartesian complex plane, while those in Fig.\ \ref{fig:stokes-phi} are in the polar complex plane, i.e. $r=\rho \me^{\mi\varphi}$. However, 
    only in Fig.\ \ref{fig:stokes-Q-0}, the monodromy of asymptotic solutions of Eq.\ (\ref{eq:master}) along the Stokes lines is non-trivial. The reason is that $r=r_0$ is the branch point in Fig.\ \ref{fig:stokes-Q-0}, but not in Figs. \ref{fig:stokes-Q-Qe} and \ref{fig:stokes-Q-0.3}.
    Therefore, we ignore the cases in Figs. \ref{fig:stokes-Q-Qe} and \ref{fig:stokes-Q-0.3}.
    \item Stokes lines cross over the point $r=r_2$ or $r=r_3$, which is the second non-trivial case, see the Cartesian Stokes portraits in Fig.\ \ref{fig:stokes-xy-2}  and polar ones in Fig.\ \ref{fig:stokes-phi-2}, respectively.
    \item Stokes lines cross over the point $r=r_1$, see the Cartesian Stokes portraits in Fig.\ \ref{fig:stokes-xy-3}  and polar ones in Fig.\ \ref{fig:stokes-phi-3}. Although it is non-trivial, the monodromy is associated with the non-physical horizon at the negative axis of $x$. Thus, we will not take into account this case, either.
\end{enumerate}

At first, we calculate the asymptotic quasinormal frequencies in the case of $Q=0$ based on the Stokes portraits, see Figs.\ \ref{fig:stokes-Q-0} and \ref{fig:stokes-ag-Q0}. 
The tortoise coordinate has the following asymptotic behavior in the limit of $r\to 0$,
\begin{equation}
    z\to -\frac{r^3}{3 Q_{\rm{ext}}^2}+O\left(r^4\right),
\end{equation}
where $z$ is the analytical continuation of $r_*$ into a complex plane.
The leading order of the effective potential reads
\begin{equation}
   V_{\rm eff}\to -\frac{2 Q_{\rm ext}^4}{r^6}+O\left(\frac{1}{r^5}\right).
\end{equation}
Thus, the master equation becomes
\begin{equation}
	  -\frac{\dif^2 \Psi}{\dif z^2} -\frac{2}{9 z^2} \Psi = \omega^2 \Psi, 
\end{equation}
which does not contain the multipole number $l$. From this equation, we solve one asymptotic solution via the combination of Bessel functions,
\begin{equation}
\label{eq:sol-r0}
    \Psi = A_1 \sqrt{2\pp \omega z} J_{\frac{1}{6}}(\omega z)+ A_2 \sqrt{2\pp \omega z} J_{-\frac{1}{6}}(\omega z),
\end{equation}
where $A_1$ and $A_2$ are two integration constants. 

Then we try to find the formula of asymptotic quasinormal frequencies by matching the monodromy of the above asymptotic solution along two contours we choose depending on the Stokes lines in Fig.\ \ref{fig:stokes-Q-0}. The first circles the outer horizon. The second contour starts at infinity in the lower left corner, rotates at the merged regular singular point, and then goes out to infinity in the upper left corner, and finally goes along a large arc to infinity in the lower left corner. The formula of asymptotic quasinormal frequencies takes the form,
\begin{equation}
    \frac{\omega}{T_{\rm H}^+} = \ln 2 + 2\pp \mi \left(n+\frac{1}{2}\right),\label{asmpqnfs}
\end{equation}
where $T_{\rm H}^+$ ($T_{\rm H}^-$) is BH temperature at the outer (inner) horizon,  $n>0$ and $n\in\mathbb{Z}$.
The details of the calculation can be found in App.\ \ref{app:monodromy}. 
Comparing with the case in the Schwarzschild BH \cite{Motl:2003cd}, we note that the real part of $\omega/T_{\rm H}^+$ is weakened.

Now let us turn to the Stokes portrait Fig.\ \ref{fig:stokes-Q-0.3-2}, where $Q\in (0, Q_{\rm ext})$.
Similarly, we can obtain the asymptotic quasinormal frequency by the monodromy method,
\begin{equation}
\label{eq:qnf-Q}
    \me^{\omega /T_{\rm H}^+}=
    -2 [\cos (\pp  \nu )+1)] \me^{-\omega /T_{\rm H}^-}
    -2 \cos (\pp  \nu )-1,
\end{equation}
which has the same form as that of Hayward BHs \cite{Lan:2022qbb}.
Here $\nu=\sqrt{1-4 \mathcal{V}_0}$ with $\mathcal{V}_0$ defined by Eq.\ \eqref{eq:leading-potential}, see App.\ \ref{app:monodromy}.
When $Q$ goes to $Q_{\rm ext}$, the inner and outer horizons are close to each other. 
When $Q$ is equal to $Q_{\rm ext}$, the inner and outer horizons merge into the extreme horizon, and the Stokes line seems to cross over the extreme horizon, see Fig.\ \ref{fig:stokes-Q-Qe-2}, but this is not the case.

To clarify whether the Stokes line crosses over the extreme horizon or not, we write down the equations of Stokes lines describing this case,
\begin{eqnarray}
\label{eq:stokes-line}
& &\frac{48-72 x}{(2-3 x)^2+9 y^2}
+\ln \left[(3 x+1)^2+9 y^2\right]
+8 \ln \left[(2-3 x)^2+9 y^2\right]
+18 x\nonumber \\
& &=5\cdot 2^{2/3}+8- 2^{4/3}+\ln \left[746496 \left(-4+ 2^{4/3}+2^{2/3}\right)\right]. 
\end{eqnarray}
The extreme horizon corresponds to the point $\{x,y\}=\{0, 2/3\}$, which is not the solution of Eq.\ \eqref{eq:stokes-line}. 
In other words, the Stokes line does not cross over the extreme horizon. More precisely, no Stokes line can be defined at that point in the complex plane.
The monodromy method is not available for the extreme case, and it is not feasible in the limit of $Q\to Q_{\rm ext}$ for Eq.\ \eqref{eq:qnf-Q}, either. The reason is that the exponential function $\me^{\omega/T_{\rm H}^+}$ blows up when $T_{\rm H}^+$ goes to zero.

\section{Black hole thermodynamics}
\label{sec:thermody}

The thermodynamics of RBHs is associated currently with many inconclusive issues, where two of the most important ones are entropy and the first law of thermodynamics.
On the one hand, the entropy of RBHs is problematic because a simple thermodynamic formula, $\widetilde{S}_+=\displaystyle\int \dif M/T$, yields  an inconsistency with that obtained by Wald's Noether-charge method \cite{Wald:1993nt} or Hawking's path integral method \cite{Gibbons:1976ue}.
On the other hand, the problem is that there are ambiguous terms in the first law of thermodynamics for RBHs if the first law is constructed in the thermodynamic phase space without redundant degrees of freedom.

In this section, we propose an alternative method to resolve the inconsistency between the  thermodynamic formula, $\widetilde{S}_+=\displaystyle \int \dif M/T$, and Wald's Noether-charge method.
In addition, using a self-consistent entropy, we also examine the thermodynamics of our BH model described by Eq.~(\ref{eq:shape}) and the relationship between the divergence of heat capacity and the divergence of thermodynamic curvatures at the end of this section.

\subsection{Thermal entropy and Wald entropy}
\label{sec:entropy}
At first, we denote $\displaystyle \widetilde{S}_+$ as the entropy obtained by the thermal formula, $\displaystyle \widetilde{S}_+\coloneqq\int\dif M/T$, and $S_{\rm{W}}$ as the Wald entropy.
Secondly, we recover the mass parameter in Eq.~(\ref{eq:shape}) by the replacements of $r\to r/(2M)$, $Q\to Q/(2M)$, and $Q_{\rm ext}\to Q_{\rm ext}/(2M)$, 
\begin{equation}
\label{eq:shape-full}
    f=1-\frac{r \left[\left(Q-Q_{\rm{ext}}\right)^2+2 M r\right]}{Q^3+r^3},
\end{equation}
where $r$, $Q$, and $M$ have the same dimension,
$[r]=[Q]=[M]$. 
Next, we represent the outer horizon $r_+$ via the following form,
\begin{equation}
   M= \frac{-\left(Q-Q_{\rm{ext}}\right)^2 r_{+}+r_{+}^3+Q^3}{2 r_{+}^2},
\end{equation}
and express the temperature in terms of $r_+$, $Q$, and $Q_{\rm ext}$,
\begin{equation}
    T=\frac{r_+ \left(Q-Q_{\rm{ext}}\right)^2-2 Q^3+r_+^3}{4 \pp  r_+ \left(Q^3+r_+^3\right)},
\end{equation}
which has an inverse dimension of mass,
$[T]=[M]^{-1}$.

Let us compute $\widetilde{S}_+$ and $S_{\rm{W}}$, respectively.
If $M$ is regarded as internal energy, 
the entropy $\widetilde{S}_+$ can be calculated by the usual thermodynamic formula,
\begin{equation}
\label{eq:entropy}
   \widetilde{S}_+\coloneqq \int^{r_+} \frac{\dif M}{T}=\pp  r_+^2-\frac{2 \pp  Q^3}{r_+}.
\end{equation}
The dimension of $\widetilde{S}_+$ is the square of mass,
$[\widetilde{S}_+]=[M]^{2}$.
The first term in Eq.\ \eqref{eq:entropy} corresponds to the entropy-area law, which equals $A/4$ in Einstein's gravity. The second term is a deviation, which
inspires us to interpret our model in terms of $F(R)$ gravity.
Moreover, the Wald entropy of $F(R)$ theory is of the form \cite{Jacobson:1993vj},
\begin{equation}
    S_{\rm{W}}=\frac{A}{4}\left.\left( \frac{\partial F(R)}{\partial R} \right)\right|_{r_+}.
\end{equation}
According to our proposal in Sec.\ \ref{sec:sources}, if the the Starobinsky form is chosen, $F(R)=R+\alpha R^2 $, see Eq.~(\ref{FRform1}),
we have
\begin{equation}
\label{eq:wald-entropy}
S_{\rm{W}}=\frac{A}{4}\left.\left(1+2\alpha R\right)\right|_{r_+},
\end{equation}
where the dimension of Ricci scalars is the square of the inverse of mass, $[R]=[M]^{-2}$, and $\alpha$ has the dimension of square of mass, $[\alpha]=[M]^2$. 
The magnitude of $R$ at the outer horizon reads
\begin{equation}
    R_{+}=\frac{6 Q^3 \left[-\left(Q-Q_{\rm{ext}}\right)^2 r_{+}-r_{+}^3+2 Q^3\right]}{r_+^2 \left(r_{+}^3+Q^3\right)^2}.
\end{equation}
If $\alpha$ is chosen to be the special form,
\begin{equation}
\alpha=\frac{\left(Q^3+r_+^3\right)^2}{6 r_+\left[-2 Q^3+(Q-Q_{\rm{ext}})^2 r_++r_+^3\right]},
\end{equation}
 $\widetilde{S}_+$ equals $S_{\rm{W}}$.

We note that many possible forms of $F(R)$ give the equality,  $\widetilde{S}_+=S_{\rm{W}}$. For instance, if we take 
\begin{equation}
    F(R)=R+\alpha \Delta(R),
\end{equation}
rather than the Starobinsky form, where $\Delta(R)$ is a non-trivial function of Ricci scalars, e.g., $\Delta(R)=1/R$, etc.,
we still obtain this equality for our model.
Such an observation can be understood when the different Lagrangians of $F(R)$ gravity lead to the same equations of motion.

\subsection{Phase transition}

In this subsection, we first analyze whether there are Davies points \cite{Davies:1977bgr} in our model, i.e., the divergent points of heat capacity, corresponding to the second-order phase transition points of our BH model. The motivation comes from the fact that our model, Eq.\ \eqref{eq:shape-full}, reduces to the Hayward model when $Q$ equals $Q_{\rm ext}$.
The heat capacity of the Hayward BH has a single Davies point \cite{Lan:2020fmn}, thus our model may have the same thermodynamic structure as the Hayward BH. 
Further, we consider the correspondence \cite{Mansoori:2013pna,Quevedo:2023vip} between the divergences of heat capacity and the divergences of different thermodynamic curvatures in our model.

\begin{figure}[!htb]
     \centering         \includegraphics[width=.6\textwidth]{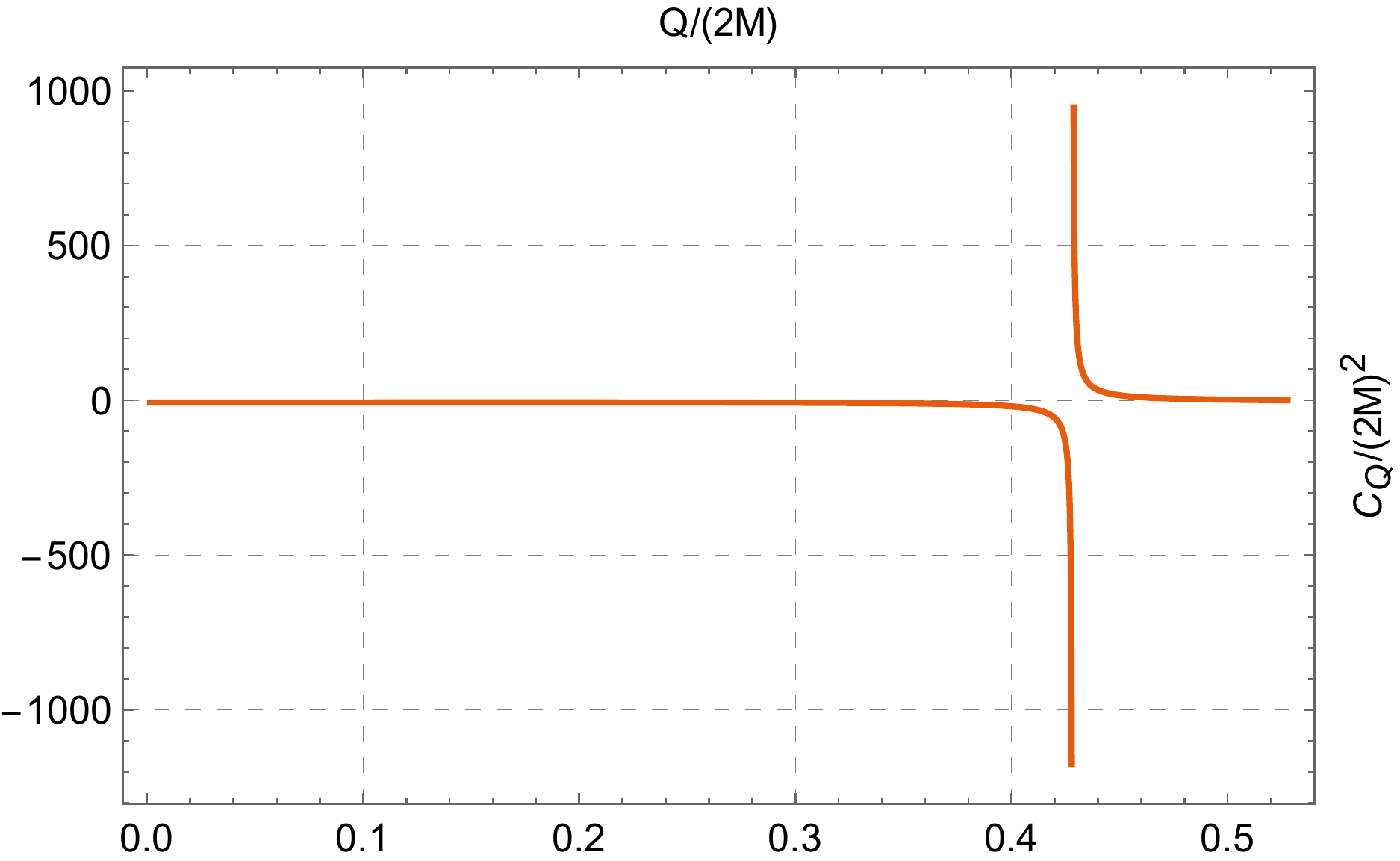}
   \captionsetup{width=.9\textwidth}
       \caption{Heat capacity.}
        \label{fig:capacity}
\end{figure}

To this end, we calculate the heat capacity when 
$Q$ is constant, i.e.,
\begin{equation}
    C_Q\coloneqq T
    \left(\frac{\partial \widetilde{S}_+}{\partial T}\right)_Q,
\end{equation}
which gives an analytical result,
\begin{equation}
    C_Q=\frac{2 \pp  \left(Q^3+r_+^3\right)^2 \left[r_+ \left(Q-Q_{\rm ext}\right)^2-2 Q^3+r_+^3\right]}{-3 r_+^5 \left(Q-Q_{\rm ext}\right)^2+10 Q^3 r_+^4+2 Q^6 r_+-r_+^7},
\end{equation}
where the divergent point (Davies point) is unique and approximately located \cite{Hendi:2015rja} at 
the zeros of  $\partial^2_{\widetilde{S}_+} M=0$, i.e.,
$Q_c/(2M)\approx0.428243$
that is less than the extreme charge $Q_{\rm ext}$, see Fig.\ \ref{fig:capacity}.
In other words, the singular state with $Q/2M<0.428243$ must undergo a second-order phase transition in order to evolve to the regular state.

\begin{figure}[!htb]
     \centering
     \begin{subfigure}[b]{0.4\textwidth}
         \centering
         \includegraphics[width=\textwidth]{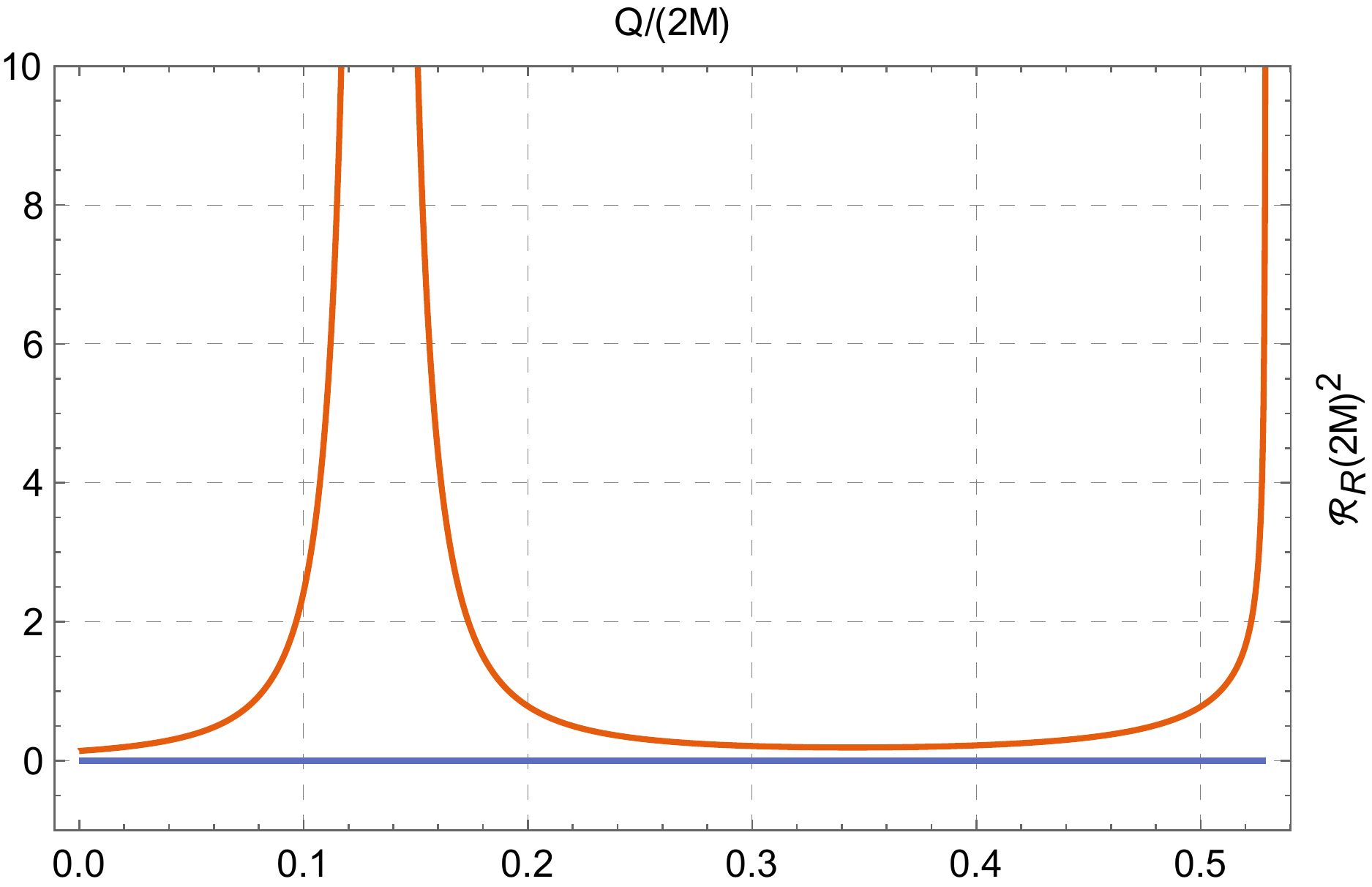}
         \caption{Ruppeiner curvature}
         \label{fig:Rth-Rupp-Q}
     \end{subfigure}
     \begin{subfigure}[b]{0.4\textwidth}
         \centering
         \includegraphics[width=\textwidth]{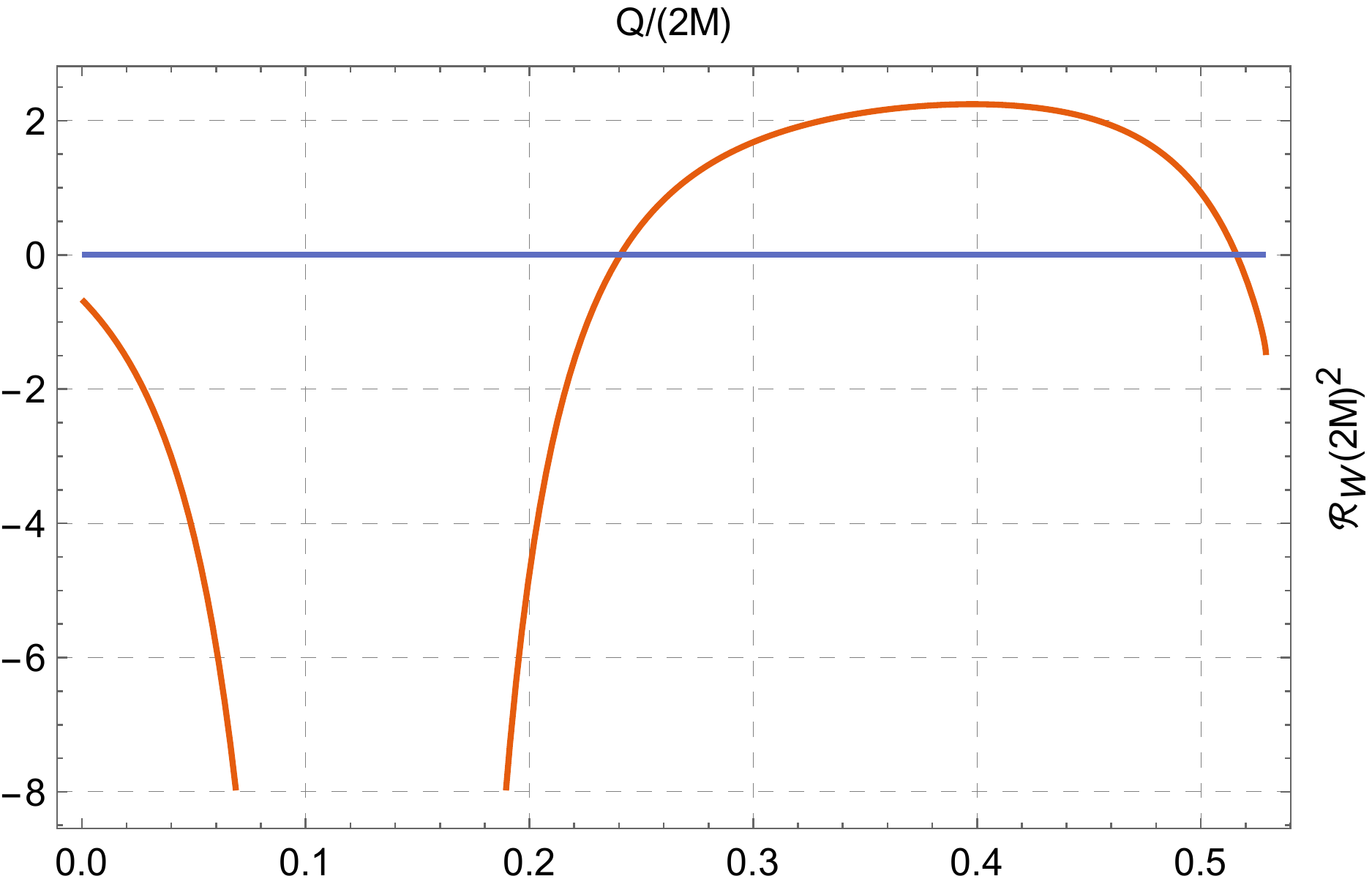}
         \caption{Weinhold curvature}
         \label{fig:Rth-W-Q}
     \end{subfigure}
      \captionsetup{width=.9\textwidth}
       \caption{Thermal curvatures of Ruppeiner and Weinhold geometries.}
        \label{fig:thermal-curvature}
\end{figure}

Now let us turn to the singularities of thermal curvatures. 
According to Ruppeiner's geometry, the thermodynamic metric is cast in the following form,
\begin{equation}
\label{eq:met-Ruppeiner}
\mathfrak{g}^{\rm R}=
 \begin{pmatrix}
 \partial^2_M \widetilde{S}_+ &  \partial_M\partial_Q \widetilde{S}_+\\
 \partial_M\partial_Q \widetilde{S}_+ &  \partial^2_Q \widetilde{S}_+
\end{pmatrix}.
\end{equation}
The corresponding thermal curvature is denoted by $\mathcal{R}_{\rm R}$ 
which is positive and  shown in Fig.\ \ref{fig:Rth-Rupp-Q},
indicating that repulsive interactions are the most prevalent 
among the microstructures \cite{Wei:2019uqg,Wei:2019yvs}.
Further, there are two divergent points in $\mathcal{R}_{\rm R}$,
\begin{equation}
Q_1/(2M)\approx 0.13358,\qquad
Q_2/(2M)= Q_{\rm ext},
\end{equation}
where $Q_1/(2M)$ is a zero of $(\partial_M\partial_Q \widetilde{S}_+)^2-(\partial^2_M \widetilde{S}_+)(\partial^2_Q \widetilde{S}_+)$
and $Q_2/(2M)$ comes from another factor in the denominator of $\mathcal{R}_{\rm R}$, see Tab.\ \ref{tab:zeros} below.
However, neither of these two singularities corresponds to $Q_c$, i.e.,
the singularities of heat capacity and Ruppeiner curvature do not have any evident relation.

A similar situation of singularity will happen in Weinhold's geometry, 
where the metric reads
\begin{equation}
\label{eq:met-Weinhold}
\mathfrak{g}^{\rm W}=
 \begin{pmatrix}
 \partial^2_{\widetilde{S}_+} M &  \partial_{\widetilde{S}_+}\partial_Q M\\
 \partial_{\widetilde{S}_+}\partial_Q M &  \partial^2_Q M
\end{pmatrix},
\end{equation}
because Eq.\ \eqref{eq:met-Weinhold} and Eq.\ \eqref{eq:met-Ruppeiner} are related to each other via a conformal factor \cite{Ruppeiner:1995zz}.
Therefore, $\mathcal{R}_{\rm W}$ has the same singularities with those of $\mathcal{R}_{\rm R}$, see Fig.\ \ref{fig:Rth-W-Q},
where $Q_1/(2M)$ now corresponds to the zero of 
$(\partial_{\widetilde{S}_+} \partial_Q M)^2-(\partial_{\widetilde{S}_+}^2 M)(\partial_Q^2 M)$.
In addition, the singularity, $Q_1/(2M)$, coincides \cite{Quevedo:2023vip} with the critical point of first-order phase transition, which corresponds to the zeros of  $(\partial_M\partial_Q {\widetilde{S}_+})^2-(\partial^2_M {\widetilde{S}_+})(\partial^2_Q {\widetilde{S}_+})$ for Ruppeiner's geometry and $(\partial_{\widetilde{S}_+} \partial_Q M)^2-(\partial_{\widetilde{S}_+}^2 M)(\partial_Q^2 M)$ for Weinhold's geometry.

\begin{figure}[!htb]
     \centering         \includegraphics[width=.6\textwidth]{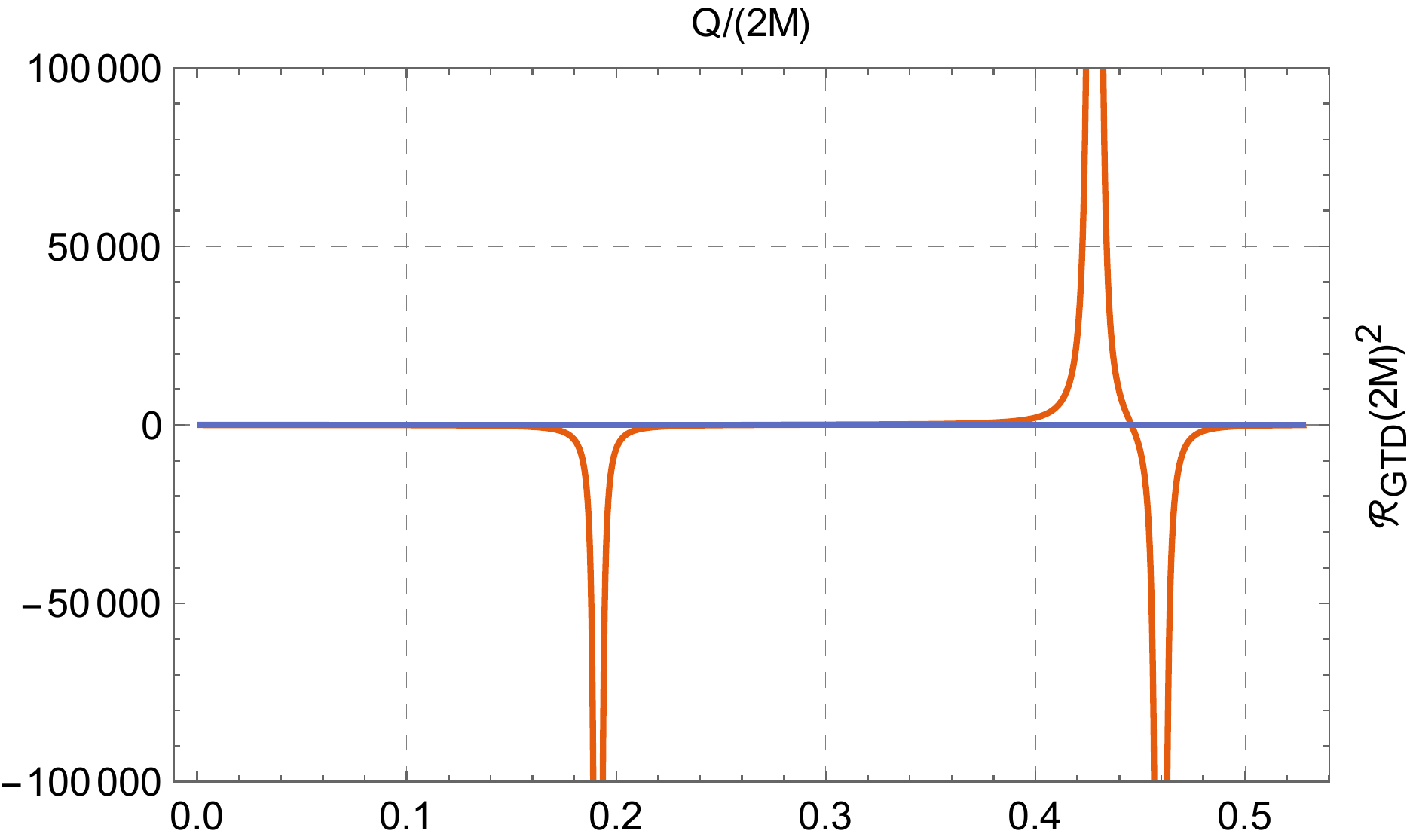}
   \captionsetup{width=.9\textwidth}
       \caption{Thermal curvature of GTD geometry.}
        \label{fig:Rth-GTD-Q}
\end{figure}

Another way to introduce a thermal metric was proposed in Refs.~\cite{Quevedo:2006xk,Quevedo:2007mj}, which is called Geometrothermodynamics (GTD).
This theory is invariant under the Legendre transformation, which leads to three metrics.
One of them is \cite{Tharanath:2014naa}
\begin{equation}
\mathfrak{g}^{\rm GTD}=
\left(
{\widetilde{S}_+} \partial_{\widetilde{S}_+} M+ Q \partial_Q M
\right)
 \begin{pmatrix}
- \partial^2_{\widetilde{S}_+} M &  0\\
 0 &  \partial^2_Q M
\end{pmatrix},
\end{equation}
which belongs to the type-II of Ref.\ \cite{Quevedo:2023vip} and 
the singularities of corresponding thermal curvatures are expected to coincide with the critical points of the second-order phase transition.
The thermal curvature, $\mathcal{R}_{\rm GTD}$, is exhibited in Fig.\ \ref{fig:Rth-GTD-Q}.
We note that there are three singularities, 
\begin{equation}
Q_3/(2M) \approx 0.191286,\qquad
Q_4/(2M) \approx 0.428243,\qquad
Q_5/(2M) \approx 0.459753,  
\end{equation}
where $Q_4/(2M)$  coincides with the singularity of heat capacity $Q_c/(2M)$ and is the zero of $\partial^2_{\widetilde{S}_+} M$, whereas
$Q_3/(2M)$ and $Q_5/(2M)$ are zeros of $\partial^2_Q M$, which have no counterparts in heat capacity.
In a word, $\mathcal{R}_{\rm GTD}$ provides more singularities than heat capacity.
This phenomenon is the same for some other thermodynamic metrics, such as in Ref.\ \cite{Hendi:2015rja}. As long as there are multiple types of singularities in the thermodynamic curvature that correspond to the zeros of formulas rather than $\partial^2_{\widetilde{S}_+} M$ (see Tab.\ \ref{tab:zeros}),  we can always reach the same conclusion as above.

\begin{table}[bht!]
\begin{center}
\begin{tabular}{l*{9}{c}r}
\hline\hline
 $\partial_{\widetilde{S}_+}^2 M$ & $\partial_Q^2 M$ & ${\widetilde{S}_+}$ & $\partial_{\widetilde{S}_+} M$ & $(\partial_M\partial_Q {\widetilde{S}_+})^2-\partial^2_M {\widetilde{S}_+}\partial^2_Q {\widetilde{S}_+}$ & ($\partial_{\widetilde{S}_+} \partial_Q M)^2-\partial_{\widetilde{S}_+}^2 M\partial_Q^2 M$  \\
\hline
 0.428243 & 0.191286 & $2^{2/3}/3$ & $2^{2/3}/3$ & 0.13358 & 0.13358   \\
 * & 0.331191 & * & * &  * & * \\
\hline
\end{tabular}
\caption{The first row displays the formulas that appear in the denominators of curvature invariants in different thermodynamic geometries. 
The second and third rows give the zeros of different formulas, where the stars in the third row imply that no real roots exist.}\label{tab:zeros}
\end{center}
\end{table}

We end this subsection with a comment on the correspondence between the divergence of heat capacity and the divergence of GTD curvature invariants.
Unlike previous works by others, we interpret the RBH in terms of $F(R)$ gravity rather than Einstein's gravity. 
This interpretation has its advantage because it guarantees the consistency of the Wald entropy $S_{\rm W}$ and the entropy calculated by the usual thermodynamic formula, $\displaystyle\widetilde{S}_+=\int \dif M/T$,
but it explains only partially the connection between the divergence of heat capacity and the divergence of curvature invariants in the GTD. The reason comes from the complicated nature of RBH thermodynamics, which is related to various factors, such as the interpretation and parameterization of spacetime metrics, as well as the construction of thermal metrics. Our model suggests that it is important to consider the poles of curvature invariants, i.e., the zeros in the denominators of curvature invariants, in relation to heat capacity when constructing a thermal metric.


\section{Conclusion and outlook}
\label{sec:conclusion}

In this paper, we propose an RBH model that can be used to simulate the final state of an SBH. 
Our motivation is to investigate the physical differences between RBHs and SBHs as a result of singularities' presence or absence. 
We also interpret this RBH model in terms of $F(R)$ gravity, 
where the matter sources are a nonlinear scalar field and a magnetic monopole in nonlinear electrodynamics. 

The interpretation by $F(R)$ gravity resolves the well-known problem of RBHs, 
namely the inconsistency between the Wald entropy and the entropy calculated by the usual thermodynamic formula.
Moreover, we find that 
the four energy conditions of two matter constituents are violated in varying degrees, 
especially the DEC which is completely violated outside the outer horizon.

Further, we investigate the dynamics and thermodynamics of our model in order to examine its physical properties.
On the side of dynamics, we study the perturbation of a massless scalar field on this BH background. 
We compute the corresponding QNFs by the 13th-order WKB approach with a Pad\'e approximation, 
and their asymptotic limit by the monodromy method.
We note from the spectrum of QNFs 
that the RBH as the final state is also the most stable dynamically,
and has the maximum oscillation frequency.
As to the asymptotic QNFs, the monodromy method cannot give any valid predictions. 
The reason for this is not the disappearance of singularity, but the merging of the inner and outer horizons on the Stokes portraits.

On the side of thermodynamics, we investigate the phase transition in addition to explaining the entropy shift by $F(R)$ gravity.
We are attempting to discover a connection between the phase transition predicted by heat capacity and three types of thermodynamic curvatures, 
particularly based on self-consistent entropy.
We draw two conclusions here.
The first is that the divergence of heat capacity does not coincide with the divergences of both Ruppeiner and Weinhold curvatures in our model, 
but that it appears as one of the divergent points in the GTD curvature.
In other words, the GTD curvature predicts more critical points of second-order phase transitions than the heat capacity.
The second conclusion is that the regular state appears as a divergent point in both Ruppeiner and Weinhold curvatures, 
indicating that it is also a critical point of first-order phase transitions according to the current study on the GTD.

Finally, we mention that the present paper focuses just on a static and spherically symmetric BH 
that evolves to its regular state when the charge or horizon radius goes to its extreme value --- a regular but non-rotating BH. 
Next, we plan to investigate the evolution of rotating BHs \cite{ghosh:2022gka}, 
especially the issue of how a singular rotating BH evolves to its regular counterpart, together with the relevant topics,  
such as the superradiance of rotating BHs \cite{Yang:2022uze}.
To this end, the Newman-Janis algorithm (NJA) \cite{Azreg-Ainou:2014pra,Azreg-Ainou:2014aqa}
will be applied to Eqs.\ \eqref{eq:metric} and \eqref{eq:shape} in order to construct the metrics of rotating BHs. 
This work is in progress.

\section*{Acknowledgement}

H.-W. Hu is grateful to C.-J. Fang, X.-H. Zhang and Y.-M. Xue for their useful discussions. This work was supported in part by the National Natural Science Foundation of China under Grant No. 12175108.


\appendix

\section{Derivations of asymptotic quasinormal frequencies (Eq. (\ref{asmpqnfs}) and Eq. (\ref{eq:qnf-Q}))}
\label{app:monodromy}

We start with Eq.\ \eqref{eq:sol-r0}
and apply the asymptotic behavior of the Bessel functions,
\begin{equation}
    J_{\pm\frac{1}{6}}(\omega z)\sim 
    \sqrt{\frac{2}{\pp \omega z}}
    \cos\left(\omega z -\alpha_\pm
    \right),\quad |\omega z|\gg 1,
\end{equation}
with
\begin{equation}
    \alpha_\pm = \frac{\pp}{4}\left(1\pm \frac{1}{3}\right),
\end{equation}
to rewrite the solution of master equations,
\begin{equation}
\begin{split}
    \Psi & \sim 2 A_1 \cos\left(\omega z -\alpha_{+}
    \right)+
    2 A_2 \cos\left(\omega z -\alpha_{-}
    \right) \\
    & = \left(A_1 \me^{-\mi\alpha_+}+A_2 \me^{-\mi\alpha_-} \right) \me^{\mi \omega z}
    +\left(A_1 \me^{\mi\alpha_+}+A_2 \me^{\mi\alpha_-} \right) \me^{-\mi \omega z}.
\end{split}
\end{equation}
One of the boundary conditions in Eq.\ \eqref{eq:boundary},
\begin{equation}
    \Psi \sim \me^{-\mi \omega z},\quad
    z \to +\infty
\end{equation}
leads to an equality 
\begin{equation}
\label{eq:cond-1}
    A_1 \me^{-\mi\alpha_+}+A_2 \me^{-\mi\alpha_-} =0,
\end{equation}
which will be applied later to determine the asymptotic quasinormal frequency.

The angle information is shown in Fig.\ \ref{fig:stokes-phi}, 
where $r$ rotates $\pp$ around $r_0=0$, i.e., 
$z$ rotates $3\pp$. 
Considering the property of the Bessel functions, 
\begin{equation}
    \sqrt{2\pp \me^{3\mi \pp}\omega z} J_{\pm \frac{1}{6}} \left(\me^{3\mi \pp}\omega z\right)
    =\me^{\frac{3\pp\mi}{2}(1\pm \frac{1}{3})}\sqrt{2\pp \omega z} J_{\pm \frac{1}{6}} \left(\omega z\right)
    \sim 2\me^{6\mi\alpha_\pm} \cos(\omega z-\alpha_\pm),
\end{equation}
we obtain the asymptotic solution after such a rotation,
\begin{equation}
\begin{split}
    \Psi & \sim 2 A_1 \me^{6\mi\alpha_+} \cos(-\omega z-\alpha_+)+
    2 A_2 \me^{6\mi\alpha_-} \cos(-\omega z-\alpha_-)\\
    &= (A_1 \me^{7\mi \alpha_+}+
    A_2 \me^{7\mi \alpha_-})\me^{\mi \omega z}
    +(A_1 \me^{5\mi \alpha_+}+
    A_2 \me^{5\mi \alpha_-})\me^{-\mi \omega z}.
\end{split}
\end{equation}
Then, closing the contour and comparing it with the rotation around the outer horizon, we arrive at
\begin{equation}
\label{eq:cond-2}
    \frac{A_1 \me^{5\mi \alpha_+}+A_2\me^{5\mi \alpha_-}}{A_1 \me^{\mi \alpha_+}+A_2\me^{\mi \alpha_-}}\me^{-\frac{\pp\omega}{\kappa}}=\me^{\frac{\pp\omega}{\kappa}}.
\end{equation}
At last, we derive the formula of asymptotic quasinormal frequencies, Eq.~(\ref{asmpqnfs}) associated with the Stokes lines in Fig.\ \ref{fig:stokes-Q-0}, from Eqs.\ \eqref{eq:cond-1} and \eqref{eq:cond-2}.

\begin{figure}[!htb]
     \centering         \includegraphics[width=.6\textwidth]{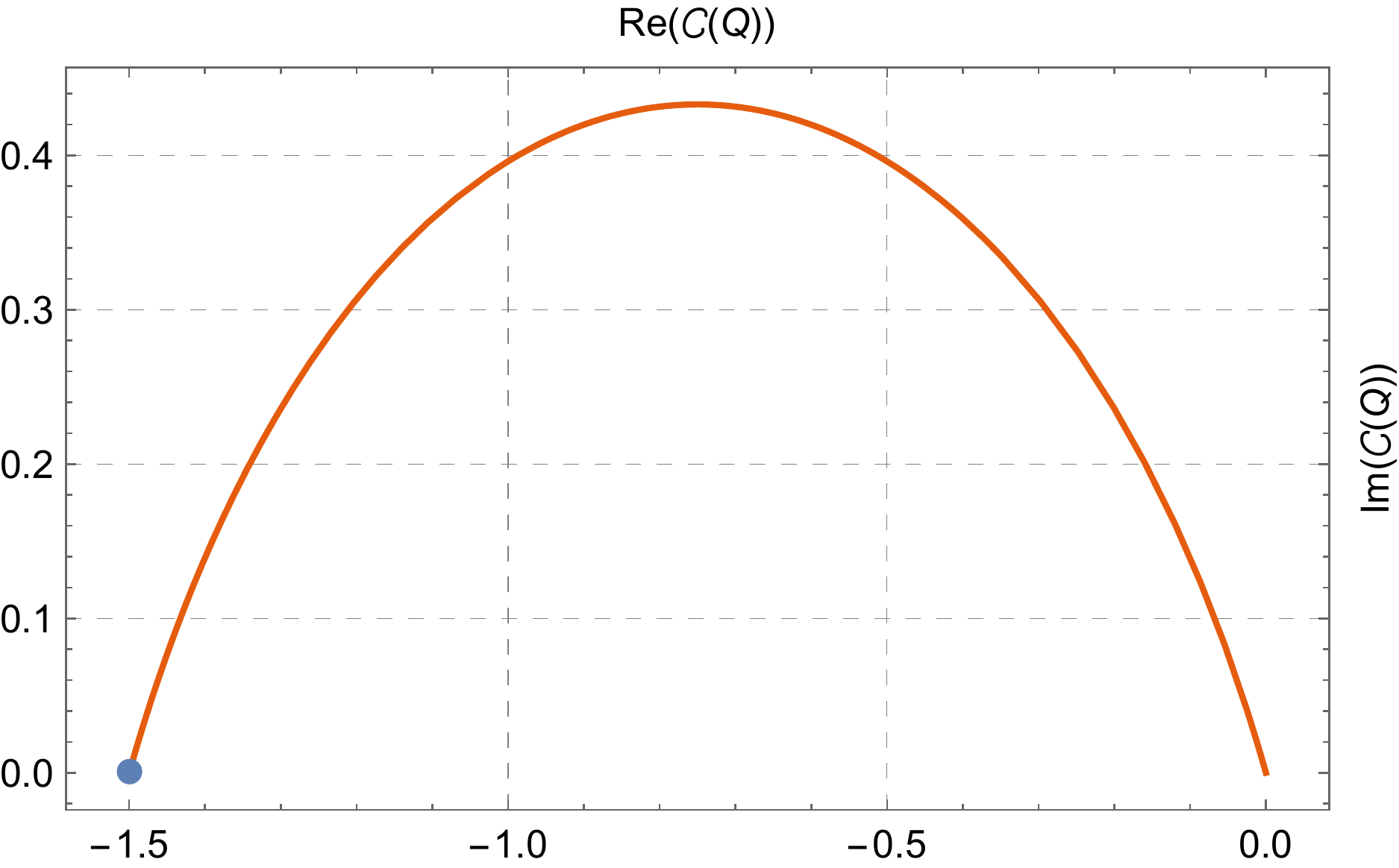}
   \captionsetup{width=.9\textwidth}
       \caption{Complex function $\mathcal{C}(Q)$. The blue point denotes the place where $Q=Q_{\rm ext}$.}
        \label{fig:c-function}
\end{figure}

Now we calculate the asymptotic quasinormal frequencies by using the monodromy method, see
Fig.\ \ref{fig:stokes-Q-0.3-2}.
At first, we compute the tortoise coordinate at $r=r_3$,
\begin{equation}
    z\sim 
    \mathcal{C}(Q)\left(r+(-1)^{2/3} Q\right)^2+
    O\left[\left(r+(-1)^{2/3} Q\right)^3\right],
\end{equation}
where $\mathcal{C}(Q)$ is a complex function of $Q$, 
\begin{equation}
   \mathcal{C}(Q)= -\frac{3 Q}{2 \left[-2 (-1)^{1/3} Q Q_{\rm{ext}}+(-1)^{1/3} Q_{\rm{ext}}^2+(-1)^{1/3} Q^2+Q\right]},
\end{equation}
see also Fig.\ \ref{fig:c-function}.
Then, we expand the effective potential at 
$r=r_3$ and obtain
\begin{equation}
  V_{\rm eff}= -\frac{(-1)^{1/3} \left[(-1)^{1/3} (Q-Q_{\rm ext})^2+Q\right]^2}{9 Q^3 \left(r+(-1)^{2/3} Q\right)^3}+O\left[\left(r+(-1)^{2/3} Q\right)^{-2}\right].
\end{equation}
Thus the master equation becomes
\begin{equation}
	  -\frac{\dif^2 \Psi}{\dif z^2} -\frac{\mathcal{V}_0}{z^{3/2}} \Psi = \omega^2 \Psi, 
\end{equation}
where the prefactor takes the form,
\begin{equation}
\label{eq:leading-potential}
    \mathcal{V}_0= -\frac{(-1)^{1/3} \mathcal{C}^{3/2} }{9 Q^3 }
    \left[(-1)^{1/3} (Q-Q_{\rm ext})^2+Q\right]^2.
\end{equation}
By imitating the Hayward BH \cite{Lan:2022qbb}, we obtain the leading order of the perturbative solution,
\begin{equation}
    \Psi_0\sim B_1 \sqrt{2\pp\omega z} J_{\frac{1}{2} \nu}(z \omega )+
    B_2 \sqrt{2\pp\omega z} J_{-\frac{1}{2} \nu}(z \omega),
\end{equation}
where $\nu=\sqrt{1-4 \mathcal{V}_0}$.
The distance between $r=r_2$ and $r=r_3$ is
$\Im[r_2-r_3]=\sqrt{3} Q$, 
i.e., the shift of $z$ is $z\to z + \sqrt{3} Q \mi$. 
After repeating a similar procedure, we finally derive Eq.~(\ref{eq:qnf-Q}) that is associated with the Stokes portrait Fig.\ \ref{fig:stokes-Q-0.3-2}.

\section{Action of matter  sources}
\label{sec:Analytical-matter}

Let us start with the equation of motion Eq.\ \eqref{eq:eq-of-motion} under the consideration of the action of matter sources described by Eq.\ \eqref{eq:action-matter}. The nontrivial components of Eq.\ \eqref{eq:eq-of-motion} for the given shape function Eq.\ \eqref{eq:shape} read
\begin{eqnarray}
\mathcal{L}+2 V&=&6 Q ^ { 3 } r ^ { 3 } \Big\{\left(Q^3+r^3\right)^4+12 r\Big[21 Q^9+Q^3(34-27 r) r^5\nonumber \\
& &\;\;\;\;\;\;\;\;\;\;\;\;+r^8(-4+3 r)-Q^6 r^2(25+9 r)\Big] \alpha\Big\}\nonumber\\
& &+24 Q^3\Big[Q^{12}+Q^3(225-77 r) r^8+7 r^{11}(-7+4 r)\nonumber\\
& &\;\;\;\;\;\;\;\;\;\;\;\;-15 Q^6 r^5(7+5 r)+Q^9 r^2(-1+31 r)\Big] \alpha \delta\nonumber\\
& &-24 Q^3 r\left(Q^9+33 Q^6 r^3-117 Q^3 r^6+40 r^9\right) \alpha \delta^2\nonumber\\
& &-2 r^2\left(-2 Q^3+r^3\right)\left(Q^3+r^3\right)^4 \delta,\label{eq:Einstein-1}\\
\frac{r}{4}  \mathcal{L}^{\prime}&=&\frac{1}{r^3\left(Q^3+r^3\right)^6}\Bigg\{9 Q ^ { 3 } r ^ { 4 } \Big[-r^2\left(Q^3+r^3\right)^3-2\Big(14 Q^9+2(3-2 r) r^8\nonumber\\
& &\;\;\;\;\;\;\;\;\;\;\;\;\;\;\;\;\;\;\;\;\;\;\;\;\;\;\;\;\;\;\;\;\;\;\;+3 Q^3 r^5(-9+2 r)+12 Q^6 r^2(1+2 r)\Big) \alpha\Big]\nonumber\\
& &\;\;\;\;\;\;\;\;\;\;\;\;\;\;\;\;\;\;\;\;\;\;\;-\Big[r^2\left(Q^3+r^3\right)^3\left(Q^6+8 Q^3 r^3-2 r^6\right)\nonumber\\
& &\;\;\;\;\;\;\;\;\;\;\;\;\;\;\;\;\;\;\;\;\;\;\;\;\;\;\;+12 Q^3\Big(Q^{12}+7(5-2 r) r^{11}+3 Q^6 r^5(4+5 r)\nonumber\\
& &\;\;\;\;\;\;\;\;\;\;\;\;\;\;\;\;\;\;\;\;\;\;\;\;\;\;\;-2 Q^3 r^8(45+7 r)+2 Q^9 r^2(1+8 r)\Big) \alpha\Big] \delta\nonumber\\
& &\;\;\;\;\;\;\;\;\;\;\;\;\;\;\;\;\;\;\;\;\;\;\;-6 Q^3 r\left(Q^9-6 Q^6 r^3-78 Q^3 r^6+64 r^9\right) \alpha \delta^2\Bigg\},\label{eq:Einstein-2}\\
0&=&-12 Q^3 \alpha\left[-21 Q^6 r^4+Q^9 \delta+15 Q^3 r^6(8 r+7 \delta)-7 r^9(3 r+8 \delta)\right]\nonumber \\
& &+r^3\left(Q^3+r^3\right)^5 W \phi^{\prime 2},\label{eq:Einstein-3}
\end{eqnarray}
where a prime denotes the derivative with respect to $r$. 

At first, we obtain directly the  Lagrangian of nonlinear electromagnetic fields as the function of $r$ from Eq.\ \eqref{eq:Einstein-2},
\begin{equation}\label{eq:L-r}
\begin{aligned}
    \mathcal{L}(r)&=\frac{2}{9}\left\{\frac{72 \alpha \delta}{Q^3 r^3}+\frac{18\left(Q^3+24 \alpha\right) \delta}{Q^6 r}+\frac{54 \alpha \delta^2}{Q^6 r^2}+\frac{20 \sqrt{3} \pp \alpha\left(18 Q^3-7 Q \delta-7 \delta^2\right)}{Q^8}\right.\\
    &\;\;\;\;\;\;\;\;\;\;+\frac{972 Q^6 \alpha\left[Q^3-r \delta(2 r+\delta)\right]}{\left(Q^3+r^3\right)^5}-\frac{27 \alpha\left[8 Q^3(-1+3 r+3 \delta)-r \delta(4 r+5 \delta)\right]}{\left(Q^3+r^3\right)^3}\\
    &\;\;\;\;\;\;\;\;\;\;+\frac{81 Q^3 \alpha\left[-13 Q^3+r \delta(10 r+9 \delta)\right]}{\left(Q^3+r^3\right)^4}-\frac{40 \alpha\left(18 Q^3+7 Q \delta-7 \delta^2\right) \ln (Q+r)}{Q^8}\\
    &\;\;\;\;\;\;\;\;\;\;+\frac{9\left[3 Q^6+22 r \alpha \delta(r+\delta)-3 Q^3\left(24 r \alpha+r^2 \delta+16 \alpha \delta\right)\right]}{Q^3\left(Q^3+r^3\right)^2}\\
    &\;\;\;\;\;\;\;\;\;+\frac{40 \sqrt{3} \alpha\left(18 Q^3-7 Q \delta-7 \delta^2\right) \arctan\left(\dfrac{Q-2 r}{\sqrt{3} Q}\right)}{Q^8}\\
    &\;\;\;\;\;\;\;\;\;-\frac{2160 \alpha \delta \ln r }{Q^6}
    +\frac{6 r \alpha \delta(68 r+61 \delta)-18 Q^3\left(60 r \alpha+r^2 \delta+44 \alpha \delta\right)}{Q^6\left(Q^3+r^3\right)}
    \\
    &\;\;\;\;\;\;\;\;\;+\frac{20 \alpha\left(18 Q^3+7 Q \delta-7 \delta^2\right) \ln (Q^2-Q r+r^2)}{Q^8}\left.+\frac{720 \alpha \delta \ln \left(Q^3+r^3\right)}{Q^6}\right\}.
\end{aligned}
\end{equation}

Then, we derive the potential of scalar fields according to Eqs.\ \eqref{eq:Einstein-1} and \eqref{eq:L-r},
\begin{equation}
	\begin{aligned}
		V(r)&=\frac{1}{2}\Bigg\{\frac { 1 } { r ^ { 3 } ( Q ^ { 3 } + r ^ { 3 } ) ^ { 6 } } \bigg\{6 Q ^ { 3 } r ^ { 3 } \left\{\left(Q^3+r^3\right)^4+12 r\left[21 Q^9+Q^3(34-27 r) r^5\right.\right.\\
		&\;\;\;\;\;\;\;\;\;\;\;\;\;\;\;\;\;\;\;\;\;\;\;\;\;\;\;\;\;\;\;\;\;\;\;\;\;\;\;\;\;\;\;\;+r^8(-4+3 r)-\left.Q^6 r^2(25+9 r)\right] \alpha\Big\}\\
  &\;\;\;\;\;\;\;\;\;\;\;\;\;\;\;\;\;\;\;\;\;\;\;\;\;\;\;\;\;\;\;-2 r^2\left(-2 Q^3+r^3\right)\left(Q^3+r^3\right)^4 \delta\\
		&\;\;\;\;\;\;\;\;\;\;\;\;\;\;\;\;\;\;\;\;\;\;\;\;\;\;\;\;\;\;\;+24 Q^3\left[Q^{12}\right.-15 Q^6 r^5(7+5 r)+Q^3(225-77 r) r^8\\
  &\;\;\;\;\;\;\;\;\;\;\;\;\;\;\;\;\;\;\;\;\;\;\;\;\;\;\;\;\;\;\;\;\;\;\;\;\;\;\;\;\;\;\;\;+7 r^{11}(-7+4 r)+Q^9 r^2(-1+31 r)\big] \alpha \delta\\
		&\;\;\;\;\;\;\;\;\;\;\;\;\;\;\;\;\;\;\;\;\;\;\;\;\;\;\;\;\;\;\;-24 Q^3 r\left(Q^9+33 Q^6 r^3-117 Q^3 r^6+40 r^9\right) \alpha \delta^2\bigg\}\\
		&\;\;\;\;\;\;\;\;\;-\frac{2}{9}\left\{\frac{72 \alpha \delta}{Q^3 r^3}+\frac{18\left(Q^3+24 \alpha\right) \delta}{Q^6 r}+\frac{54 \alpha \delta^2}{Q^6 r^2}+\frac{20 \sqrt{3} \pp \alpha\left(18 Q^3-7 Q \delta-7 \delta^2\right)}{Q^8}\right.\\
		&\;\;\;\;\;\;\;\;\;\;\;\;\;\;\;\;\;\;\;-\frac{27 \alpha\left[8 Q^3(-1+3 r+3 \delta)-r \delta(4 r+5 \delta)\right]}{\left(Q^3+r^3\right)^3}\\
		&\;\;\;\;\;\;\;\;\;\;\;\;\;\;\;\;\;\;\;+\frac{9\left[3 Q^6+22 r \alpha \delta(r+\delta)-3 Q^3\left(24 r \alpha+r^2 \delta+16 \alpha \delta\right)\right]}{Q^3\left(Q^3+r^3\right)^2}\\
		&\;\;\;\;\;\;\;\;\;\;\;\;\;\;\;\;\;\;\;+\frac{6 r \alpha \delta(68 r+61 \delta)-18 Q^3\left(60 r \alpha+r^2 \delta+44 \alpha \delta\right)}{Q^6\left(Q^3+r^3\right)}\\
        &\;\;\;\;\;\;\;\;\;\;\;\;\;\;\;\;\;\;\;+\frac{972 Q^6 \alpha\left(Q^3-r \delta(2 r+\delta)\right)}{\left(Q^3+r^3\right)^5}+\frac{81 Q^3 \alpha\left(-13 Q^3+r \delta(10 r+9 \delta)\right)}{\left(Q^3+r^3\right)^4}\\
		&\;\;\;\;\;\;\;\;\;\;\;\;\;\;\;\;\;\;\;+\frac{40 \sqrt{3} \alpha\left(18 Q^3-7 Q \delta-7 \delta^2\right) \arctan\left(\dfrac{Q-2 r}{\sqrt{3} Q}\right)}{Q^8}-\frac{2160 \alpha \delta \ln r}{Q^6}\\
		&\;\;\;\;\;\;\;\;\;\;\;\;\;\;\;\;\;\;\;-\frac{40 \alpha\left(18 Q^3+7 Q \delta-7 \delta^2\right) \ln(Q+r)}{Q^8}+\frac{720 \alpha \delta \ln \left(Q^3+r^3\right)}{Q^6}\\
		&\;\;\;\;\;\;\;\;\;\;\;\;\;\;\;\;\;\;\;+\left.\frac{20 \alpha\left(18 Q^3+7 Q \delta-7 \delta^2\right) \ln \left(Q^2-Q r+r^2\right)}{Q^8}\right\}\Bigg\}.
	\end{aligned}
\end{equation}

At last, we solve $W \phi^{\prime 2}$ easily from Eq.~(\ref{eq:Einstein-3}),
\begin{equation}
W \phi^{\prime 2}=-\frac{12 Q^3 \alpha}{r^3\left(Q^3+r^3\right)^5}\left[-21 Q^6 r^4+Q^9 \delta+15 Q^3 r^6(8 r+7 \delta)-7 r^9(3 r+8 \delta)\right].\label{wphiprimesqrt}
\end{equation}

Moreover, 
we establish the relationship between $\mathcal{L}$ and the contraction of strength tensors $\mathcal{F}$ by using Eqs.\ \eqref{confieldten} and \eqref{eq:L-r},
\begin{equation}
    \begin{aligned}
        \mathcal{L}(\mathcal{F})&=\frac{2}{9 Q^8} \left.\Bigg\{36 \times 2^{1 / 4} |\mathcal{F}|^{3 / 4} Q^{7 / 2} \alpha \delta+9 \times 2^{3 / 4} |\mathcal{F}|^{1 / 4} Q^{3 / 2}\left(Q^3+24 \alpha\right) \delta\right.\\
        &\;\;\;\;\;\;\;\;\;\;\;\;\;\; +20 \sqrt{3} \pp \alpha\left(18 Q^3-7 Q \delta-7 \delta^2\right)+\dfrac{972 Q^{14} \alpha\left(Q^3-\dfrac{2 \sqrt{2} Q \delta}{\sqrt{|\mathcal{F}|}}-\dfrac{2^{1 / 4} \sqrt{Q} \delta^2}{|\mathcal{F}|^{1 / 4}}\right)}{\left(\dfrac{2^{3 / 4} Q^{3 / 2}}{|\mathcal{F}|^{3 / 4}}+Q^3\right)^5}\\
        &\;\;\;\;\;\;\;\;\;\;\;\;\;\;+\dfrac{81 Q^{11} \alpha\left(-13 Q^3+\dfrac{10 \sqrt{2} Q \delta}{\sqrt{|\mathcal{F}|}}+\dfrac{92^{1 / 4} \sqrt{Q} \delta^2}{|\mathcal{F}|^{1 / 4}}\right)}{\left(\dfrac{2^{3 / 4} Q^{3 / 2}}{|\mathcal{F}|^{3 / 4}}+Q^3\right)^4}+27 \sqrt{2} \sqrt{|\mathcal{F}|} Q \alpha \delta^2\\
        &\;\;\;\;\;\;\;\;\;\;\;\;\;\; -\frac{6 |\mathcal{F}|^{1 / 4} Q\left[\sqrt{2} \sqrt{Q}\left(3 Q^3-68 \alpha\right) \delta+132 \sqrt{|\mathcal{F}|} Q^{5 / 2} \alpha \delta\right]}{2^{3 / 4}+|\mathcal{F}|^{3 / 4} Q^{3 / 2}} \\
        &\;\;\;\;\;\;\;\;\;\;\;\;\;\;+\frac{6 \times 2^{1 / 4} |\mathcal{F}|^{1 / 2} Q   \alpha\left(180 Q^3-61 \delta^2\right)}{2^{3 / 4}+|\mathcal{F}|^{3 / 4} Q^{3 / 2}}
        \\
        &\;\;\;\;\;\;\;\;\;\;\;\;\;\;+\frac{27 |\mathcal{F}|^{7 / 4} Q^4 \alpha\left[8 \sqrt{|\mathcal{F}|} Q^{5 / 2}(1-3 \delta)+4 \sqrt{2} \sqrt{Q} \delta+2^{1 / 4} |\mathcal{F}|^{1 / 4}\left(-24 Q^3+5 \delta^2\right)\right]}{\left(2^{3 / 4}+|\mathcal{F}|^{3 / 4} Q^{3 / 2}\right)^3}\\
        &\;\;\;\;\;\;\;\;\;\;\;\;\;\;+\frac{9\sqrt{2} |\mathcal{F}| Q^3\left(-3 Q^3+22 \alpha\right) \delta+3 |\mathcal{F}|^{3 / 2}\left(Q^8-16 Q^5 \alpha \delta\right)}{\left(2^{3 / 4}+|\mathcal{F}|^{3 / 4} Q^{3 / 2}\right)^2}\\
        &\;\;\;\;\;\;\;\;\;\;\;\;\;\;+\frac{18 \times 2^{1 / 4} |\mathcal{F}|^{5 / 4} Q^{5 / 2} \alpha\left(-36 Q^3+11 \delta^2\right)}{\left(2^{3 / 4}+|\mathcal{F}|^{3 / 4} Q^{3 / 2}\right)^2}\\
        &\;\;\;\;\;\;\;\;\;\;\;\;\;\;+40 \sqrt{3} \alpha\left(18 Q^3-7 Q \delta-7 \delta^2\right) \arctan\left(\frac{1-\frac{22^{1 / 4}}{|\mathcal{F}|^{1 / 4} \sqrt{Q}}}{\sqrt{3}}\right)\\
        &\;\;\;\;\;\;\;\;\;\;\;\;\;\;+540 Q^2 \alpha \delta\left[\ln \left(\frac{|\mathcal{F}|}{2}\right)-2 \ln Q\right]+720 Q^2 \alpha \delta \ln \left(\frac{2^{3 / 4} Q^{3 / 2}}{|\mathcal{F}|^{3 / 4}}+Q^3\right)\\
        &\;\;\;\;\;\;\;\;\;\;\;\;\;\;+20 \alpha\left(18 Q^3+7 Q \delta-7 \delta^2\right) \ln \left[Q\left(\frac{\sqrt{2}}{\sqrt{|\mathcal{F}|}}-\frac{2^{1 / 4} \sqrt{Q}}{|\mathcal{F}|^{1 / 4}}+Q\right)\right]\\
        &\;\;\;\;\;\;\;\;\;\;\;\;\;\;-40 \alpha\left(18 Q^3+7 Q \delta-7 \delta^2\right) \ln \left(\frac{2^{1 / 4} \sqrt{Q}}{|\mathcal{F}|^{1 / 4}}+Q\right) \Bigg\}.
    \end{aligned}
\end{equation}

For the scalar field, we derive it in terms of the ansatz Eq.\ \eqref{eq:scalar-field},
\begin{equation}
    \left|\phi(r)\right|=\int \dif r \sqrt{\frac{1}{\left(Q^3+r^3\right)^{4 / 3}}+\frac{\delta}{r^2\left(Q^3+r^3\right)^{4 / 3}}}+\phi_0,
\end{equation}
where $\phi_0$ is an integration constant 
which can be fixed by  $\phi\to 0$ as $r$ approaches infinity.
In addition, we separate $W$ from $W\phi^{\prime 2}$ according to Eqs.\ \eqref{eq:scalar-field} and \eqref{wphiprimesqrt}, 
\begin{equation}
W(r)=\frac{12 Q^3 \alpha\left(-21 Q^6 r^4+Q^9 \delta+15 Q^3 r^6(8 r+7 \delta)-7 r^9(3 r+8 \delta)\right)}{r\left(Q^3+r^3\right)^{11 / 3}\left(r^2+\delta\right)}.
\end{equation}
Taking radical coordinate $r$ as a parameter, we solve the dependences of $V$ and $W$ on $\phi$ numerically and draw them in Fig.\ \ref{fig:sep-scalar-2}.

\bibliographystyle{utphys}

\bibliography{references}

\end{document}